%% file: gk_mdf_alone_referee2.tex
\documentclass[12,preprint]{aastex}
\usepackage{graphicx}
\usepackage{natbib}
\usepackage{float}
\usepackage{lscape}
\usepackage{calc}
\usepackage{multirow}
\usepackage{color}

\shorttitle{METALLICITY DISTRIBUTION FUNCTION OF G AND K DWARFS} 
\shortauthors{SCHLESINGER ET AL.}

\newcommand{\msun}{M$_\odot$}

\begin{document}
\title{The Metallicity Distribution Functions of SEGUE G and K dwarfs:
Constraints for Disk Chemical Evolution and Formation}

\author{Katharine J.~Schlesinger\altaffilmark{1}
Jennifer A.~Johnson\altaffilmark{2,3},
Constance M.~Rockosi\altaffilmark{4},
Young Sun Lee\altaffilmark{5},
Heather L.~Morrison\altaffilmark{6},
Ralph Sch{\"o}nrich\altaffilmark{7},
Carlos Allende Prieto\altaffilmark{8,9},
Timothy C. Beers\altaffilmark{10,11},
Brian Yanny\altaffilmark{12},
Paul Harding\altaffilmark{6},
Donald P. Schneider\altaffilmark{13},
Cristina Chiappini\altaffilmark{14,15}, 
Luiz N. da Costa\altaffilmark{14,16}, 
Marcio A.G. Maia\altaffilmark{14,16}, 
Ivan Minchev\altaffilmark{15}, 
Helio Rocha-Pinto\altaffilmark{14,17},  
Bas{\'i}lio X. Santiago\altaffilmark{14,17}
}

\altaffiltext{1}{Research School of Astronomy and Astrophysics, The Australian National University, Weston, ACT 2611, Australia}
\altaffiltext{2}{Department of Astronomy, The Ohio State University, 140 W 18th Ave, Columbus, OH 43210, USA}
\altaffiltext{3}{Center for Cosmology and AstroParticle Physics, The Ohio State University, 191 West Woodruff Ave, Columbus, OH 43210, USA} 
\altaffiltext{4}{UCO/Lick Observatory, University of California, Santa Cruz, CA 95064, USA}
\altaffiltext{5}{Department of Astronomy, New Mexico State University, Las Cruces, NM 88003}
\altaffiltext{6}{Department of Astronomy, Case Western Reserve University, Cleveland, OH 44106, USA}
\altaffiltext{7}{Max Planck Institute for Astrophysics, Garching Karl-Schwarzschild-Strasse 1, Postfach 1317, D-85741 Garching, Germany}
\altaffiltext{8}{Instituto de Astrof\'{i}sica de Canarias, 38205 La Laguna, Tenerife, Spain}
\altaffiltext{9}{Departamento de Astrof\'{i}sica, Universidad de La Laguna, 38206 La Laguna, Tenerife, Spain} 
\altaffiltext{10}{National Optical Astronomy Observatory, Tucson, AZ 85719, USA}
\altaffiltext{11}{Department of Physics and Astronomy and JINA: Joint Institute for Nuclear Astrophysics,
Michigan State University, East Lansing, MI 48824, USA}
\altaffiltext{12}{Fermi National Accelerator Laboratory, P.O. Box 500, Batavia, IL 60510, USA}
\altaffiltext{13}{Department of Astronomy and Astrophysics, Penn State University, 408A Davey Laboratory, University Park, PA 16802} 
\altaffiltext{14}{Laborat{\'o}rio Interinstitucional de e-Astronomia - LIneA, Rua Gal. Jos{\'e} Cristino 77, 20921-400 Rio de Janeiro, Brazil}
\altaffiltext{15}{Leibniz-Institut f{\"u}r Astrophysik Potsdam, An der Sternwarte 16, 14482 Potsdam, Germany} 
\altaffiltext{16}{Observat{\'o}rio Nacional, Rua Gal. Jos{\'e} Cristino 77, 22460-040 Rio de Janeiro, Brazil} 
\altaffiltext{17}{Universidade Federal do Rio de Janeiro, Observat{\'o}rio do Valongo, Lad. Pedro Ant{\^o}nio 43, 20080-090 Rio de Janeiro, Brazil}

\begin{abstract}

We present the metallicity distribution function (MDF) for 24,270 G
and 16,847 K dwarfs at distances from 0.2 to 2.3 kpc from the Galactic
plane, based on spectroscopy from the Sloan Extension for Galactic
Understanding and Exploration (SEGUE) survey. This stellar sample is
significantly larger in both number and volume than previous
spectroscopic analyses, which were limited to the solar vicinity,
making it ideal for comparison with local volume-limited samples and
Galactic models. For the first time, we have corrected the MDF for the
various observational biases introduced by the SEGUE target selection
strategy. The SEGUE sample is particularly notable for K dwarfs, which
are too faint to examine spectroscopically far from the solar
neighborhood. The MDF of both spectral types becomes more metal-poor
with increasing $|Z|$, which reflects the transition from a sample
with small [$\alpha$/Fe] values at small heights to one with enhanced
[$\alpha$/Fe] above 1 kpc. Comparison of our SEGUE distributions to
those of two different Milky Way models reveals that both are more
metal-rich than our observed distributions at all heights above the
plane. Our unbiased observations of G and K dwarfs provide valuable
constraints over the $|Z|$-height range of the Milky Way disk for
chemical and dynamical Galaxy evolution models, previously only
calibrated to the solar neighborhood, with particular utility for
thin- and thick-disk formation models.

\end{abstract}


\section{Introduction}
\label{sec:intro} 

Measuring the metallicity distribution of stars in the Milky Way disk
is imperative for understanding its chemical and dynamical
evolution. Cool stars, such as G and K dwarfs, have lifetimes
comparable to the age of the Galaxy, providing a complete fossil
record of chemical development. By measuring the metallicity
distribution of these cool stars, we provide constraints on the disk's
star-formation history, and how it varies with respect to time and
location. The metal-poor end of the metallicity distribution function
(MDF) reveals information about the earliest era of star formation in
the Galaxy, such as the mass function of the earliest stars, the
extent of chemical pre-enrichment, and the rates and yields of core
collapse supernovae. The metal-rich end of the MDF reflects recent
Galaxy conditions, such as the present-day mass function and the
frequency and yields of both type Ia and II supernovae. The MDF
provides information about the regulation of star formation and the
merger and accretion history of the Galaxy.

There is a wide range of chemical and dynamical evolution simulations
that model the structure of the Galactic disk. These vary with respect
to properties such as the star-formation history, initial mass
function, prominence of inflows and outflows, and the likelihood of
mergers. All use observed samples to test their predictions of the
metallicity distribution. Until recently, we lacked an adequate
observational sample to quantitatively test these models beyond the
solar neighborhood. In this work, we use the Sloan Extension for
Galactic Understanding and Exploration (SEGUE, \citealt{yanny09})
survey to determine an unbiased MDF of cool dwarfs over a large volume
of the disk, from 0.2 to 2.3 kpc from the plane of the Galaxy.

Initial examination of the chemistry of local G dwarfs revealed the
``G-dwarf problem'' \citep{vandenbergh62, pp75, wg95, rpm96,
  favata97}. Early models of star formation and chemical evolution,
such as the simple closed box model of \citet{schmidt}, predicted many
more low-metallicity G dwarfs than were actually observed. It was
initially suspected that the G-dwarf problem arose from observational
biases, namely that metal-rich stars are brighter and thus likely to
be over-represented in a magnitude-limited sample. However, the
deficiency of low-metallicity stars persisted in later observational
samples that corrected for these biases. Work such as the
Geneva-Copenhagen survey (GCS) of the solar neighborhood
\citep{jorgensen00, nordstrom04, holmberg07, holmberg09, casagrande11}
indicated that for a large volume-complete and kinematically-unbiased
sample, the simple closed box model over-predicted the number of cool
metal-poor dwarfs even more than originally reported.

We know that many of the assumptions in the simple closed box model
are unphysical, in particular instantaneous recycling and the absence
of gas flows. More recent models have updated and experimented with
the model parameters to better match the observed metallicity
distribution, for example, adding inflows of low-metallicity material
\citep{larson72, chiappini97}, varying the formation time scales of
different Galaxy components \citep{chiappini01}, or commencing star
formation from pre-enriched material \citep{truran71}. Models have
also experimented with the initial mass function (IMF)
\citep{chiappini00, romano05}; a proposed metallicity-dependent IMF
will produce more high- than low-mass stars at early times, resulting
in fewer cool metal-poor stars.

This exploration of the G-dwarf problem prompted observations of the
MDF of cooler stars. K and M dwarfs have even longer lifetimes than
G-dwarf stars. There is a chance that more metal-rich G dwarfs have
evolved off of the main sequence, leading to a metallicity bias in the
MDF, which will not occur for cooler spectral types. In addition,
comparing the MDFs of different spectral types is particularly useful
for constraining the variation in the IMF. For example, if more
low-mass stars are created as the metallicity of the environment
increases, there should be relatively fewer metal-poor K dwarfs than G
dwarfs. Work such as \citet{mould82}, \citet{favata97},
\citet{flynnmorell97}, \citet{rpmaciel98}, \citet{kotoneva02}, and
\citet{woolf12}, found that, just as with G-dwarf stars, the simple
closed box model predicted more metal-poor stars than were observed
for both K and M dwarfs, implying that the IMF and star-formation
history for these different spectral types are similar to one another.

Measurements of the G- and K-dwarf MDF have been mostly confined to
the solar neighborhood, limiting our understanding of disk properties
and evolution processes to the local volume. However, the chemical and
dynamical structure of the disk does not appear to be uniform with
respect to Galactic height, but rather composed of two components, a
thin and thick disk, of unclear interdependence. \citet{gilmore83}
first detected the thick disk in the Milky Way when they determined
that the stellar number density as a function of height above the
plane was best fit by two components\footnote[1]{The likely existence
  of a second stellar component close to the Galactic plane was
  previously noted by \citet{yoshii82}, who referred to it as a halo
  component, even though its inferred density was 10 times that of the
  local halo.  The \citet{yoshii82} normalization relative to the
  local thin disk (0.01--0.02) and scale height ($\sim$2 kpc) were
  commensurate with the values later determined by \citet{gilmore83}
  (0.02, and 1.5 kpc, respectively). See also \citet{yoshii87}.}, one
with a scale height of approximately 300 pc, and the second with a
scale height of 1350 pc.  Analyses by \citet{gilmore85, wg95}, and
\citet{chiba00} established that the two populations were distinct in
      [Fe/H]. The so-called thick disk was metal-poor, with a peak
      metallicity around [Fe/H]$=-$0.6, in contrast to the thin disk,
      with a peak metallicity typically around [Fe/H]$=-$0.2. Further
      observation of solar neighborhood samples revealed that the two
      populations were distinct in kinematics \citep{soubiran03}, age
      \citep{fuhrmann98}, and $\alpha$-abundance \citep{fuhrmann98,
        prochaska00, bensby03, bensby05, reddy06}. However,
      \citet{norris_ryan} and recent work by \citet{bovy11, bovy11b}
      and \citet{liu12}, focusing on [$\alpha$/Fe] vs. [Fe/H] and
      kinematics, questioned whether or not the two components were
      actually separable from one another, proposing that the Milky
      Way disk has a ``thicker disk component,'' rather than two
      distinct structures. The debate over the basic structure of the
      Milky Way disk emphasizes the need for an unbiased spectroscopic
      sample that extends over a large volume of the disk, such that
      we can investigate the metallicity structure of both of the
      proposed components with a uniform large data set. Furthermore,
      it is critical to use our own Galaxy to disentangle the
      mechanisms behind disk development because the thick disk
      appears to be a regular feature in numerous galaxies
      \citep{burstein79, dalcanton02, yoachim08a, yoachim08b},
      including those at redshifts as high as z$\sim$3
      \citep{elmegreen06}. In this work, we investigate and constrain
      the chemical structure of the Milky Way as a whole, including
      both thin- and thick-disk components.

\subsection{Previous Analyses of the Chemical Structure of the Disk} 

Past observations of the MDF have been limited in accuracy, sample
size, and, most importantly, volume. Due to the low luminosity of cool
stars, analyses such as \citet{pp75}, \citet{gilmore85}, \citet{wg95},
\citet{rpm96}, \citet{flynnmorell97}, \citet{rpmaciel98},
\citet{jorgensen00}, and \citet{kotoneva02} relied on photometric
calibrations to determine metallicities. Photometric metallicity
determinations are susceptible to errors from reddening corrections,
have reduced sensitivity at low metallicity, and depend strongly on
the adopted calibration to spectroscopic estimates, which vary from
work to work. Using spectroscopic measurements increases the accuracy
and precision of metallicity determinations, in addition to providing
kinematic information such as radial velocities, albeit with the
significant added cost of increased observing time. Previous
spectroscopic analyses of these low luminosity targets were limited to
hundreds of stars along individual lines of sight \citep{wg95,
  favata97, fuhrmann98, fuhrmann04, allendeprieto04, luck05, luck06,
  luck07, fuhrmann08, arnadottir09, fuhrmann11, katz11}.

Two recent surveys have improved analyses of the MDF of the Milky Way
disk by increasing the sample size and sky coverage. The GCS is a
magnitude-limited survey, with $\sim$14,000 F and G dwarfs with
metallicities estimated from Str{\"o}mgren photometry
\citep{jorgensen00, nordstrom04, holmberg07, holmberg09,
  casagrande11}. The RAdial Velocity Experiment (RAVE) further expands
the sample of cool stars, with around 17,000 F and G dwarfs and
spectroscopic metallicities \citep{siebert11}. Although they have
uniform data sets over a large region of the sky, neither of these
have dwarf stars far beyond the plane of the Galaxy. The GCS sample is
limited to 200 pc from the Sun \citep{casagrande11}. The dwarfs in the
RAVE sample probe to a maximum distance of approximately 1 kpc; the
typical distance for their cool dwarf sample ranges from 50--250 pc
\citep{zwitter10, steinmetz12}. Previous analyses of individual line
of sights were similarly limited, extending no further than 50 pc from
the Sun. Both the GCS and RAVE samples exhibit MDFs that peak at
around solar metallicity, despite their different methods for
estimating [Fe/H] \citep{casagrande11, coskunoglu11}, indicating that
they are dominated by thin-disk stars, as expected for a solar
neighborhood sample. Without a large sample of cool stars far beyond
the solar neighborhood, we cannot accurately constrain the chemical
and dynamical development of the Milky Way disk as a whole.

In contrast to GCS and RAVE, the early work on G dwarfs of
\citet{gilmore85} reached Galactocentric height above the plane,
$|Z|$, of around 1.6 kpc, beyond the local volume. This analysis,
however, consisted of only ten lines of sight and estimated
metallicity from ultraviolet excess, based on broadband
photometry. More recent work by \citet{katz11} examines subgiants and
giants as high as 5 kpc above the plane of the Galaxy. Although they
use spectroscopic metallicities, this sample probes only $\sim$400
stars along two lines of sight, limiting their statistical power to
measure the extremes of the MDF or its variation over their sample
volume.

SDSS and SEGUE provide a uniform sample over a large portion of the
Milky Way; previous analyses have sought to use this data to constrain
the chemical structure beyond the solar neighborhood. \citet{ivezic08}
and \citet{bond10} examined over 2 million F and G dwarfs in SDSS,
utilizing broadband \emph{ugr} photometry to estimate both the
metallicity and distance for each star. This large sample probes far
beyond the local volume, ranging from $|Z|$ of 0.5 to 7 kpc, and is
dominated by thick-disk and halo stars. They find that the disk
metallicity decreases from around [Fe/H] of $-$0.6 close to the plane
of the Galaxy to plateau at around $-$0.8. They also detect a clear
break between the disk and halo stars, which have a mean metallicity
of around $-$1.4 \citep{bond10}. Although they have a large and
unbiased stellar sample, uncertainties in the photometric metallicity
calibration affect the absolute scale of the MDF and manifest as
errors in the derived photometric parallax relationships. They also
cannot exclude binaries from the sample; an undetected companion has a
significant effect on stellar photometry
\citep{schlesinger10}. Finally, although they analyze a range of
stellar types, they focus their analysis on F stars, specifying a
color range of $(g-r)$ from 0.2 to 0.4. F dwarfs have shorter
lifetimes than G dwarfs, and are thus an incomplete sample of the disk
chemical evolution.

Avoiding reddening and calibration uncertainties, past studies
utilized SDSS spectroscopy to measure the MDF. \citet{allendeprieto06}
examined a sample of 22,770 F and G dwarfs from SDSS Data Release 3,
ranging from 1 to 8 kpc in $|Z|$. They focus their analysis on more
metal-poor stars, examining the metallicity differences between the
thick disk and halo. \citet{allendeprieto06} find that the MDF of G
dwarfs between 1 and 3 kpc of the plane, associated with the thick
disk, have a peak around [Fe/H]$=-$0.7, significantly more metal-rich
than those above this height, which are halo stars and exhibit a peak
[Fe/H] of $-$1.6. With the release of SEGUE in Data Release 6, the
sample size and coverage in Galactocentric height and radius have
improved significantly since this work, allowing us to more accurately
determine the MDF of G dwarfs with respect to spatial
position. Furthermore, \citet{allendeprieto06} do not account for
observational biases that originate from the SDSS target selection
algorithm, which did not sample the stellar range of metallicity and
age equally.

The recent work of \citet{lee11b} utilize the SEGUE spectroscopic
G-dwarf sample to examine the properties of the $\alpha$-separated
thin- and thick-disk components. Similar to \citet{ivezic08}, and
\citet{bond10}, they find that the thick-disk component is more
metal-poor than the thin-disk component, and dominates far from the
plane of the Galaxy. Although they use up-to-date SSPP parameters,
they do not examine the entire G-dwarf sample available in SEGUE,
selecting only stars specifically targeted by SEGUE as G dwarfs,
rather than all stars that fulfill the criteria. \citet{lee11b} also
do not account for the target selection biases in SEGUE, and are thus
biased towards metal-poor stars, although these observational biases
are less significant in their sample than those of
\citet{allendeprieto06}.  Most importantly, rather than examining the
MDF of the disk as a whole, they examine the relative metallicity
structure of two $\alpha$-separated components. We examine the MDF of
the SEGUE sample without making any chemical or kinematic separation
of the populations.

\subsection{This Work}

With a large number of stars and extensive sky and volume coverage,
the complete SEGUE survey provides an ideal sample to examine the
chemical abundance distribution in the Galaxy \citep{yanny09}. The
SEGUE data consist of SDSS $ugriz$ photometry and spectroscopy for
240,000 stars over a range of 14$<g<$20.3 in $\sim$3500 square degrees
on the sky. There are 50,210 and 26,834 SEGUE stars that fulfill the
G- and K-dwarf photometric criteria, reaching Galactic distances from
the plane, $|Z|$, of around 3.5 kpc and ranging from Galactic radius,
R, of 5 to 13 kpc. Not only is this the largest spectroscopic sample
available, it covers a much more extensive volume of the Milky Way
disk than all previous analyses. Furthermore, it is the largest and
most extensive spectroscopic sample of K-dwarf stars, whose faint
magnitudes make it difficult to probe far from the solar
neighborhood. Previous samples of K-dwarfs were limited in volume,
spatial-coverage, sample size, and atmospheric parameter accuracy.

We utilize this sample to determine an unbiased MDF of both G- and
K-dwarf stars throughout a large volume of the disk and determine how
they change with respect to height above the plane. By using both
spectral types, we examine a larger volume of the disk, constraining
the disk chemical structure both close and far from the plane of the
Galaxy, whereas analyses of hotter stars in SEGUE are limited to
thick-disk dominated space. In addition, we compare the MDF of the two
spectral types to analyze whether or not the star-formation and
chemical-evolution history vary with respect to stellar mass.

SEGUE selects stars as spectroscopic targets using a series of
photometric and proper motion criteria to isolate stars of different
spectral types. Because the SDSS and SEGUE survey addressed many
different scientific questions, it had a large variety of target
types. Each of these isolate a different portion of parameter space,
resulting in samples which do not sample the stellar range of
metallicity and age equally. The net effect for the entire sample is
an observational bias in favor of metal-poor stars. This is the first
work to systematically determine and account for the various
observational biases in the SEGUE G- and K-dwarf sample. By comparing
SDSS photometry and SEGUE spectroscopy for each line of sight, we
assign weights to each SEGUE star to account for the various
target-selection biases. Computing the MDF with this weighted sample
allows us to examine how the underlying population of stars throughout
a large volume of the disk varies with respect to spatial position.

In this work, we first discuss how we extract the G and K dwarf stars
from the SEGUE database (\S\,\ref{sec:sample}) and estimate the
distance to each star using isochrones (\S\,\ref{sec:distances}). We
then constrain the different observational biases that may arise in
the SEGUE sample (\S\,\ref{sec:contamination}) and detail our
technique to correct for those biases that arise from the SEGUE target
selection algorithm (\S\,\ref{sec:accounting_biases}). Utilizing these
target selection weights, we determine complete and unbiased
metallicity distribution functions for both G and K dwarfs that
accurately reflect the underlying chemical structure
(\S\,\ref{sec:mdf}). In \S\,\ref{sec:discussion}, we compare the
observed metallicity structure to previous analyses and two Galaxy
models, the stellar population synthesis model TRILEGAL 1.4
\citep{girardi05} and the radial migration model of
\citet{schonrich09a, schonrich09b}. These comparisons reveal how
critical an accurate picture of the disk chemical structure is for
understanding how the disk formed and evolved and how valuable our
observations are for understanding how the Milky Way developed.

\section{The SEGUE Stellar Sample}
\label{sec:sample} 

\subsection{Basic Survey Design}
 
The SEGUE survey combines the extensive uniform data set of photometry
from SDSS with medium-resolution (R$\sim$\,1800) spectroscopy over a
broad spectral range (3800-9200\AA) for $\sim$\,240,000 stars over a
range of spectral types \citep{yanny09}.  Technical information about
the Sloan Digital Sky Survey is published on the survey design
\citep{york00, eisenstein11}, telescope and camera \citep{gunn06,
  gunn98}, astrometric \citep{pier03} and photometric \citep{ivezic04}
accuracy, photometric system \citep{fuku96}, and photometric
calibration \citep{hogg01, smith02, tucker06, pad08}. Beyond its large
size, this sample has a homogeneous data set, substantial area
coverage, and spectroscopically-determined stellar atmospheric
parameters. This work utilizes photometry from Data Release 7 (DR7,
\citealt{abazajian09}). The atmospheric parameters are a modified
version of those released as part of DR8 (\citealt{aihara11},
Appendix\,\ref{app:sspp_optimize}).

Each SEGUE plug-plate\footnote[2]{Throughout the rest of this work, we
  will refer to plug-plates simply as plates.} covers a circular
region of 7 square degrees, probing the sky with 640 spectroscopic
fibers. SEGUE selects targets from SDSS for spectroscopic observation
based on photometric and proper-motion cuts. For an individual SEGUE
line of sight, approximately 375 and 95 fibers are allotted to G and K
dwarfs, respectively. The stars assigned spectroscopic fibers are
randomly selected from all of the stars in SDSS photometry that meet
the target-selection criteria. These criteria, and the number of
fibers devoted to each target type, changed over the course of the
SEGUE survey observations to improve the efficiency of some of the
target selections. We use the target identifications from Data Release
7 applied uniformly to all of the photometric and spectroscopic
data. We also limit ourselves to pointings which fall under the SEGUE
program and not the extragalactic SDSS survey or the SEGUE-2 program,
as neither of these targeted the G and K dwarfs explicitly. Finally,
we eliminate any pointings that do not have both bright
($r_0\leq$17.8) and faint ($r_0\geq$17.8) plates; this requirement
ensures we probe the same magnitude range for all lines of sight.

The resulting spectra are processed through the SSPP, an automated
system that determines atmospheric parameters, such as effective
temperature, surface gravity, and metallicity \citep{lee08_I}. The
SSPP employs 6 primary methods for the estimation of T$_{\rm{eff}}$,
10 for the estimation of $\log g$, and 12 for the estimation of
[Fe/H]. For an in-depth description of all of the different SSPP
calculations and techniques, see \citet{lee08_I,lee08_II}. This
program's outputs have been checked against high-resolution spectra of
stars within globular and open clusters, as well as in the field
\citep{lee08_II, allendeprieto08, smolinski10}. The uncertainties of
the SSPP for targets with $S/N$=25 per pixel, where each pixel is
$\approx$1\AA, are $\sigma(\rm{T_{eff}})$=200 K, $\sigma(\log g)$=0.4
dex, and $\sigma(\rm{[Fe/H]})$=0.3 dex.  These uncertainties increase
as the $S/N$ decreases: for $S/N$=10, $\sigma(\rm{T_{eff}})$=260 K,
$\sigma(\log g)$=0.6 dex, and $\sigma(\rm{[Fe/H]})$=0.45 dex
\citep{lee08_I}.

Not all of the techniques employed by the SSPP are accurate for
analyzing cool dwarf stars. We have examined each method for
estimating stellar parameters for our spectral types, refining and
optimizing the SSPP for G and K dwarfs (see
Appendix\,\ref{app:sspp_optimize}). Specifically, for determinations
of surface gravity and metallicity, we eliminate techniques which were
designed for hotter and/or evolved stars and refine existing
techniques to improve their accuracy. This revised SSPP was tested on
open and globular clusters (see Table\,\ref{tab:clusters_info}). The
modified techniques result in negligible shifts to the overall
metallicity determined by the SSPP for each cluster, well within the
uncertainties. Our work on the revised SSPP will be part of the
improvements to the SSPP released with SDSS DR9.

\subsection{Extracting G- and K-dwarf Stars from SEGUE}
\label{sec:extraction}

SEGUE G and K dwarfs are selected using a simple color and magnitude
cut. The SEGUE ``G dwarfs'' are defined as having 14.0$<r_0<$20.2 and
0.48$<(g-r)_0<$0.55, while the ``K dwarfs'' have 14.5$<r_0<$19.0 with
0.55$<(g-r)_0<$0.75 \citep{yanny09}\footnote[3]{We find that these
  color cuts are equivalent to a spectral type of G5 or G6 for G
  dwarfs and G7 through K2 for K dwarfs \citep{johnson63}.}. Note that
the the subscript 0 indicates dereddening and absorption correction
using \citet{sfd98} values. For [Fe/H] from $-$0.5 to $-$2.5, the Yale
Rotation Evolution Code isochrones (YREC, \citealt{an09}) indicate
that these $(g-r)$ colors correspond to a temperature range of
$\approx$4800$-$5300 K for K dwarfs and $\approx$5000$-$5600 K for G
dwarfs. In this paper, we will refer to the SEGUE ``G and K dwarf''
categories simply as G and K dwarfs. These target categories were
designed to isolate stars at a range of ages. Stars of hotter spectral
types that formed early in the disk's history have evolved past the
main-sequence turnoff, biasing samples towards younger, more
metal-rich, stars. As we do not have parallax information or precise
surface gravities, we can not estimate the ages for our stellar
sample, as done in \citet{edvardsson93}, \citet{nordstrom04}, and
\citet{casagrande11}. Fortunately, the effect of age on the MDF will
be minimized for these cool stars.

\begin{figure}
\centering \includegraphics[width=\textwidth]{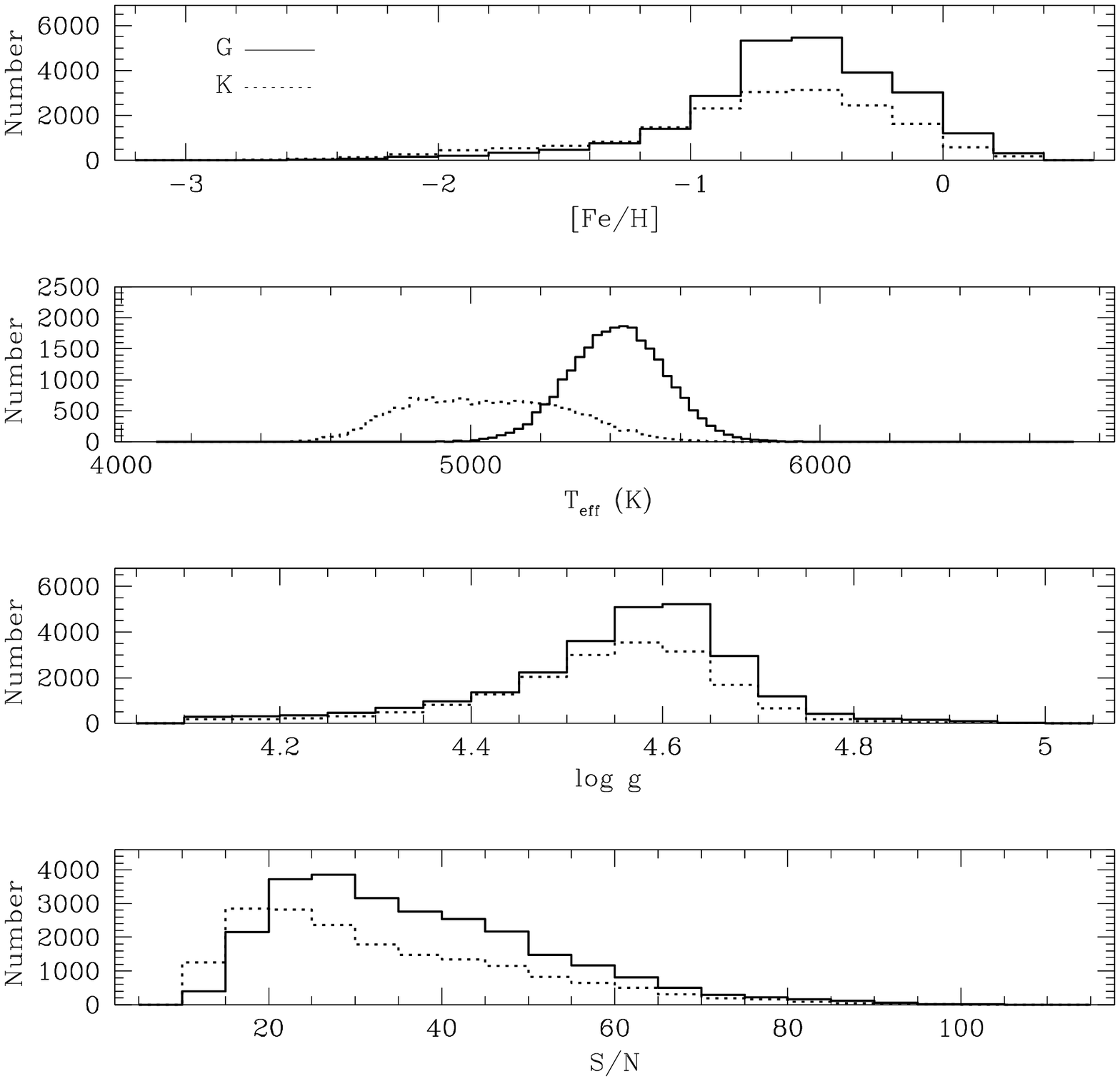}
\figcaption[sample_param_0624.eps]{The atmospheric parameters of our
  sample of around 26,600 G and 18,500 K dwarfs with appropriate $S/N$
  and surface gravity. The solid line represents the G dwarfs, and the
  dotted line the K dwarfs. All of these stars have $\log g \geq$4.1,
  $S/N\geq$10 per pixel, where each pixel is $\approx$1\AA, and a
  measured temperature and metallicity.
\label{fig:prelim}}
\end{figure}

Extracting every star with SEGUE spectroscopy that matches these color
and magnitude criteria from SDSS Data Release 7 photometry and the
SSPP for Data Release 8 results in 50,210 and 26,834 stars in the G
and K categories, respectively. We then restrict our sample to targets
with $S/N \geq$10, because these spectra have better-constrained and
better-understood uncertainties than those with lower $S/N$
\citep{lee08_I}. We also eliminate targets where, for various reasons,
the SSPP was unable to determine the temperature, metallicity, and/or
surface gravity. In addition to catastrophic failures, we remove
targets that the SSPP flags due to temperature or noise issues. For
example, if the temperature determined for a star by the SSPP and that
from a ($g-z$) relationship differ by more than 500 K, we eliminate it
from our sample. Similarly, if a spectrum is flagged as noisy by the
SSPP, we remove it from the sample, even if its reported $S/N$ is
greater than 10. Finally, we select all stars with $\log g \geq$4.1 to
isolate dwarf stars. This surface-gravity cut is discussed in depth in
\S\,\ref{sec:subgiants}. We also adjust the magnitude limits of our
sample. The SDSS saturation limit varies over the instrument. To
ensure that the bright end of the sample is complete, we set a uniform
bright limit of $r_0\geq$15. In addition, for the sake of our
observational bias corrections, we removed the faintest G-dwarf
targets, with $r_0$ greater than 18.45 mag (see
\S\,\ref{sec:r_weights}). With these parameter criteria, in
conjunction with the color and updated magnitude limits, we have
around 26,600 G and 18,500 K dwarfs. The distribution of our G- and
K-dwarf spectroscopic sample in various atmospheric parameters is
shown in Figure\,\ref{fig:prelim}. The values shown for each target
are slightly different than those included in DR8 \citep{aihara11};
specifically, the [Fe/H] and surface-gravity determinations have been
optimized for the sample, as discussed in
Appendix\,\ref{app:sspp_optimize}.

\section{Distance Determinations} 
\label{sec:distances}  

\begin{figure} 
\centering \includegraphics[width=\textwidth]{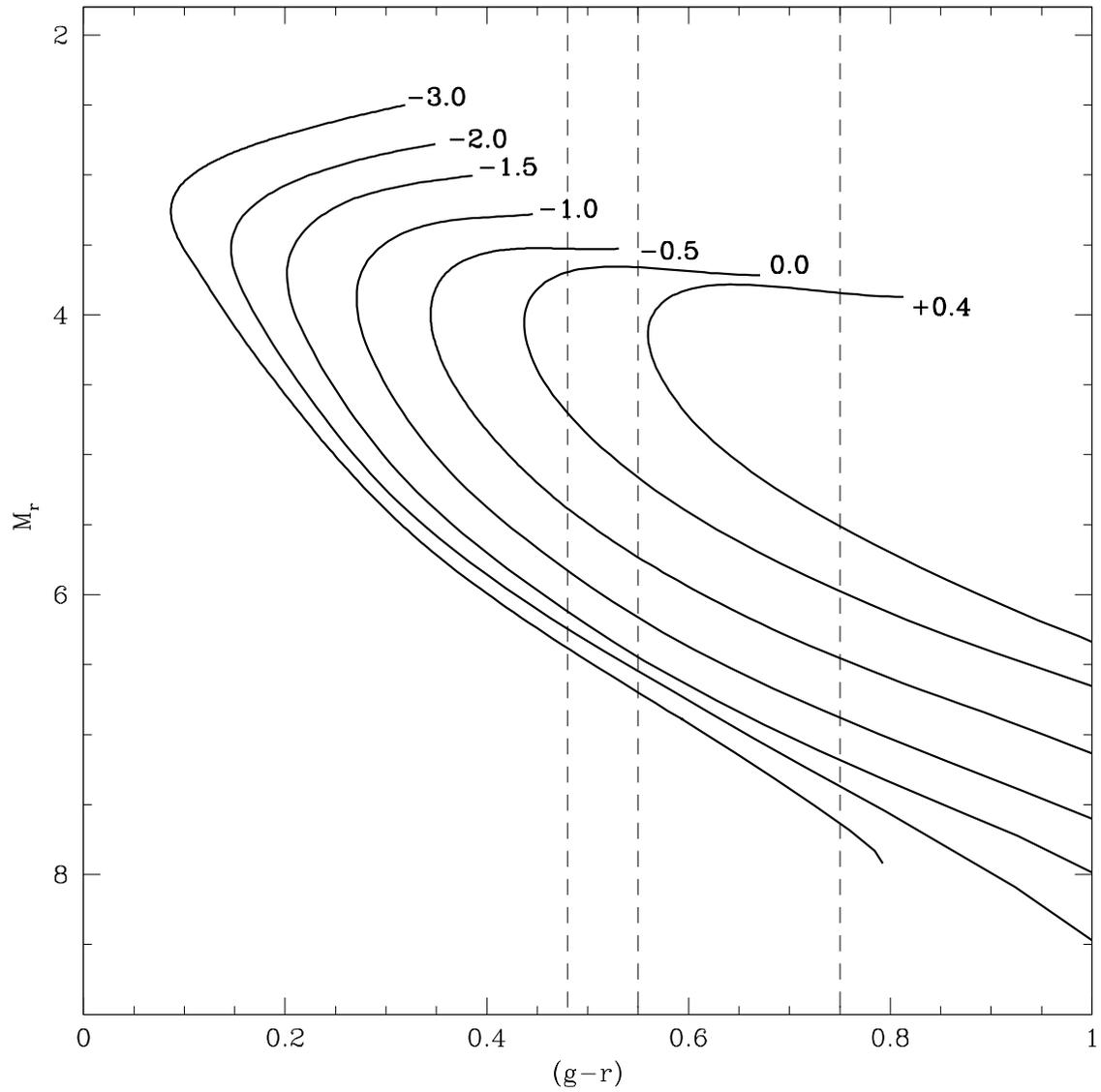}
\figcaption[yrec_v_dart_1115.eps]{The YREC isochrone set for a range
  of metallicities at an age of 10 Gyr. Each isochrone is labeled with
  its specific [Fe/H]. The vertical dashed lines show the $(g-r)$
  color range for SEGUE G and K dwarfs. These isochrones extend
  slightly past the main-sequence turnoff to the subgiant branch.
\label{fig:isochrones}}
\end{figure}

Distances are critical both for determining a volume-complete MDF and
for examining the behavior of these stars over different regions of
the Galaxy. As these stars lack trigonometric parallaxes, we estimate
distances by matching each star to YREC isochrones in color and
metallicity (Fig.\,\ref{fig:isochrones}, \citealt{an09}). There are
numerous isochrone sets available, such as the Dartmouth
\citep{dartmouth}, BaSTI \citep{pietr04}, and Padova isochrones
\citep{girardi04}. Each isochrone set will predict slightly different
behavior along the main sequence, where our stars lie, resulting in
systematic changes in distance between sets
(\S\,\ref{sec:iso_choice}). We also estimate the distances using the
photometric parallax relationship of \citet{ivezic08}
(\S\,\ref{app:pp_comp}).

We utilize the YREC isochrones for three main reasons. First, the YREC
set is directly calibrated on open and globular clusters, while the
other isochrones are not. The \citet{ivezic08} photometric parallax
relationship is calibrated on the open and globular cluster distances
from \citet{harris96}. Analysis in the interim has found significant
variation in these calculated distances, increasing the uncertainty of
the calibration, e.g., Table 2 in \citet{an09}. Second, as they are
specifically designed to work with SEGUE observations, the YREC
isochrones are guided by clusters observed in the SDSS {\it ugriz}
bands, whereas the \citet{ivezic08} photometric parallax relationship
must convert from Johnson-Cousins to {\it ugriz}, contributing
additional distance uncertainty. Finally, the YREC isochrones cover
the widest metallicity range of all the different options, from [Fe/H]
of $-$3.0 to $+$0.4. Both the other isochrones and the
\citet{ivezic08} photometric parallax relationship extend to only
      [Fe/H] of $-$2.5.

\subsection{Isochrone Matching} 
\label{sec:yrec_iso} 

\begin{figure}
\centering
\includegraphics[width=0.8\textwidth]{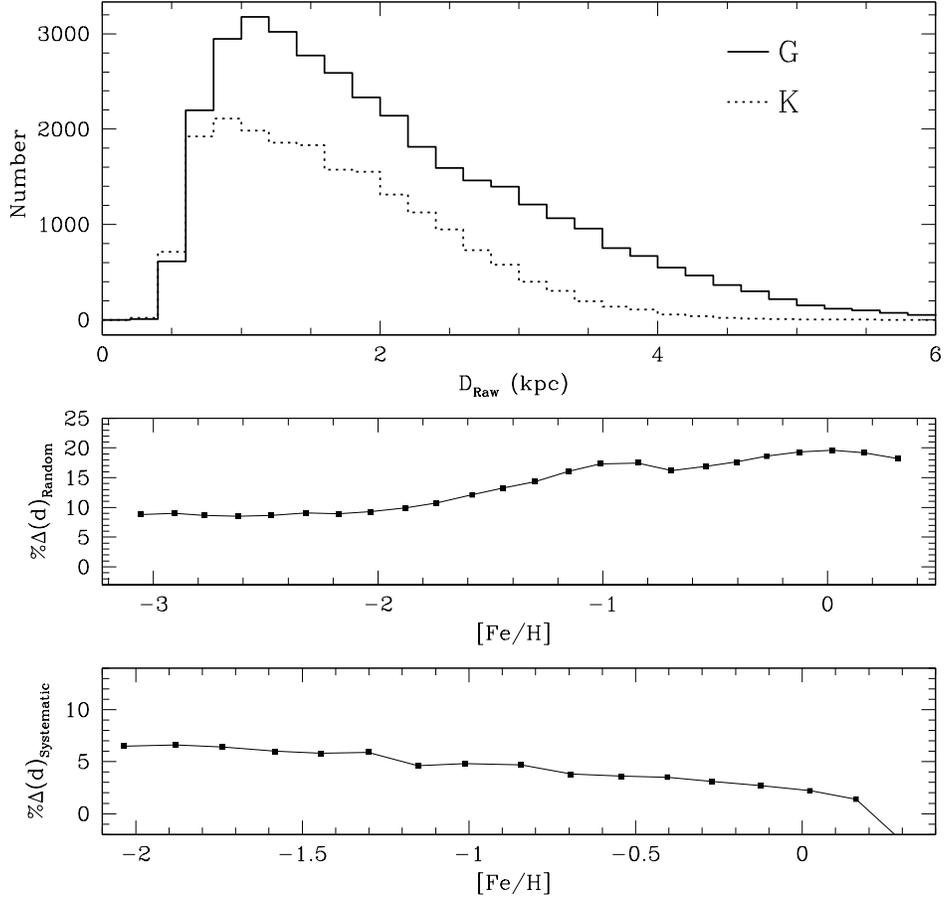}
\figcaption[dist_plots_0713.eps]{The top plot shows the distribution
  of G- (solid) and K-dwarf (dashed) distances, calculated by matching
  to YREC Isochrones. The K dwarfs tend to be closer in distance than
  the G dwarfs, which we expect as they are typically less
  luminous. The K dwarfs are peaked at a distance of around 0.95 kpc;
  the G dwarfs peak around 1.25 kpc. The overlapping distance range
  for K dwarfs and G dwarfs ranges from 1.19 to 1.84 kpc and 1.59 to
  2.29 kpc, respectively. The middle plot shows the distance error due
  to random errors with respect to [Fe/H]. This includes uncertainty
  from SSPP [Fe/H], photometry, [$\alpha$/Fe], and isochrone
  choice. We divide the sample into 25 bins of [Fe/H]; the points
  represent the mean uncertainty in each bin. The bottom figure shows
  the systematic uncertainty with respect to [Fe/H], plotted in the
  same way. This combines the systematic underestimate in distance
  from assuming all stars are alone and the overestimate from
  assigning an age of 10 Gyr to each star. Although we use TRILEGAL
  for the majority of these uncertainty determinations, we use the SB
  Galaxy model to constrain the age uncertainty. Although this model
  covers a wider range of ages, it covers a smaller range of
  metallicity, going no lower than around $-$2. At this metallicity,
  there is a negligible change in distance with a change in age.
\label{fig:prelim_dist}}
\end{figure}

For each star, we select the closest YREC isochrone both above and
below the target in [Fe/H]. We limit our isochrones to the main
sequence, as we expect all of our targets to be in this evolutionary
stage. We also assume an age of 10 Gyr; uncertainties associated with
this value are discussed in \S\,\ref{sec:age_effects}. We then match
the target's $(g-r)_0$ color to each of the bracketing isochrones,
extracting the predicted parameters when the $(g-r)_0$ color matches
within 0.001 mag. Using linear interpolation, we determine
relationships between stellar parameters, such as the absolute
magnitude in $ugriz$, with respect to [Fe/H]. We then extract
isochrone parameters using these relationships and the SSPP estimate
of [Fe/H], applying the distance modulus to derive a distance in each
filter. Our final value is the mean distance over all of the SDSS
filters. We also estimate distances using cubic, rather than linear,
interpolation, finding little difference between the stellar
parameters determined from these two
schemes. Figure\,\ref{fig:prelim_dist} shows the distribution of
distances for G and K dwarfs.

\subsection{Distance Uncertainties}
\label{sec:paper_distunc}

There are a number of random and systematic uncertainties, such as the
uncertainty in SSPP measurements and undetected binarity, that must be
taken into account in our distance estimates. These uncertainties vary
with respect to metallicity; small changes in metallicity result in
large changes in absolute magnitude at the metal-rich end, whereas
there is a smaller effect at the metal-poor end. We use the TRILEGAL
1.4 \citep{girardi05} and \citet{schonrich09a,schonrich09b} Galaxy
models (\S\,\ref{sec:trilegal_models}
and\,\ref{sec:schoenrich_models}) to estimate the effect of errors in
different parameters on our sample. These are discussed at length in
Appendix\,\ref{app:uncertainties}; the total random and systematic
distance errors are shown in Fig.\,\ref{fig:prelim_dist}. The random
distance uncertainty originates in the photometric, SSPP [Fe/H], and
[$\alpha$/Fe] errors. There is an additional random distance
uncertainty that originates from isochrone choice. Although
differences between each isochrone set will create a systematic shift
in distance, comparison between the YREC \citep{an09}, Dartmouth
\citep{dartmouth}, Padova \citep{girardi04}, and BaSTI \citep{basti}
isochrones reveals that each shifts the distance in a different
way. While distances determined with the Dartmouth set finds larger
distances than YREC estimates, those from Padova and BaSTI are
smaller. Thus, we treat this variation as a random uncertainty. We
find that the total random distance uncertainty ranges from around
18\% for stars with [Fe/H]$>-$0.5 to 8\% for more metal-poor stars,
dominated by uncertainty in the SSPP [Fe/H] determination.

The systematic uncertainties in distance stem from our age assumptions
and the possibility of undetected binarity. Both of these are
discussed at length in Appendix\,\ref{app:uncertainties}. The most
metal-rich stars are likely younger than our assumed age of 10
Gyr. This leads to a systematic shift in distance of $-$3\% for the
most metal-rich stars, while the metal-poor stars are largely
unaffected. An undetected companion has a comparable effect but goes
in the opposite direction; each pair will be systematically shifted by
approximately $+$5\% in distance over the entire metallicity
range. Conservatively, 65\% of all G and K dwarfs are expected to have
companions \citep{dm91}, leading to a total systematic shift of around
$+$3\% in distance due to undetected binarity.

We adjust our distances to account for these systematic changes. We
assign an offset to each star from age and binarity based on its
optimized SSPP [Fe/H], which we then convolve with a Gaussian
distribution. As our age assumptions will cause us to overestimate the
distances, we subtract the estimated offset from our estimated
distance to each star. Assuming each star is single will cause us to
underestimate the distance to approximately 65\% of the sample. We
randomly select 65\% of the sample and increase their distance by the
assigned systematic offset from binarity. The systematic uncertainties
are much smaller than the random uncertainties at the metal-rich end
of the distribution, whereas, due to binarity, they are comparable at
the metal-poor end.

\subsection{Testing our Calculated Distances on Globular and Open Clusters} 
\label{sec:dist_cluster} 

None of our targets have measured parallaxes; thus, we cannot directly
confirm that our calculated distances are correct. We instead test our
methods against SDSS spectra of stars in open and globular clusters,
observed for testing the SSPP determinations and SDSS photometry
\citep{lee08_II, an09, smolinski10}. Unfortunately, many of these
studies focused on stars past the main-sequence turnoff because they
are considerably brighter than dwarfs. We have targets along the main
sequence for the clusters M13, M67, NGC 2420, and NGC 6791. This
limits us to a metallicity range from [Fe/H] of $-$1.54 to +0.30.

We select stars from the samples of \citet{lee08_II} and
\citet{smolinski10} with well-determined [Fe/H], $S/N \geq$10, and
$\log g \geq$4.1 to ensure that our cluster members are on the main
sequence. We do not cut on $(g-r)_0$ color for the globular clusters
because this would severely limit our sample size. Furthermore, our
distance calculations should hold for any star on the main
sequence. As a check, we determined the average distance to the
cluster derived from a color-constrained sample to that from the
larger main-sequence sample; they match within the expected errors in
distance. Thus, a color cut would not significantly affect our
distance measurements to these clusters.

\begin{figure} 
\centering
\includegraphics[width=\textwidth]{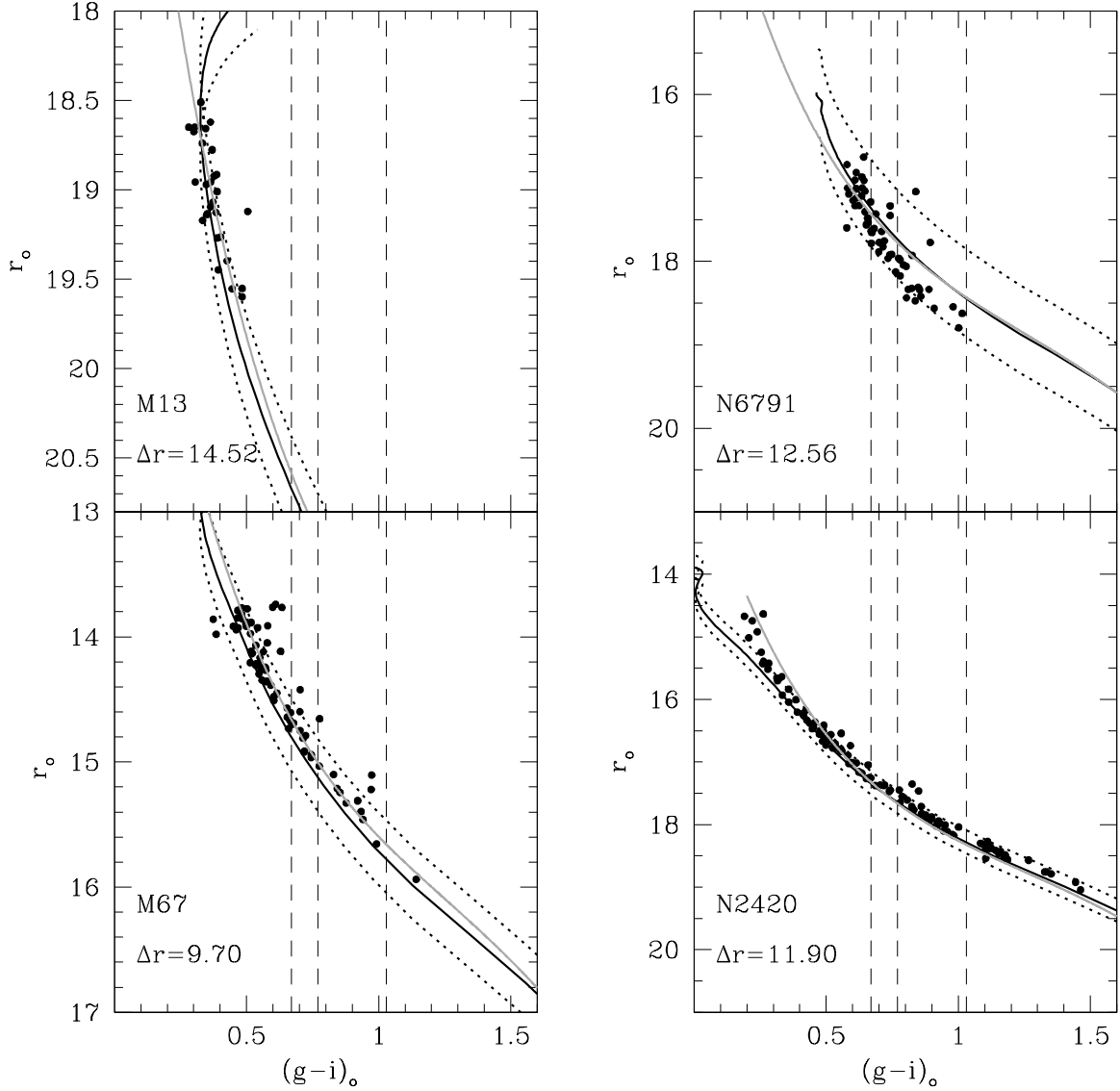}
\figcaption[clusters_pprd_paper_1016.eps]{Dwarf targets in various
  open and globular clusters compared to shifted YREC isochrones of
  comparable age (listed in Table\,\ref{tab:cluster_dist}). For each
  cluster, we extract all targets with $\log g\geq$4.1 and calculate
  their distance using a set of different methods. We have shifted the
  isochrones according to the mean distance for each cluster from the
  YREC isochrone technique (in black). The short dashed lines shift
  the isochrone by the 1$\sigma$ dispersion of the distance
  measurements for each cluster. The gray lines show the photometric
  parallax relationship using the mean SSPP metallicity of the
  clusters (see \S\,\ref{app:pp_comp}). The vertical long-dashed lines
  show the SEGUE G- and K-dwarf color cut converted to ($g-i$). Each
  plot is labeled with the name of the cluster and the distance
  modulus used to shift the isochrones. The cluster distances
  determined using all of our methods are listed in
  Table\,\ref{tab:cluster_dist}.
\label{fig:cluster_dist}}
\end{figure}

The calculated cluster distances are listed in
Table\,\ref{tab:cluster_dist}, in addition to other parameters from
\citet{harris96}, \citet{lee08_II}, \citet{smolinski10}, and the WEBDA
database \citep{paunzen}. A comparison of the clusters and YREC
isochrones is shown in Figure\,\ref{fig:cluster_dist}. The observed
stars agree well with the shifted YREC isochrones. The published
distances to these clusters vary considerably \citep{harris96,
  kraft03, an09, smolinski10}; thus, we are pleased with the general
agreement we observe from our isochrone-matching method.

\section{Constraining Possible Biases in the SEGUE Sample} 
\label{sec:contamination} 

The selection criteria for G and K dwarfs create numerous biases which
affect the sample. For our SEGUE G- and K-dwarf sample to accurately
reflect the chemical properties of these spectral types in the
Galactic disk, we must ensure that we constrain and account for these
observational biases. In this section, we investigate and constrain
the possible sources of contamination in the sample and the
observational biases induced by the SEGUE target-selection algorithm.
We also constrain the effect of each of these errors on our MDF.

\subsection{Photometric Errors} 

Although photometric errors for SDSS photometry are small, typically
2-3\% for each filter, these uncertainties may bump targets in and out
of the G- and K-dwarf color and magnitude range. As targets become
fainter, their photometric errors increase, making it more likely that
a fainter star will be shifted out of the SEGUE color and magnitude
criteria due to photometric errors.

\citet{newby11} determine that the photometric uncertainties are
constant up to $r_0$ of 19.7. The SEGUE target-selection criteria for
K dwarfs selects stars as faint as $r_0$ of 19, and are thus
unaffected by the increasing photometric errors. The G-dwarf magnitude
range extends to $r_0$ of 20.2; however, we trim our G-dwarf sample to
$r_0<$18.45 in order to limit our sample to a magnitude range where
the faint end is complete (\S\,\ref{sec:r_weights}). The photometric
uncertainties are thus consistent over the full magnitude range of our
G- and K-dwarf sample.

\begin{figure}
\centering
\includegraphics[width=\textwidth]{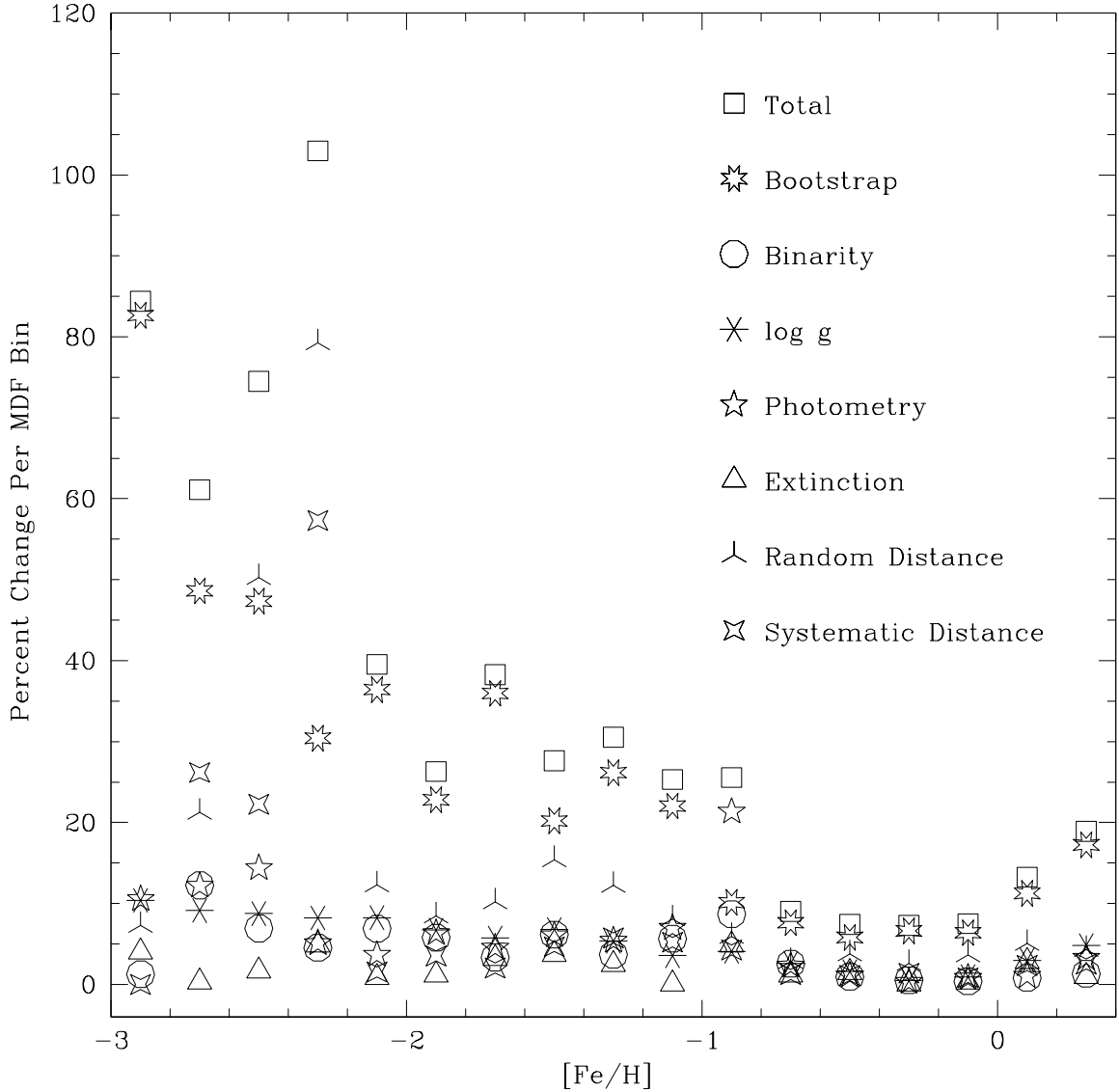}
\figcaption[mdferr_plotter_update.eps]{The change in each [Fe/H] bin
  of the MDF due to different uncertainties. We run a Monte Carlo
  analysis on the TRILEGAL simulation convolving the expected
  uncertainty in each of these parameters with a Gaussian to examine
  how the structure of the MDF changes. At the metal-poor end, the
  number of stars in each bin is quite small. Thus, the uncertainty
  from different factors is very large. Typically, uncertainties from
  our bootstrap analysis and random distance errors dominate. The
  random distance uncertainties shown here neglect the correlated
  [Fe/H]-distance error. The MDF uncertainties presented in the
  following figures and tables reflect the total uncertainty per bin,
  combining all of the shown errors in quadrature.
\label{fig:mdf_per_error}}
\end{figure}

We examine the effect of photometric uncertainty on target selection
using TRILEGAL models along SEGUE lines of sight
(\S\,\ref{sec:trilegal_models}). We estimate the photometric errors in
$g_0$ and $r_0$ from the exponential functions of \citet{newby11}. We
then convolve these errors with a Gaussian over multiple iterations to
examine how the sample of G- and K-dwarf stars change with changes in
photometry. Approximately 2\% percent of stars are shifted in and out
of the sample; those stars removed and inserted into the sample by
photometric uncertainties cover a similar range in metallicity
space. We examine how the MDF changes over the full metallicity range
with this photometric uncertainty, comparing each iteration to the
underlying MDF of the model itself. We find that the uncertainty in
each bin of the MDF due to photometric errors is on average 6\% and
should not create any [Fe/H] biases in our derived MDF (see
Fig.\,\ref{fig:mdf_per_error}).

\subsection{Undetected Binarity}
\label{sec:undetected_companions} 

As the target-selection criteria for G and K dwarfs is based purely
upon a color and magnitude cut, quantifying the effect of an
undetected companion on photometry is critical. Work by
\citet{schlesinger10} utilized numerical modeling of the SEGUE G- and
K-dwarf stars to constrain the effect of binarity on the photometry of
the sample. A companion will typically change the $(g-r)$ color
measured for a target by 0.01$\pm$0.02 magnitudes, less than the
uncertainty in SDSS photometry. Undetected companions will also change
the photometry in individual filters, affecting distance estimates
(Appendix\,\ref{sec:distance_binaries}), which we include in the total
systematic distance uncertainty.

Although this is a small change in color, it can still shift stars in
and out of the $(g-r)$ range for G and K dwarfs. Furthermore, as
companions are typically cooler than the primary star, they will make
targets appear redder, which may result in undetected binarity
preferentially removing the coolest stars from the
sample. \citet{schlesinger10} investigated the effect of binarity on
the properties of SEGUE G and K dwarfs, finding that, assuming 65\% of
all G and K dwarfs are in binaries \citep{dm91}, the addition of a
companion will bump around 1\% of stars into and 2\% of stars out of
the $(g-r)$ color criteria. Not only is this a minimal effect, but the
amount of stars shifted in and out are roughly comparable, indicating
that binarity will have little effect on the G- and K-dwarf target
selection. Using TRILEGAL modeled lines of sight and a modeled
population of companions (Appendix\,\ref{sec:distance_binaries}), we
run a Monte Carlo analysis to estimate the uncertainty in the MDF due
to undetected companions. Binarity does not preferentially affect
stars of a particular metallicity, with a typical uncertainty of 4\%
for each MDF bin.

Beyond photometry, \citet{schlesinger10} combine their numerical model
with synthetic spectra processed through the SSPP to constrain how a
secondary will affect the estimates of atmospheric
parameters. Effective temperature, which we do not use in our
analysis, is the most altered by undetected companions; around 18\% of
G- and K-dwarf primaries will be shifted by more than 150 K, the
reported SSPP uncertainty for $S/N$ of 50. However, undetected
binarity has very little effect on the SSPP [Fe/H]
estimates. Approximately 99\% of the modeled G and K dwarfs undergo
shifts of less than 0.2 dex in [Fe/H], less than the expected
uncertainty in the SSPP estimates over all $S/N$. Thus, the effect of
binarity on SSPP atmospheric parameters is well within the expected
errors, and we do not take it into account.

\subsection{Subgiant Contamination} 
\label{sec:subgiants} 

The color and magnitude cuts used to identify our cool dwarfs do not
exclusively target main-sequence stars. Subgiants and giants,
specifically K giants, can fall into our sample and drastically affect
the accuracy of our distance estimates, which assume all of the
targets are on the main sequence. To isolate dwarf stars we apply a
cut on the SSPP $\log g$, limiting it to 4.1 and above. However, the
uncertainty of the SSPP surface gravity estimates for stars with $S/N$
of 25 is $\pm$0.4 dex, which may result in evolved stars contaminating
the dwarf sample.

We use a three-fold approach to examine the extent of the subgiant
contamination. First, we employ Galaxy models to estimate the number
of subgiants expected to fall in our SEGUE G- and K-dwarf
sample. Second, we manufacture a series of synthetic spectra at
various evolutionary stages and process them through the SSPP. This
allows us to determine if there is a particular region of parameter
space where the SSPP has difficulty distinguishing dwarf and giant
stars, and also constrain how many dwarfs may be lost with a stringent
cut in $\log g$.  Finally, we use the Mg index to distinguish between
giants and dwarfs in our sample \citep{morrison03}.

For each line of sight in the SEGUE sample, we model the distribution
of stars using TRILEGAL 1.4 (\S\,\ref{sec:trilegal_models}). We
examine every target in these model distributions with $(g-r)$ in the
G/K dwarf range and $\log g<$4.2 to determine the size of possible
contamination\footnote[4]{The surface-gravity limit for these models
  is slightly higher than that applied to our sample, as these models
  are based on Padova isochrones \citep{girardi04}, which have the
  main-sequence turnoff at $\log g\,\approx$4.2 whereas YREC
  isochrones place it at 4.1}. Combining the proportion of subgiants
along the TRILEGAL line of sight with the uncertainty in the SSPP
surface-gravity measurements, we find that, over all lines of sight,
the SEGUE color and magnitude criteria will include a mean of
2$\pm$1\% subgiants incorrectly identified as dwarfs.

We then combine our population analysis with a study of the SSPP
surface-gravity determination using synthetic spectra. As the
empirically-corrected YREC isochrones used in this study do not extend
beyond the main-sequence turnoff, we adopt atmospheric parameters for
the giant and dwarf models from the Dartmouth Stellar Evolution
isochrones \citep{dartmouth} over a range of metallicity (see
\S\,\ref{sec:iso_choice} for more details about the
isochrones). Breaking the isochrones down into 0.01 magnitude blocks
in color, we manufacture synthetic spectra for both giants and dwarfs
using MARCS model atmospheres processed through TurboSpectrum
\citep{gustaffson08, ap98}. Our MARCS model atmospheres assume
solar-scaled abundances, with the solar composition from
\citet{grevesse07}, and plane-parallel geometry. They cover a range in
effective temperature from 4700-5800 K, $-$2.5 to $+$0.5 in [Fe/H],
and 1.2 to 5 in $\log g$. Each of these spectra was adjusted to a
range of $S/N$: 50, 25, and 10. We also analyzed the non-degraded
spectra. The spectral synthesis and noise-modeling processes are
discussed in more detail in \citet{schlesinger10}.

\begin{figure}
\centering \includegraphics[width=0.8\textwidth]{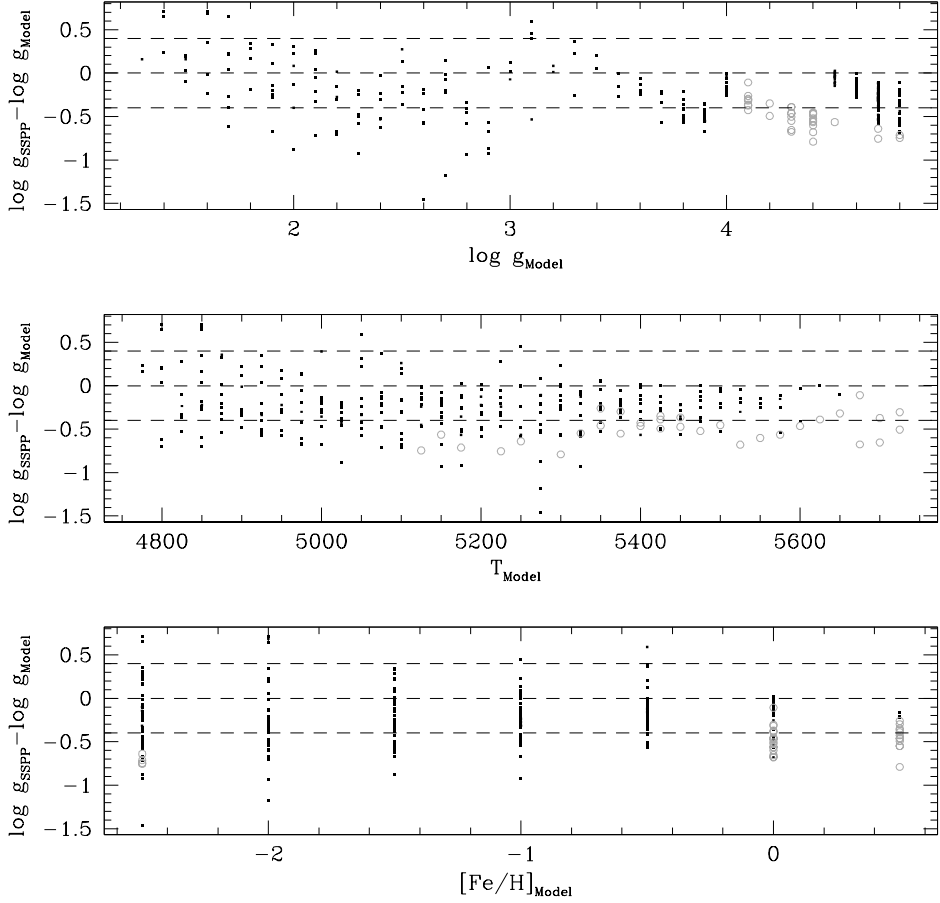}
\figcaption[lcd4_sn025_0722.eps]{A comparison of the surface gravity
  values for the grid of synthetic spectra of dwarfs and subgiants to
  those determined by the SSPP. These synthetic spectra were degraded
  to be at a $S/N$ of 25, where the expected SSPP uncertainty in $\log
  g$ is $\pm$0.40 dex. The top panel shows the difference between the
  model and SSPP-calculated surface gravities with respect to that of
  the model. The gray open circles show the grid points where modeled
  stars with dwarf surface gravities have SSPP $\log g$ of less than
  4.1. The dashed lines show a change in $\log g$ of 0 and $\pm$0.4
  dex. The SSPP tends to underestimate the surface gravity of these
  synthetic spectra. There is some structure in $\Delta \log g$ at the
  high $\log g$ end; this is likely due to the continuum removal
  processes in the SSPP and is being examined for the SEGUE DR9 public
  data release. The middle figure shows the change in surface gravity
  with respect to model effective temperature. Dwarfs misidentified as
  subgiants tend to be at hotter temperatures. Finally, the bottom
  panel shows the distribution of $\Delta \log g$ with respect to
  model [Fe/H]. Misidentified dwarfs tend to be the most metal-rich
  stars, although there are a few occurences at the metal-poor end,
  where weak lines make it difficult to measure atmospheric
  parameters. The most metal-rich stars tend to have surface gravities
  close to the cutoff; even small underestimates can force them from
  the sample.
\label{fig:logg_comp_logg} }
\end{figure}

Each of these simulated spectra is then processed through the SSPP
using the parameter estimates as described in
Appendix\,\ref{app:sspp_optimize}. A comparison of the calculated
parameters and those input in the models for $S/N\sim$25, the most
common $S/N$ for the G- and K-dwarf sample, is shown in
Figure\,\ref{fig:logg_comp_logg}. At a $S/N$ of 25, the surface
gravity tends to be underestimated by 6\%, less than the listed SSPP
uncertainties. No giants or subgiants will be identified as dwarfs by
their $\log g$ values. As $S/N$ decreases, the uncertainties in the
SSPP parameters increase. At the lowest signal to noise ratio in our
sample, there is a 2\% chance of a subgiant being identified as a
dwarf by the SSPP and included in our G- and K-dwarf sample. From our
population studies of the galaxy models, we expect approximately 2\%
of stars that fall into our G- and K-dwarf color cuts to be subgiants,
and $\sim$2\% of these will be misidentified as dwarfs by the SSPP,
resulting in a less than 1\% chance that a subgiant will be counted as
a dwarf in our sample.

Our final check on the extent of evolved-star contamination uses the
Mg index \citep{morrison03}. At a given [Fe/H], giants will have a
smaller Mg index value (i.e., less atomic Mg and MgH absorption) than
dwarfs. As SEGUE targets K-giant stars in addition to K dwarfs, we
focus on this spectral type, calculating the Mg index directly from
the SEGUE spectrum. While this index has been calibrated for giants
using known open and globular cluster members (Ma et al, in
preparation), we have almost no calibrating observations of known
metal-poor dwarfs. Therefore, we identified stars from our sample
which were very likely to be dwarfs using a reduced proper-motion
criterion \citep{lepine11}. Binning in metallicity, we isolate a
control sample of ``true'' dwarf stars by specifying that the total
proper motion must be greater than 20 mas/yr, and the $r_0$ magnitude
reduced proper motion is greater than 13. We then examine the spectra
of targets where the Mg index is lower than expected for the true
dwarfs, e.g., they are in the parameter space of evolved stars. This
visual inspection consisted of three tests:
\begin{itemize} 
\item The strength of the Mg feature in non-continuum-corrected SEGUE 
  spectra \citep{morrison00}.
\item A comparison of the strength of Ca I $\lambda$4227 and Ca II K
  feature \citep{morrison00}.
\item The ratio of the strength of Sr II $\lambda$4077 to three nearby Fe I
  lines \citep{morgan43, rose84}. 
\end{itemize}
These three tests are valuable luminosity discriminants and isolate
subgiant and giant stars masquerading as dwarfs in the sample. Our
visual analysis indicates that less than 1\% of the total sample of
SEGUE K dwarfs are actually evolved stars, confirming that subgiant
and giant contaminants have a negligible presence in our sample for
both spectral types. We do not take this negligible contamination into
account.

Our analysis of synthetic spectra indicates that the SSPP tends to
underestimate surface gravity, which prevents subgiants from entering
the sample. However, it also means that a number of dwarf stars will
fall out of the sample when it is selected using a surface-gravity
cut. In particular, we will preferentially lose high-metallicity
stars, as these have surface gravities closer to the boundary of $\log
g$. We run a Monte Carlo analysis to examine how uncertainties in
surface gravity will affect the MDF.  We model each SEGUE line of
sight with the TRILEGAL Galaxy model (\S\,\ref{sec:trilegal_models}),
selecting G- and K-dwarf ``spectroscopic'' targets based on SEGUE
photometric criteria.  We then determine the ``true'' MDF for this
sample. Next, we vary the modeled surface gravity values, convolving
them with a Gaussian error distribution with $\sigma$ of 0.6 dex, the
largest possible SSPP uncertainty. We then compare the MDF of these
modified samples to the original, to estimate the MDF uncertainty from
SSPP $\log g$ errors. For [Fe/H]$\geq-$1.0, the change in each MDF bin
is around 3\%. Below this metallicity, the percent change is higher,
around 10\%, due to the small-number statistics of this portion of the
simulated distribution (Fig.\,\ref{fig:mdf_per_error}). These are much
smaller than the expected MDF uncertainties from bootstrapping alone
(\S\,\ref{sec:monte_carlo}) and do not induce a bias in the
metallicity distribution.

\subsection{Extinction} 

Although the lines of sight in SEGUE typically probe above and below
the plane of the Galaxy, undergoing small amounts of extinction, this
small amount of reddening can affect SEGUE target selection, which is
based on color and magnitude. We determine the extinction per plate
using the \emph{r}-band extinction from the \citet{sfd98} maps for
every object listed as a star\footnote[5]{These reddening values were
  extracted from the DR7 table PhotoObjAll.}. All but 11 of the plates
(1888, 2052, 2179, 2300, 2334, 2335, 2623, 2669, 2679, 2680, and 2805)
have extinction less than 0.5 mags in \emph{r} and E$(g-r)$ of less
than 0.2 mag. Reddening cuts remove around 6\% of the sample.

Although many of the lines of sight have minimal extinction, there is
still a danger that reddening will bias the G- and K-dwarf
sample. Closer stars will have undergone less extinction than those
farther away; using the extinction corrections from \citet{sfd98} may
make close-in, metal-rich, stars appear artificially blue, leading to
inaccurate distance estimates and bumping stars in and out of the G-
and K-dwarf SEGUE sample. This should be a minimal effect, as our
stars have a minimum $|Z|$ of 0.2 kpc, beyond the expected scale
height of 0.14 kpc for dust in the disk \citep{mendez}.

We use the sample of \citet{cheng11} to examine the accuracy of
\citet{sfd98} extinction estimates with respect to
distance. \citet{cheng11} calculated the extinction for a sample of
SEGUE main-sequence turnoff stars by matching the isochrone-predicted
$(g-r)$ color, found using the SSPP parameters, to the
observed. Unlike the \citet{sfd98} values, this reddening calculation
does not assume that each star lies behind the full amount of
line-of-sight dust. For their four lines of sight with E$(g-r)$ less
than 0.2 mag, similar to our own, the \citet{cheng11} isochrone
extinction is comparable to the \citet{sfd98} values over the full
distance range (Cheng, private communication). Thus, in areas of
generally low extinction, using \citet{sfd98} extinction estimates
will not induce a bias in our sample against metal-rich stars by
preferentially removing nearby stars.

As with photometric errors and undetected binarity, extinction can
also artificially shift stars in and out of the G- and K-dwarf color
and magnitude range. Specifically, if extinction is overestimated,
bluer stars will be scattered in and red stars out of the color
range. To constrain these effects, we simulate the distribution of
reddening values at a range of extinctions, using the main-sequence
turnoff stars from \citet{cheng11}, on TRILEGAL Galaxy models along each
SEGUE line of sight. Modeling the extinction with respect to distance
along each SEGUE line of sight, we examine how the G- and K-dwarf sample
changes. Although reddening does shift stars in and out of the G- and
K-dwarf color range, it does not preferentially remove stars from a
certain metallicity space. The uncertainty per metallicity bin in the
MDF due to extinction is typically around 2\% (see
Fig.\,\ref{fig:mdf_per_error}).

\subsection{Volume Completeness} 
\label{sec:vol_comp} 

Our G- and K-dwarf sample covers a wide range of
metallicities. Metal-rich stars are brighter than metal-poor ones for
a given $(g-r)$ color. Thus, the volume coverage of the sample varies
with respect to metallicity for the same magnitude range. At the
bright end, metal-rich stars will saturate; at the faint end,
metal-poor stars are not sufficiently luminous to be observed by
SEGUE. To ensure that our SEGUE sample and MDF is not biased with
respect to chemistry due to the metallicity-dependent volume coverage,
we specify distance limits for our sample.

We examine the YREC isochrones for a metallicity of [Fe/H] of +0.4 and
$-$3.0, extracting the $r$-band magnitudes for targets with $(g-r)$ of
0.48, 0.55, and 0.75. Using the distance modulus, we calculate the
maximum and minimum distance for both spectral types over a range of
metallicity. For K dwarfs, the distance range for SEGUE volume
completeness is from 1.19 to 1.84 kpc; for G dwarfs, the distance
range is from 1.59 kpc to 2.29 kpc. Note that these distances utilize
adjusted $r$-magnitude limits for the G and K dwarfs, as described in
\S\,\ref{sec:r_weights}. In this paper, we refer to the distance range
from 1.59 to 1.84 kpc as the overlap distance range. These distance
limits are used in comparisons between the G- and K-dwarf sample. For
analysis of an individual spectral type, we use the larger distance
range associated with either G or K dwarfs. We refer to this range as
the spectral-type distance range.

Random and systematic uncertainties in distance will shift stars in
and out of the overlap range. We estimate the effect of these changes
in distance on our MDF with a Monte Carlo analysis of the TRILEGAL
simulation (\S\,\ref{sec:trilegal_models}). Each star is assigned a
typical percent change in distance from random and systematic effects,
which is then convolved with a Gaussian over multiple iterations. We
use a random distance uncertainty from errors in photometry,
[$\alpha$/Fe], and isochrone choice; the systematic uncertainties
originate in age and binarity assumptions. For [Fe/H]$\geq-$2.0,
variation in the systematic distance will change each MDF bin by
around 3\%. The number of stars at the metal-poor end is very small,
leading to larger uncertainties, on average 35\% (see
Fig.\,\ref{fig:mdf_per_error}). The random distance uncertainties have
a larger effect than systematic uncertainties, changing each bin by
approximately 7\% and 40\% at the metal-rich and metal-poor ends
respectively. These distance errors are included in our total MDF
uncertainties.

\begin{figure} 
\centering \includegraphics[width=\textwidth]{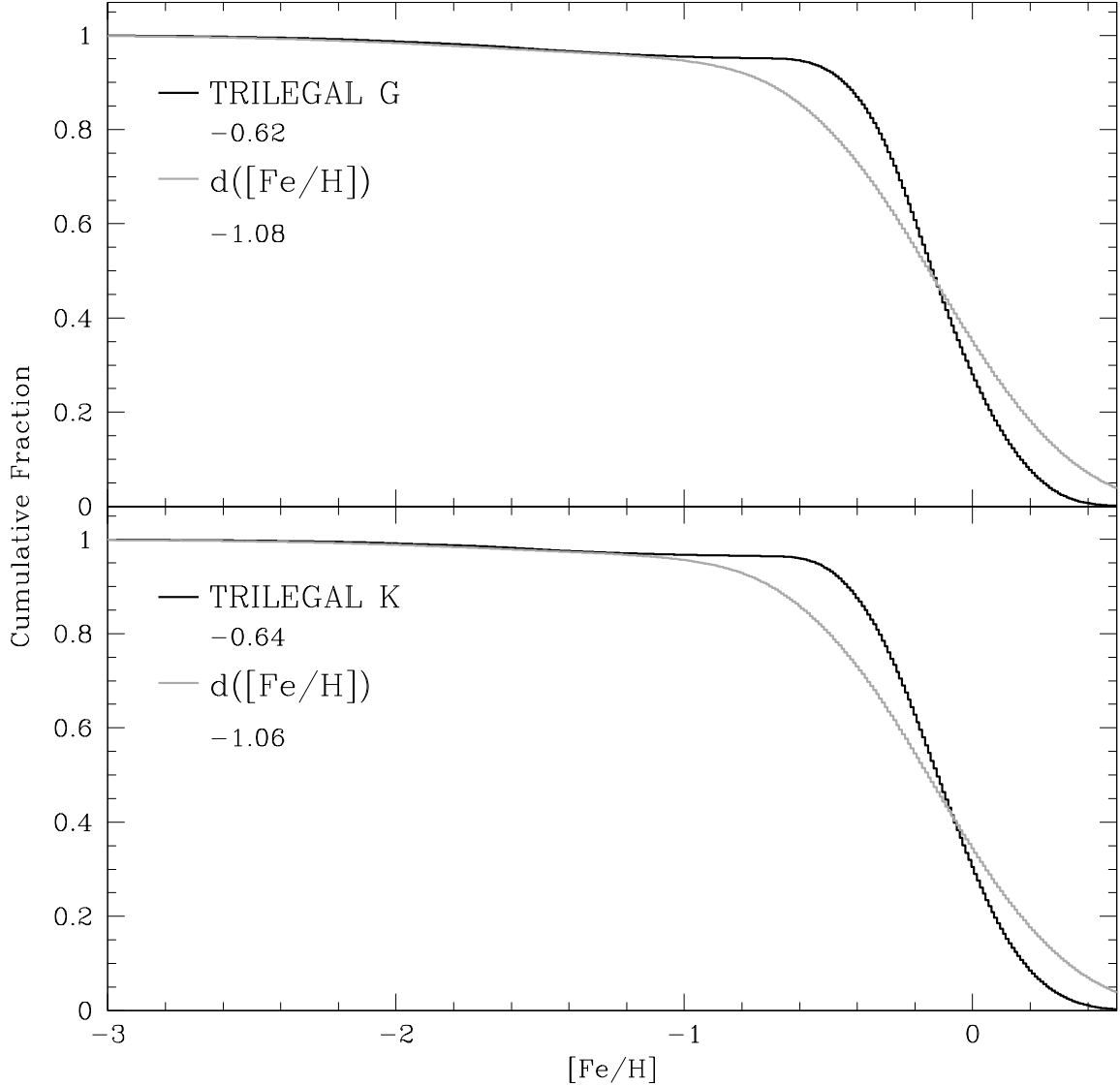}
\figcaption[]{Comparison of the true cumulative metallicity
  distribution from the TRILEGAL model to a distribution which
  includes the correlated uncertainties in [Fe/H] and distance. We
  simulate the G (top, black) and K (bottom, black) dwarfs in SEGUE
  using a TRILEGAL model. Using a Monte-Carlo analysis, we factor in
  the uncertainties in [Fe/H] and the resulting change in
  distance. This broadens the metallicity distribution a significant
  amount, as indicated by the gray d([Fe/H]) lines. We estimate the
  slopes of these distributions between a cumulative fraction of 0.25
  and 0.75; they are listed on the figure. This change estimates the
  amount of broadening in our MDFs caused by errors in metallicity.
\label{fig:correlated_mdf_err}} 
\end{figure}

We do not include the change in distance due to SSPP [Fe/H]
uncertainties as part of our random distance uncertainty, as these
properties are strongly correlated. Overestimating the metallicity
will lead to an overestimate in distance, and vice
versa. Uncertainties from [Fe/H] will both broaden the true
metallicity distribution and shift stars in and out of the sample due
to distance cuts. To estimate the size of these effects, we perform a
Monte-Carlo analysis along the TRILEGAL-modeled lines of sight. For each
modeled G- and K-dwarf star, we vary the [Fe/H], convolving the SSPP
errors with a Gaussian, and adjust the corresponding distance to
account for this change in metallicity. We then compare the different
Monte Carlo iterations to the true distribution (see
Fig.\,\ref{fig:correlated_mdf_err}). The cumulative distributions of G
and K dwarfs indicate that the [Fe/H] and distance error broadens the
MDF of the TRILEGAL model. To quantify the extent of this effect, we
estimate the slope of the cumulative distribution between fractions
0.25 and 0.75. Larger uncertainties in [Fe/H] will manifest in larger
negative slopes in this portion of the cumulative distribution. For
the true model distribution, the slope is around $-$0.6 dex$^{-1}$,
whereas it increases to $-$1.0 dex$^{-1}$ when we factor in
$\sigma_{[Fe/H]}$. We list the broadened slopes for both
spectral types in Figure\,\ref{fig:correlated_mdf_err}. Thus,
uncertainties will cause us to measure a broader, by slightly less
than a factor of 2, version of the true underlying distribution.

\subsection{Evolutionary Effects} 
\label{sec:evolutionary} 

To derive a complete chemical history of the Galactic disk, we must
ensure that no stars have evolved off of the main sequence, as this
would bias our SEGUE sample against metal-rich
stars. Figure\,\ref{fig:isochrones} shows the YREC isochrone set at an
age of 10 Gyr for a range of metallicities \citep{an09}. The G- and
K-dwarf SEGUE color-cut regions are indicated. The most metal-rich 10
Gyr isochrones, calibrated to NGC 2862 and NGC 6791, have the
main-sequence turnoff before reaching the red edge of the color range
for the SEGUE G dwarfs. At younger ages, these metal-rich isochrones
cover the entire G ($g-r$) range, indicating that there is a chance we
are losing old, metal-rich G dwarfs due to evolutionary effects.

To investigate the extent of this evolutionary effect, we compare the
metal-rich samples of the two spectral types. Over the overlapping
distance range for both G and K dwarfs, approximately 3$\pm 1$\% of G
dwarfs have metallicities greater than 0. Around 10$\pm 3$\% of K
dwarfs are this metal rich. Between [Fe/H] of $-$0.3 and 0, the G- and
K-dwarf distributions are consistent with one another. This indicates
that our selection criteria loses metal-rich ([Fe/H]$\geq$0) G dwarfs
older than 5 Gyr.

\subsection{Correcting for Metallicity Biases from Target Selection} 
\label{sec:accounting_biases} 

For our G- and K-dwarf sample to accurately reflect the chemical
properties of these spectral types in the Galactic disk, we must
ensure that we constrain and account for the different metallicity
biases induced by SEGUE target selection.

There are three metallicity biases in the SEGUE G and K dwarf sample
that we correct for:
\begin{itemize} 
\item Other SEGUE target categories that are biased towards metal-poor
  stars overlap the color range of the SEGUE G and K dwarfs, biasing
  the color-selected sample in metallicity.
\item SEGUE has the same limited number of spectroscopic fibers for
  each line of sight, regardless of its stellar density.
\item The G- and K-dwarf color cuts select a range of stellar masses,
  and thus a subset of the mass function, that varies with
  metallicity.
\end{itemize}

To correct for the effects of target selection, we compare the
spectroscopic sample along each individual line of sight to the
photometric sample. For each pointing and target category, we scale
the spectroscopic distribution in $r_0$ such that it matches the
underlying $r_0$ magnitude distribution of all of the stars of that
\emph{could have} been observed spectroscopically. Our sample
consists of 152 different lines of sight, each with a different
distribution of spectral types and number of stars. Thus, we must
correct for the target-selection biases on a plate-by-plate basis.

For each bright and faint plate combination, we extract every possible
photometric stellar target within the plate radius that passes the G-
and K-dwarf color and magnitude criteria, whether or not it was
observed spectroscopically\footnote[6]{The photometric information is
  from PhotoObjAll database in DR7}. We then match each of these stars
with SEGUE spectroscopy. Matching photometry and spectroscopy is a
non-trivial task because some plates were observed multiple times and
many stars were repeated, some unintentionally, as geometric overlaps
from multiple plates, and others as purposeful
re-observations. Additionally, targets from SEGUE were occasionally
observed again as part of SEGUE-2. We removed all duplicate
spectroscopic observations, selecting the highest $S/N$ observations
using the sciencePrimary parameter. We also trimmed our spectroscopic
sample to eliminate any poor observations, removing all targets that
have $S/N<$10, an incalculable [Fe/H], or specific warning flags
(\S\,\ref{sec:extraction}). These targets remain within the
photometric sample, as they are simply counted as stars that do not
have spectra. For each spectroscopic target that fulfills the G- or
K-dwarf color and magnitude cut, we determine three weights:
target-type, r-magnitude, and mass-function. As we do not know the
$\log g$ of targets assigned SEGUE fibers in advance, we calculate
these weights for all usable spectroscopic G and K stars. We trim our
sample in surface gravity when we calculate our unbiased MDFs.

\subsubsection{Target-Type Weights} 
\label{sec:t_weights} 

\begin{figure} 
\centering
\includegraphics[width=\textwidth]{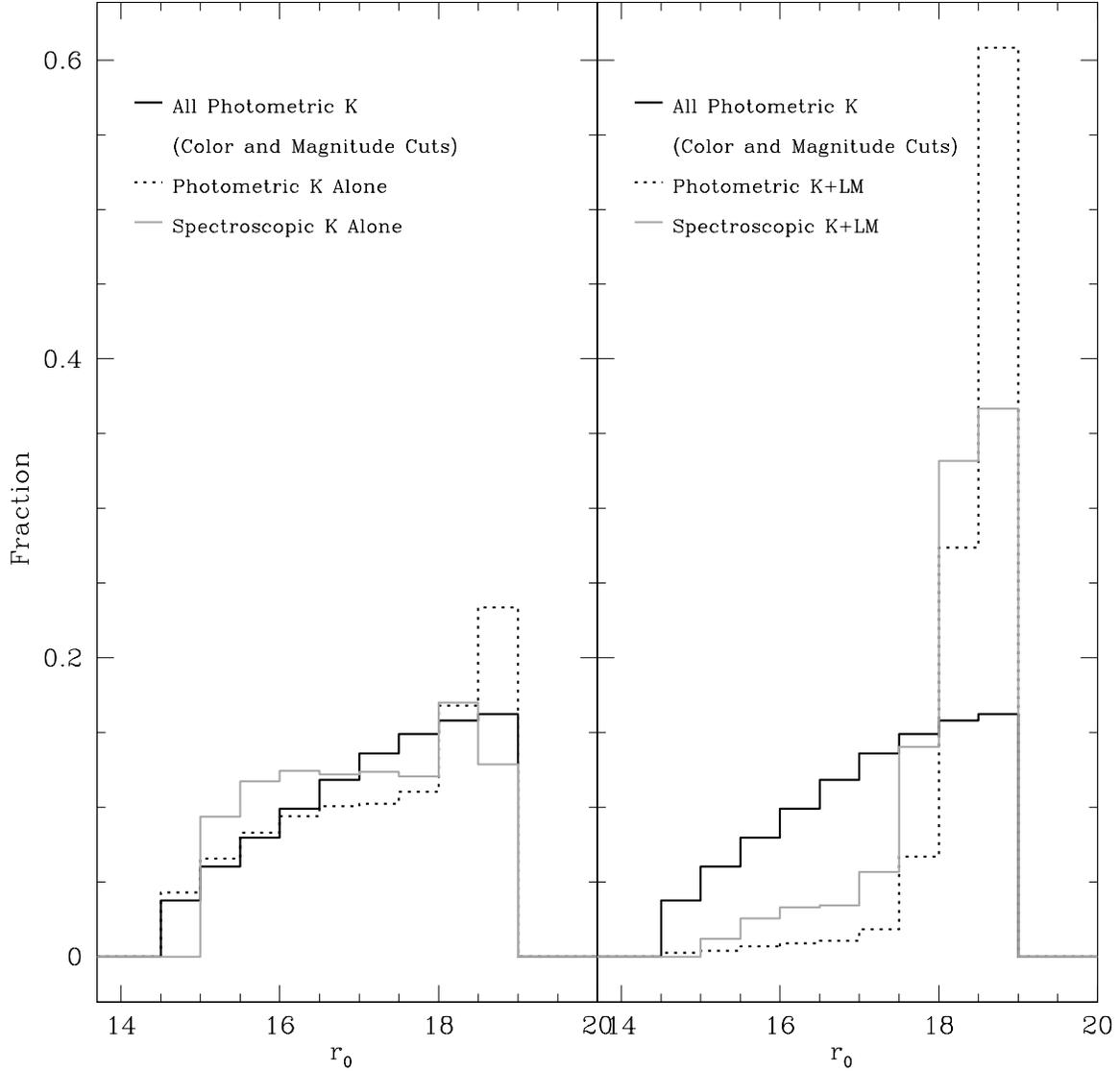}
\figcaption[target_type_categories]{The SEGUE G- and K-dwarf sample
  overlap with other SEGUE target categories, biasing them in
  metallicity space. The solid black line shows the $r_0$ distribution
  of all photometric stars along the SEGUE lines of sight that meet
  the color and magnitude criteria of a K dwarf. On the left hand
  side, we show the distribution of all photometric stars that meet
  \emph{only} the criteria of a K dwarf as the dotted line. We also
  present the $r_0$ distribution of all SEGUE spectroscopic
  observations that meet \emph{only} the criteria of a K dwarf. In
  contrast, on the right hand side we show the photometric and
  spectroscopic sample of stars that meet the criteria of a K dwarf
  \emph{and} a low-metallicity target. It has a significantly
  different distribution than the K-dwarf targets alone.
\label{fig:target_type_categories}}
\end{figure}

The simple selection criteria of SEGUE G and K dwarfs overlap with the
criteria for other different SEGUE targets. SEGUE categories often
focus on specific ranges in parameter space, such as stars with low
metallicity or small proper motions. Targets that fulfill multiple
target-type criteria have multiple opportunities to be assigned a
spectroscopic fiber. This leads to an overabundance of other stellar
categories in the G- and K-dwarf sample. For example, the SEGUE
low-metallicity star and K-giant categories use a $ugr$ selection
based around UV excess to identify targets, overlapping with the G and
K dwarf criteria in $(g-r)$. These target categories will bias the
SEGUE G- and K-dwarf sample towards metal-poor stars
(Figure\,\ref{fig:target_type_categories}).

Every stellar target observed photometrically by SDSS is labeled with
the SEGUE categories it fulfills. Any individual star may pass the
selection criteria for more than one target type and is then labeled
as being a candidate for more than one SEGUE spectroscopic sample. We
compare the number of stars identified as photometric candidates for
each target type and each combination of target types (i.e., a K-dwarf
and a low-metallicity target) with the number of those candidates that
have good spectra, as defined in \S\,\ref{sec:sample}. We assign a
``target-type weight'' such that the two distributions match. This
removes biases due to the overlapping target types in SEGUE, be they
over- or under-sampled by the spectroscopic fibers, ensuring that the
types of stars probed by SEGUE reflects that of the underlying
photometric sample. 

\begin{figure} 
\centering \includegraphics[width=\textwidth]{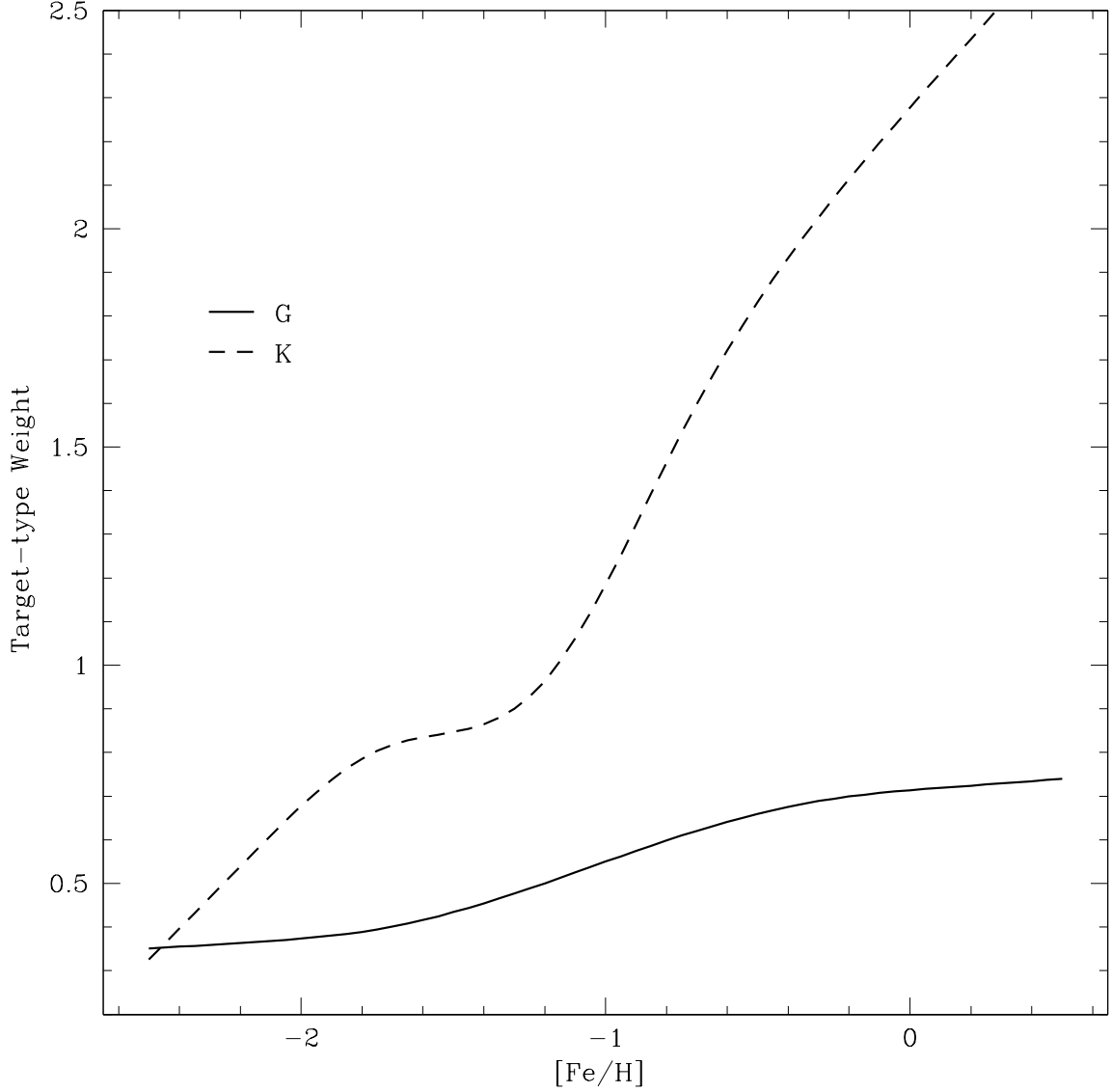}
\figcaption[type_weight_plotter]{A comparison of the mean target-type
  weight with respect to [Fe/H] for G (solid line) and K dwarfs
  (dashed line). There are fewer SEGUE fibers devoted to K-dwarf
  targets along each line of sight, making the sample more
  contaminated by stars assigned fibers for other stellar
  categories. The target-type weights account for this by scaling down
  the proportion of metal-poor stars and scaling the metal-rich end of
  the distribution. The variation in target-type weights for G-dwarf
  stars is much smaller, as it suffers less from target-selection
  biases.
\label{fig:type_weight_plots}}
\end{figure}

Figure\,\ref{fig:type_weight_plots} shows the mean target-type weight
with respect to [Fe/H] for both spectral types. This correction
reduces the proportion of metal-poor stars for both spectral
types. Otherwise, the K-dwarf weights are typically much higher than
those for G-dwarf stars. There is also more variation with
[Fe/H]. While the selections of metal-poor and K-giant stars in the
G-dwarf color region is relatively inefficient, the K-dwarf color
selection is more contaminated with stars from other categories,
biasing the metallicity structure. At [Fe/H]$\geq-$0.5, the K dwarfs
have a target-type weight of around 2.2, indicating that metal-rich K
dwarfs are under-represented by a factor of two. In contrast, the
target-type weight for metal-rich G dwarfs is approximately 0.7,
suggesting that they are over-sampled in SEGUE.  Below [Fe/H] of
$-$1.5, the target-type weights are around 0.7 and 0.4, for K and G
dwarfs respectively. Our calculated target-type weights will be
available with SDSS Data Release 9.

\citet{lee11b}, \citet{bovy11}, and \citet{liu12} determine their
stellar sample by extracting only stars assigned spectroscopic fibers
as G-dwarfs. This greatly diminishes the size of the spectroscopic
sample. Furthermore, although this method does not explicitly include
stars targeted as low-metallicity, that does not make it unbiased in
[Fe/H]. For example, if all of the low-metallicity stars along a
particular line of sight receive SEGUE fibers as low-metallicity
targets, none will be included in the G-dwarf sample, regardless of
whether or not they fulfill the criteria, creating a bias towards
metal-rich stars. Fortunately, the G-dwarf sample is significantly
less affected by the metallicity bias of SEGUE target selection
(Figure\,\ref{fig:type_weight_plots}).

\subsubsection{$r$-magnitude Weights}
\label{sec:r_weights}

SEGUE probes each line of sight with a limited number of fibers; the
survey cannot spectroscopically observe every target in a field. Lower
latitude pointings are closer to the plane of the Galaxy, they have
many more stars and tend to be more metal rich than high latitude
lines of sight. SEGUE assigns the same number of fibers to each line
of sight, regardless of latitude and stellar density. This leads to a
bias against metal-rich stars in the observed MDF of the full
survey. Furthermore, to accurately measure the properties of the Milky
Way, we must ensure that our spectroscopic sample reflects the
apparent-magnitude distribution of the underlying photometry along
each line of sight.

To calculate these ``$r$-magnitude weights'', we separate the
spectroscopic and photometric sample into G and K dwarfs based on
$(g-r)_0$ color. We examine the distribution in $r_0$ magnitude for
each individual spectral type, binning up the spectroscopic and
photometric targets in 0.5 magnitude bins from $r_0=$13 to 21.5
magnitudes. These bin sizes ensure that we typically have at least one
spectroscopic target associated with each group of photometric
stars. While the $r$-magnitude weights change when specifying a
smaller bin size in $r_0$, the metallicity distribution function
changes very little, well within the expected uncertainties for each
bin of [Fe/H]. We compare the number of spectroscopic targets in each
bin to the number of potential targets in the photometric sample. The
inverse of this ratio is the weight assigned to each spectroscopic
target with a measurable [Fe/H] and $S/N\geq$10 to recreate the parent
photometric distribution. This quantity, which we refer to as the
$r$-magnitude weight, accounts for the fact that SEGUE does not have
unlimited fibers.

Examining the faint end of the G-dwarf sample reveals that there are
very few usable spectroscopic targets with $S/N \geq$10 at the
faintest magnitudes. However, there are often many photometric targets
at these faint magnitudes, leading to large $r$-magnitude weights and
anomalous spikes in the metallicity distribution function. To
determine the true magnitude range of the usable spectroscopic sample,
we examine the cumulative distribution of spectroscopic targets in
$r_0$ for each plate. We find that 85\% of all spectroscopic G dwarfs
have $r_0 \leq$18.45. By setting 18.45 as our faint magnitude limit
for G dwarfs, we avoid anomalous weighting. The magnitude criteria for
K dwarfs in SEGUE are significantly more conservative than that for
the G dwarfs, extending to $r_0$ of 19; thus, the K-dwarf sample does
not have the same faint-end problem.

As K-dwarf stars are more frequent than G-dwarf stars, and sampled
more sparsely in SEGUE, these targets tend to have higher
$r$-magnitude weights. Both G- and K-dwarf stars also have high
$r$-magnitude weights for lines of sights that probe lower Galactic
latitudes.  For G dwarfs with [Fe/H]$\geq-$0.5, the typical
$r$-magnitude weight is around 13; the typical weight for K dwarfs is
around 15. This value decreases to 8 and 11 for G and K dwarfs below
[Fe/H] of $-$1.5, respectively. As with the target-type weights, the
$r$-magnitude weights will be available to the public as part of SDSS
Data Release 9.

\subsubsection{Mass Function Weights} 
\label{sec:m_weights}

The $(g-r)_0$ cut that defines the G- and K-dwarf targets corresponds
to different mass ranges for each metallicity. For example, the
G-dwarf color cut isolates stars with mass near 0.7\,\msun for [Fe/H]
of $-$1.0 and near-solar mass stars for solar metallicity. Mass
functions for the Galaxy predict a larger number of less-massive
stars; thus, the color cut of SEGUE will result in a metallicity bias
in our sample. Previous studies (e.g., \citealt{jorgensen00})
discarded stars to find a mass range over which their samples were
reasonably complete, but, due to our narrow color range, this would
drastically limit our sample. The masses probed by our color and
metallicity range are typically from 0.5 to 1.0\,\msun, but do not
necessarily overlap for each metallicity bin. As this is not a large
range in mass, we expect the metallicity bias introduced by our color
cuts to be a small effect compared to other observational biases that
affect the sample.

We employ the TRILEGAL 1.4 model (\S\,\ref{sec:trilegal_models}),
which utilizes a Chabrier mass function, to estimate this effect. We
extract two samples from the model for each individual line of
sight. The first sample has the same color and magnitude range as the
G- or K-dwarf sample and is restricted to stars with $\log g \geq$4.1,
to identify main-sequence stars. We bin this sample in metallicity for
each spectral type and examine the distance range the sample covers at
each [Fe/H].  The second sample is also limited to stars on the main
sequence and in the G or K magnitude range. However, rather than a
color cut, we extract all stars with masses between 0.5 to 0.6\,\msun
that fall within the same distance range of stars in the color
cut. This ensures we are comparing the two samples over the same
volume of space. For each metallicity bin, we compare the number of
stars that fulfill the color criteria to the number of targets with
masses between 0.5 and 0.6\,\msun within the same volume. This ratio
indicates how much to scale each metallicity bin of the G or K dwarfs
to simulate sampling a consistent portion of the mass function over
the entire metallicity range. As the mass function varies slightly
from plate to plate, we calculate these weights for each SEGUE
pointing.

We also calculate mass weights based on the
\citet{schonrich09a,schonrich09b} Galaxy models (SB), which utilize a
Salpeter IMF \citep{salpeter55}, to examine how much these values will
vary with different assumptions about the mass function
(\S\,\ref{sec:schoenrich_models}). The distribution of weights from
the SB models is comparable to the mass-function weights derived from
the TRILEGAL simulation; using SB values rather than TRILEGAL has a
negligible effect on the metallicity distribution. The TRILEGAL models
cover a wider metallicity range than those of
\citet{schonrich09a,schonrich09b}, reaching lower metallicities, so we
use these weights throughout our analysis.

As with the other two weights, the mass-function weights decrease the
proportion of metal-poor stars. However, in contrast to the weights
for overlapping categories and spectroscopic sampling, the
mass-function weights are not large; above [Fe/H] of $-$0.5, both G
and K dwarfs have a typical mass weight of around 1.0. For metal-poor
stars ([Fe/H]$\leq-$1.5), the typical mass weight is 0.7 for G dwarfs
and 0.9 for K dwarfs. These values will not be released as part of
SDSS Data Release 9, as other groups will have their own preferred
mass function.

\subsubsection{Testing the Target-Selection Weighting on Galaxy Models} 
\label{sec:galaxy_modeling} 

\begin{figure} 
\centering
\includegraphics[width=\textwidth]{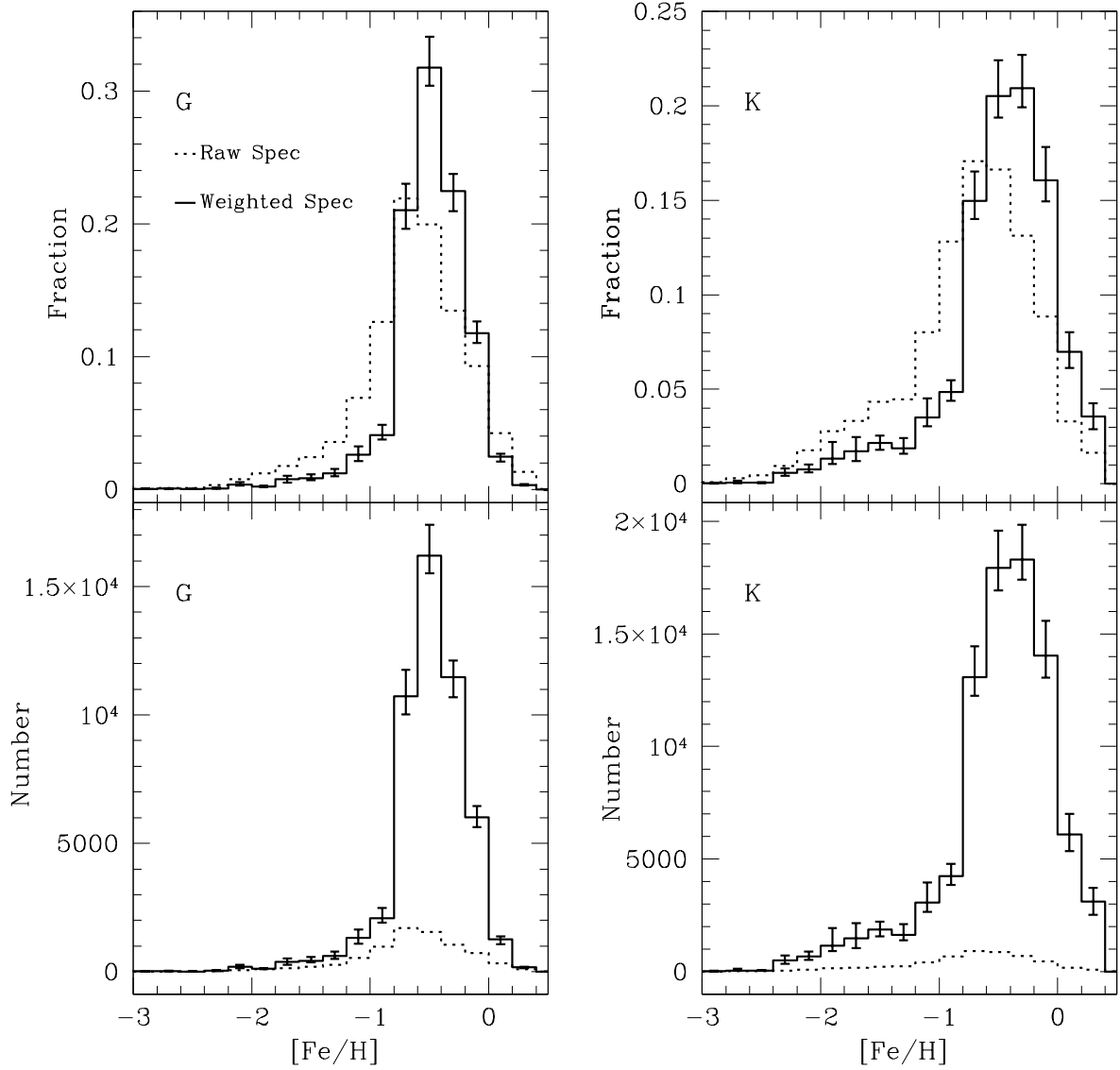}
\figcaption[tp_wcompst_0.2_all_500_050312.eps]{The raw and weighted
  MDF for the G (left) and K (right) dwarfs. The raw spectroscopic
  sample is the dashed line. The sample adjusted for target-type,
  $r$-magnitude, and mass-function weights is the solid line. The
  error bars on the solid line reflect the error in each bin based on
  our total MDF uncertainty, combining our bootstrap analysis with
  other errors (\S\,\ref{sec:monte_carlo}). Both samples are limited
  to their spectral-type distance ranges, as described in
  \S\,\ref{sec:vol_comp}. Note that the corrections associated with
  the target selection scheme boost the metal-rich end of the
  distribution.
\label{fig:overall} }
\end{figure}

Figure\,\ref{fig:overall} shows the original and adjusted MDF for G
and K dwarfs over all the SEGUE lines of sight for the spectral-type
distance range. As explained in
\S\,\ref{sec:t_weights},\,\ref{sec:r_weights},
and\,\ref{sec:m_weights}, the corrections for SEGUE target-selection
biases significantly affect the MDF, increasing the metal-rich end.

\begin{figure} 
\centering
\includegraphics[width=\textwidth]{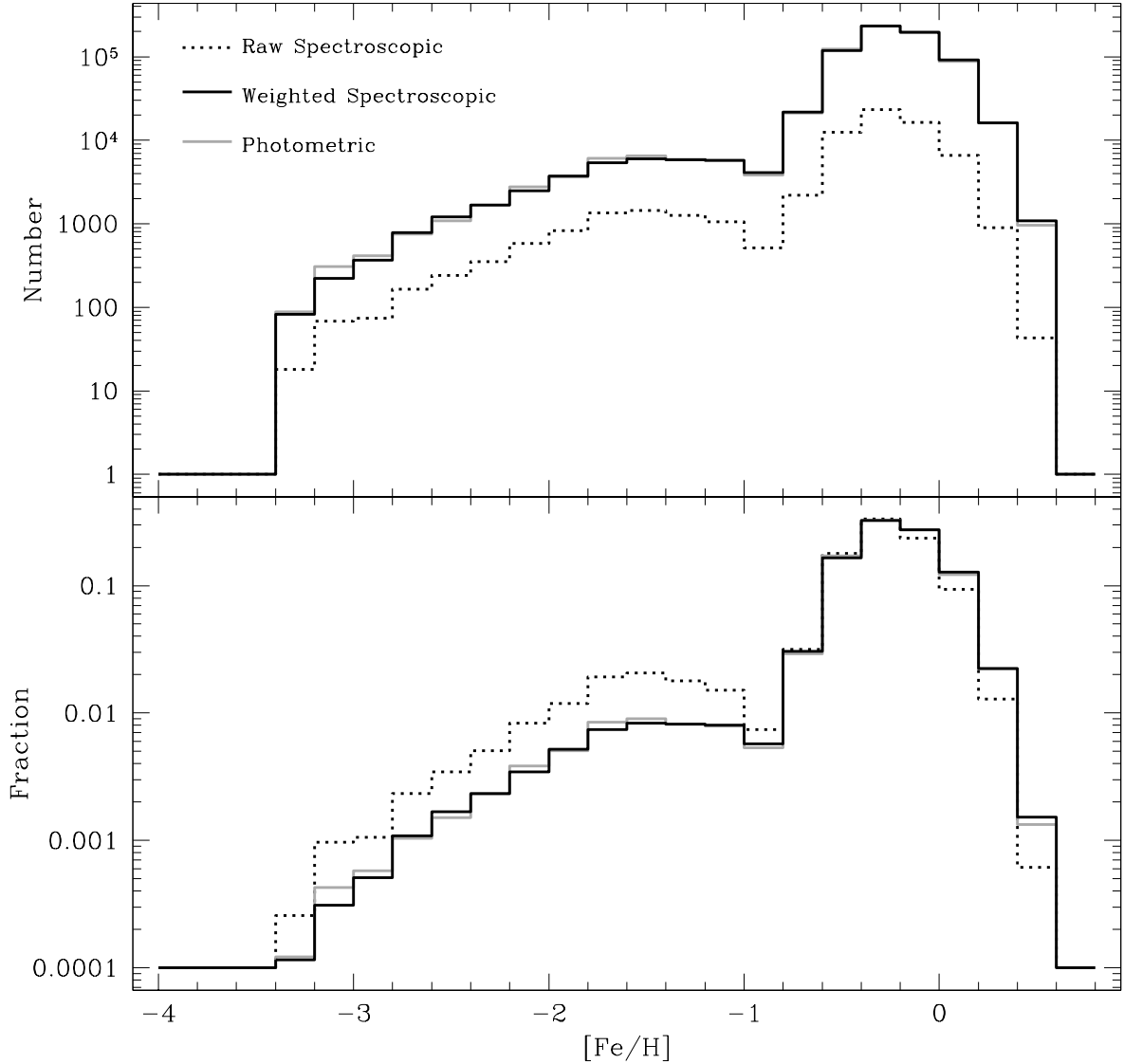}
\figcaption[trilegal_Test_fig1_1117.eps]{A test of the target-type and
  $r$-magnitude weighting system on the TRILEGAL Galaxy model. The
  solid black line represents the distribution of a randomly selected
  subsample of G and K dwarfs representing the SEGUE spectroscopic
  sample, which we refer to as the ``raw spectroscopic.'' The gray
  line is the distribution of every modeled star that meets the G- and
  K-dwarf photometric criteria. Finally, the dashed line is the
  distribution of the spectroscopic sample once we have weighted it in
  $r$-magnitude and target type. The top figure shows the numerical
  distribution of the three samples; the good match between the
  weighted spectroscopic and photometric distributions indicates that
  our technique accurately corrects for SEGUE target selection
  biases. The bottom figure shows the fractional distribution of the
  three samples. The ``raw'' spectroscopic sample is biased towards
  metal-poor stars, but our weighting scheme corrects for
  this. \label{fig:tri_rmagtype_all} }
\end{figure}

To ensure that our weighting algorithm appropriately removes biases
from target selection, we test it on a series of modeled lines of
sight from the TRILEGAL 1.4 theoretical Galaxy model
\S\,\ref{sec:trilegal_models}. We analyze each TRILEGAL line of sight
as a photometric sample for SEGUE target selection, labeling all
modeled stars which meet the color and magnitude criteria of the SEGUE
G- and K-dwarf categories.  We randomly select 375 G and 95 K dwarfs
for each line of sight as our ``spectroscopic'' observations,
weighting this mock SEGUE sample according to the methodology
discussed in \S\ref{sec:r_weights} and \S\ref{sec:m_weights}. We then
compare the weighted parameter distribution in [Fe/H] in two ways to
that of the underlying sample.  First, we compare the
``spectroscopic'' sample, weighted in target type and $r$-magnitude,
to the modeled photometric sample (see
Fig.\,\ref{fig:tri_rmagtype_all}). The combination of these two
weights recreates the metallicity distribution for all modeled stars
that fulfill the G- and K-dwarf criteria. This suggests that our
photometry-based weighting techniques allow us to examine the
metallicity structure of the underlying stellar populations in the
disk using our spectroscopic sample.

\begin{figure} 
\centering
\includegraphics[width=\textwidth]{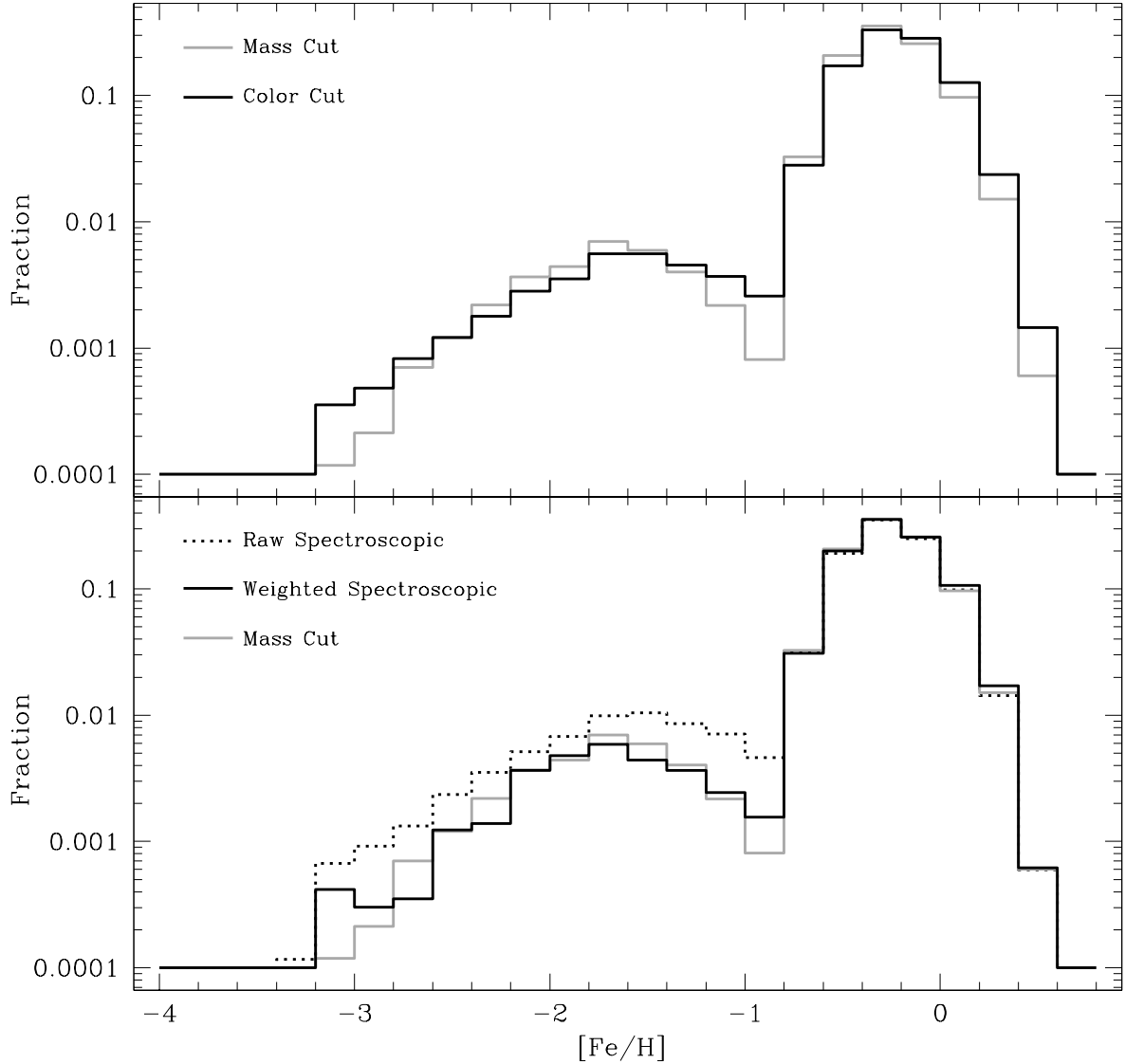}
\figcaption[trilegal_Test_fig2_1117.eps]{A test of the mass-function
  weights on the SEGUE lines of sight modeled with TRILEGAL. The top
  figure compares the effect of a color cut on the metallicity
  distribution to a mass cut. The color cut extracts all stars within
  the magnitude and $(g-r)$ range of the SEGUE G and K dwarfs, shown
  in black. The mass cut selects all stars within the magnitude range
  with masses between 0.5 and 0.6\,\msun, shown in gray. Each
  metallicity bin of the two samples probes the same volume of space.
  The two distributions are quite similar, although there are some
  discrepancies at the metal-poor end. The bottom row tests our
  weighting scheme. The gray distribution remains the same as in the
  top plot. The additional distributions represent a randomly chosen
  G/K dwarf subsample to represent SEGUE spectroscopy (black) and this
  distribution weighted in target type, $r$-magnitude, and mass
  function (dashed). The weighted distribution scales the raw
  spectroscopic down at the metal-poor end and boosts the metal-rich
  end, such that it simulates isolating a uniform portion of the mass
  function. \label{fig:tri_massfunc_all} }
\end{figure}

Our second comparison, shown in Fig.\,\ref{fig:tri_massfunc_all}, uses
the target-type, $r$-magnitude, and mass-function weight to recreate
the distribution of all stars that meet the G- and K-dwarf magnitude
criteria and fall in a uniform mass range. The underlying photometric
sample that we tested the target-type and $r$-magnitude weights on is
slightly biased in mass, as explained in \S\,\ref{sec:m_weights}. The
top plot compares a $(g-r)$ color cut to a mass cut in metallicity
space, where both samples cover the same volume for each individual
metallicity bin. There is little difference between the two
distributions. The bottom plot indicates that the color cut manifests
as a bias towards metal-poor stars. Our target-selection weights
account for this, removing the color-cut effect such that it appears
our sample is extracted from a uniform portion of the mass
function. We also tested our target-selection corrections on the
Galaxy model of Sch{\"o}nrich \& Binney (SB, 2009a,b,
\S\,\ref{sec:schoenrich_models}), with similar results.

\subsection{Summary of the G- and K-dwarf Sample} 
\label{sec:gk_sample_summary}

Our SEGUE sample consists of 16,847 K dwarfs with $S/N\geq$10, surface
gravities larger than $\log g=$4.1, appropriate $r_0$ magnitudes, and
E($g-r$) less than 0.2 mag. This sample has
$-$3.4$\leq$[Fe/H]$\leq$0.6, 4.1$\leq \log g \leq$5.0, and
4150$\leq$T$_{eff}\leq$6315 K. It contains stars with distances from
0.3 to 5.8 kpc and covers Galactic radii from 4.8 to 13.4 kpc and
heights from the plane up to $\pm$4.0 kpc.  We have 24,270 G dwarfs
with the appropriate $S/N$, surface gravity, extinction, and
$r_0$. This sample ranges from [Fe/H] of $-$3.1 to 0.6, surface
gravities up to 5.0, and effective temperatures from 4210 to 6300
K. It extends to larger distances than the K-dwarf sample, from
$\sim$0.4 to 10.2 kpc, covering from 3.7 to 16.5 kpc in Galactic
radius and -7.7 to 9.7 kpc in height above the Galactic plane.

We have performed the first complete analysis of the different
possible biases in this SEGUE sample.  The color and magnitude
selection criteria of the two categories result in a sample that is
quite clean, with few stars of inappropriate color being shifted into
the sample by photometric errors or undetected binarity. We also
remove lines of sight from our sample with E$(g-r)$ greater than 0.2
mag and establish that uncertainties in the \citet{sfd98} extinction
measurements have little effect on our distance estimates.

We also constrain how stars of different evolutionary phase affect our
sample. Combining an analysis of $\log g$ with MgH and Galaxy models,
we determine that the SEGUE G and K dwarfs will have a negligible
amount of contamination from evolved stars. We will lose old,
metal-rich stars from the G-dwarf sample due to evolution; our $(g-r)$
color cut biases the G-dwarf sample against stars with [Fe/H]$\geq$0.

We estimated the effect of different uncertainties in the G- and
K-dwarf sample on the MDF (see Fig.\,\ref{fig:mdf_per_error}). The
uncertainties are dominated by distance errors, both random and
systematic. None of the various uncertainties, from photometric errors
to undetected binarity, induce a bias in the metallicity of the
sample. In our MDF, we include every different source of uncertainty
in the structure, combining them in quadrature
(\S\,\ref{sec:monte_carlo}).

Although issues ranging from surface gravity to extinction have little
effect on our derived MDF, the target selection algorithm of SEGUE
does bias the sample in metallicity space, favoring metal-poor
stars. The same number of stars are observed in fields at low Galactic
latitude, which have a high stellar density and are dominated by
high-metallicity stars, as in those at high Galactic latitude, where
the stellar density is much lower and the stars are primarily
metal poor. Thanks to the quantitative target-selection algorithm of
SEGUE, we have both quantified and corrected for these different
biases using a series of weights. The uncertainties from these values
are integrated into our MDF uncertainties via a bootstrap analysis
(\S\,\ref{sec:monte_carlo}).

\section{The Metallicity Distribution Functions of G and K Dwarfs over the Disk} 
\label{sec:mdf} 

\subsection{MDF of the Disk as a Whole}

\begin{figure} 
\centering
\includegraphics[width=\textwidth]{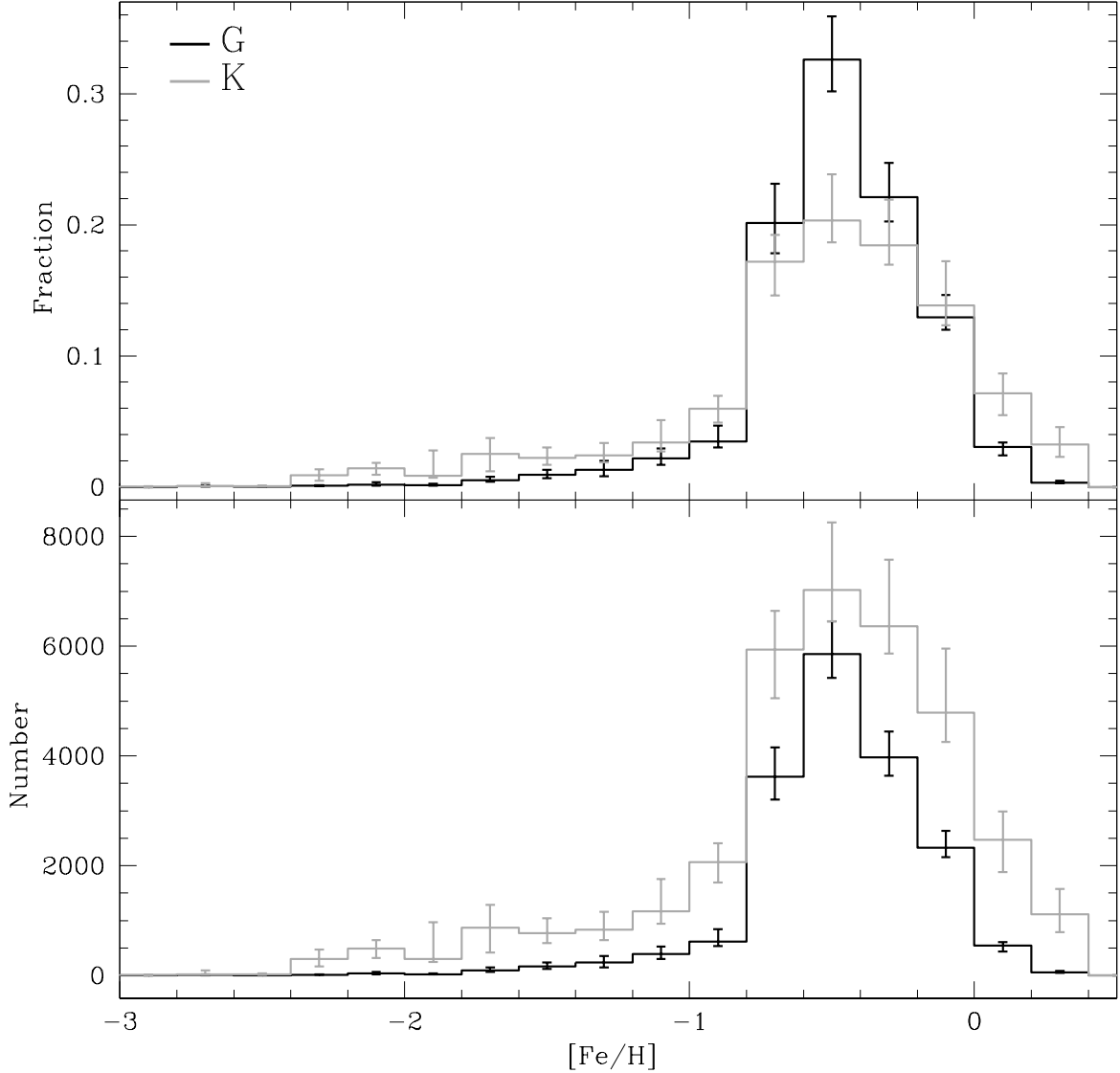}
\figcaption[tp_vc_0.2_all_500_050312.eps]{Comparison of the G- (black)
  and K-dwarf (gray) MDFs. Both samples are limited to distances
  between 1.59 and 1.84 kpc. The error bars are based upon our total
  MDF uncertainty, combining a bootstrap analysis over 500 iterations
  with other sample uncertainties (\S\,\ref{sec:monte_carlo}). The two
  samples cover a similar metallicity range and peak at approximately
  the same value. At low $|Z|$, K dwarfs have more metal-rich stars,
  due to evolutionary effect. At high $|Z|$, they also have more
  metal-poor stars. This results in a broader distribution than
  observed for the G dwarfs over all $|Z|$.
\label{fig:overall_comp} }
\end{figure}

For each spectral type, we determine the distribution of stars with
respect to metallicity and then apply our target selection weights
such that the distribution reflects the chemical structure of the
underlying disk population. This allows us to improve on previous
analyses, namely, photometric metallicity estimates, limited volume
coverage, observational biases, and/or small sample
size. Figures\,\ref{fig:overall} and\,\ref{fig:overall_comp} show the
total MDF for G and K dwarfs over the spectral type and overlapping
distance ranges, respectively. This is the first unbiased MDF using
spectroscopic metallicities over a substantial volume of the disk for
both G and K dwarfs.

\begin{figure} 
\centering 
\includegraphics[width=\textwidth]{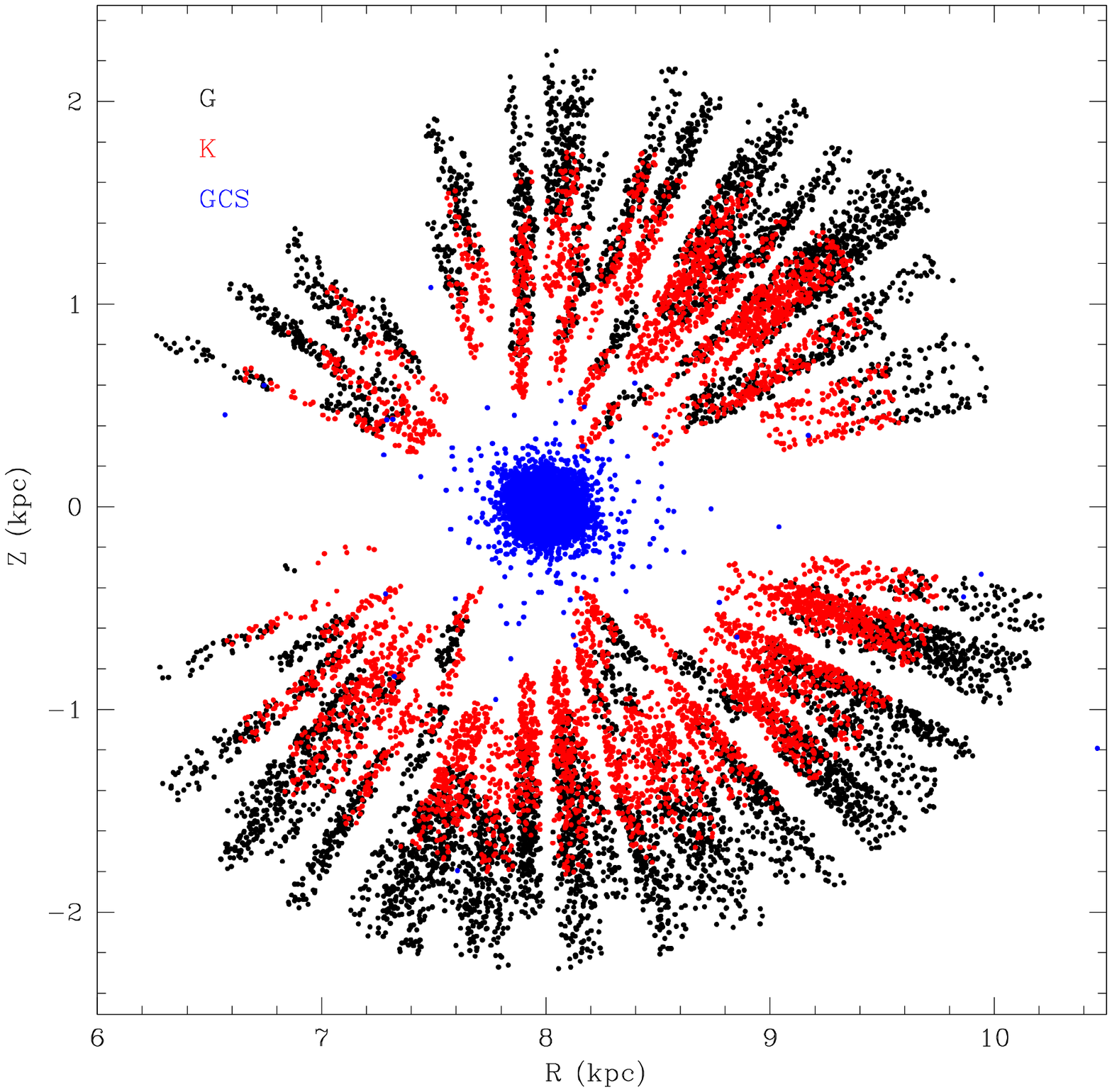}
\figcaption[spatial_gk_gcs_1014.eps]{The spatial positions of our G
  (black) and K (red) dwarfs. The Geneva-Copenhagen Survey (GCS)
  sample of stars is shown in blue \citep{nordstrom04,
    haywood08,casagrande11}. The G dwarfs are limited to distances
  between 1.59 and 2.29 kpc, whereas K dwarfs fall between 1.18 and
  1.84 kpc, based on the magnitude limits of SEGUE. This reduces our
  stellar sample to 5,407 K- and 7,834 G-dwarfs. Our sample probes the
  disk both towards and away from the Galactic center and above and
  below the Galactic plane, with the line-of-sight structure of SEGUE
  clearly visible. The GCS sample is limited to the solar
  neighborhood, with distances typically less than 200 pc.
\label{fig:spatial_gkstd} } 
\end{figure}

These MDFs provide information about the chemical structure of the
disk from approximately 0.2 to 2.3 kpc above and below the plane of
the Galaxy, and from Galactic radii of around 6 to 11 kpc
(Figure\,\ref{fig:spatial_gkstd}). The MDF of both G and K dwarfs
range from metallicities of around $-$2.5 to $+$0.4. The distributions
peak at around [Fe/H] of $-$0.4 to $-$0.6, indicating that the sample
has a large population of thick-disk stars, in addition to substantial
contributions from the thin disk with [Fe/H]$=-$0.2 \citep{gilmore85,
  wg95, chiba00}. Unlike the dwarf stars in the GCS and RAVE surveys,
the SEGUE sample has coverage of both the \emph{in-situ} thin and
thick disks. Thus, our total MDFs reflect the overall disk properties,
rather than an individual component of the Milky Way disk. Detailed
information about these distributions for further analysis is
available in Table\,\ref{tab:total_gkmdf}.

\subsection{Uncertainties in the Metallicity Distributions} 
\label{sec:monte_carlo} 

We use a two-pronged approach to estimate the uncertainties in our
MDF. First, we run a bootstrap analysis on our data set, which
provides information about the variation in the sample, with respect
to atmospheric parameters, photometry, and target-selection weights
while taking into account that we do not know the true underlying
metallicity distribution. We randomly select G and K dwarfs from the
original spectroscopic sample until our modified version has the same
number of stars of each spectral type, sampling with replacement,
e.g., an individual target can show up multiple times in the bootstrap
sample. We then examine the variation in the metallicity distributions
of these iterations, comparing with our actual result. As the
lower-metallicity bins have a smaller number of stars, their
uncertainties are typically larger (Fig.\,\ref{fig:mdf_per_error});
below [Fe/H] of $-$1, the uncertainty in the fraction in each bin is
around 40\%. Above this metallicity, the uncertainty in each bin is
typically around 8\%.

Errors from photometry, undetected binarity, $\log g$, and extinction
typically manifest as an additional uncertainty per bin of around 5\%
each and induce no significant metallicity biases in the distribution
(\S\,\ref{sec:contamination}). However, random and systematic distance
errors contribute a significant amount of uncertainty to our MDFs,
around 40\% for low-metallicity bins
(\S\,\ref{sec:paper_distunc}). The total MDF uncertainties are a
combination of the measured variation from all of these different
sources, combined with those from our bootstrap analysis. These
account for both the parameter and selection uncertainties and the
fact that we do not know the true underlying distribution. We combine
the different uncertainties in each MDF metallicity bin in quadrature;
they are listed in Table\,\ref{tab:total_gkmdf} for both G and K
dwarfs over different distance ranges. As the lower-metallicity bins
have a smaller number of stars, their uncertainties are typically
larger; below [Fe/H] of $-$2, the uncertainty in the fraction in each
bin is around 70\%. Between [Fe/H] of $-$2 and $-$1, the uncertainties
drop to 24\%; above this metallicity, the uncertainty in each bin is
typically around 10\%. In addition, the correlated [Fe/H] and distance
uncertainty will broaden the overall derived MDF
(\S\,\ref{sec:vol_comp}, Fig.\,\ref{fig:correlated_mdf_err}).

\subsection{Variation with Height above the Plane} 
\label{sec:mdf_wrt_z} 

The chemical structure of the disk in Galactic height is a useful
diagnostic for dynamical formation models of the Milky Way. It also
reveals the interplay of the proposed thin- and thick-disk by probing
both components with an \emph{in-situ} spectroscopic sample.

\begin{figure} 
\centering
\includegraphics[width=\textwidth]{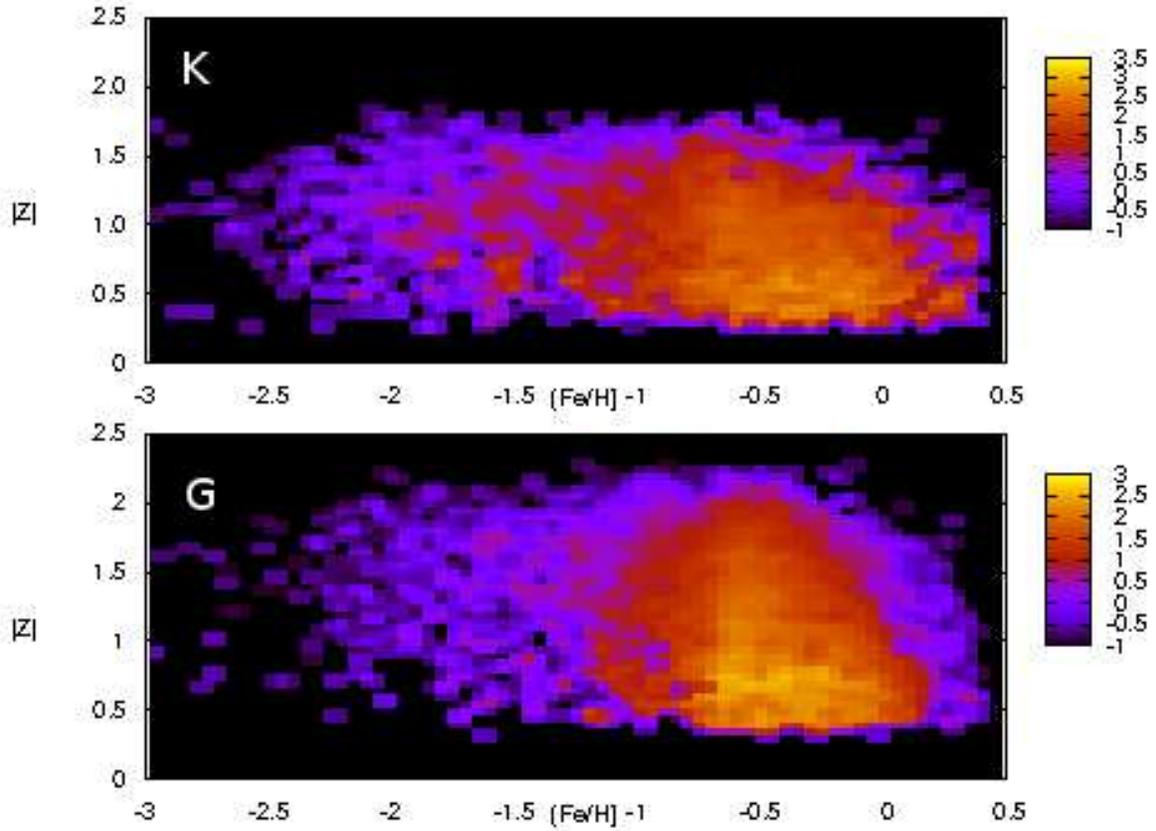}
\figcaption[hess_medians_fehcut_1014.eps]{The distribution of G
  (bottom) and K dwarfs (top) in [Fe/H] with respect to $|Z|$. Stars
  are sorted into bins of [Fe/H] and $|Z|$. The number of stars in
  each bin is then adjusted using the target-type, $r$-magnitude, and
  mass-function weights. The logarithm of the weighted number of
  targets is indicated by color, as labeled on the right
  side. \label{fig:hess_diagram}}
\end{figure}

\begin{figure} 
\centering
\includegraphics[width=\textwidth]{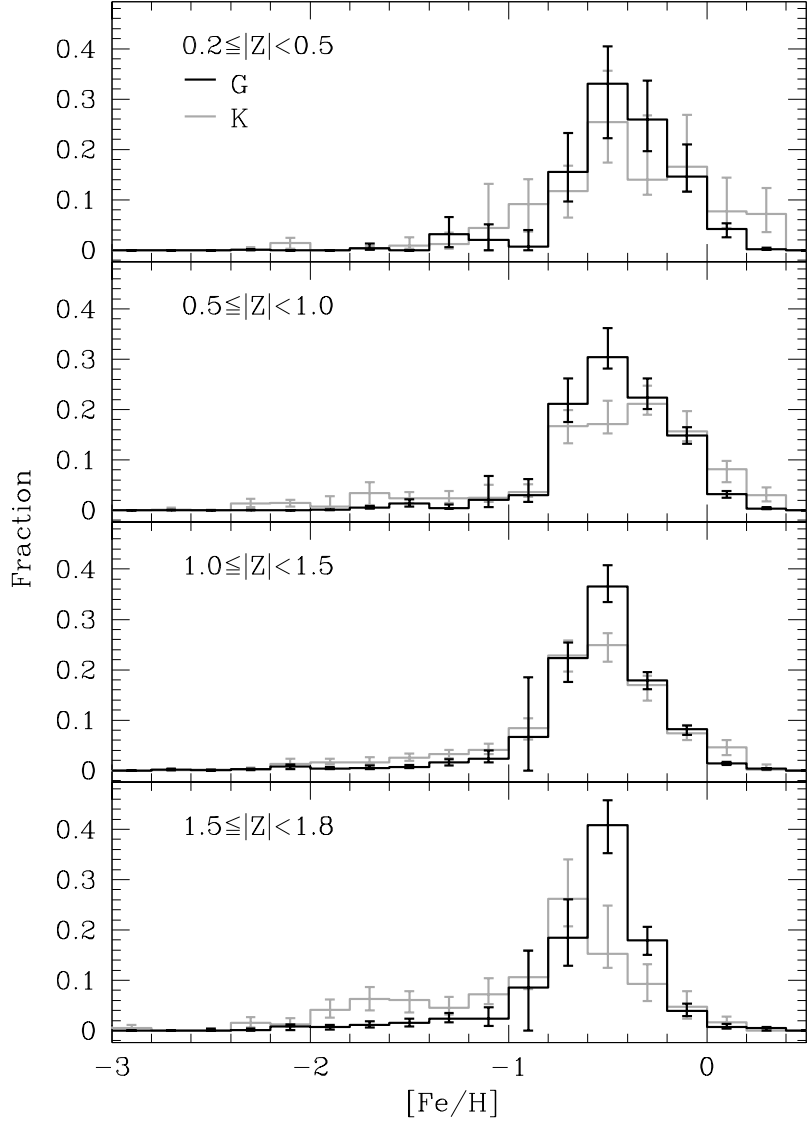}
\figcaption[zbp_frac_vc_0.2_all_500_050312.eps]{The weighted MDF over
  the overlap distance range for the G (black) and K (gray) dwarfs at
  a range of $|Z|$, where $|Z|$ is the current distance from the plane
  in kpc. The uncertainty in each bin was determined from our
  bootstrap analysis over 500 iterations combined with additional
  measured uncertainties (\S\,\ref{sec:monte_carlo}). These
  distributions are listed in Table\,\ref{tab:gd_zbin_vc} and
  \ref{tab:kd_zbin_vc}.
\label{fig:mdf_z_kdgd_dcut} } 
\end{figure}

Our G- and K-dwarf samples probe far above and below the Galactic
plane; Figure\,\ref{fig:hess_diagram} shows the distribution of stars
with respect to [Fe/H] and distance from the plane, using the
spectral-type distance ranges for G and K dwarfs. We bin the sample in
$|Z|$ and [Fe/H], adjusting each bin with the target-selection weights
for each individual star. The colors represent the logarithm of the
weighted number of stars in each bin. We compare the MDFs of the G and
K dwarfs over the overlapping distance range at different heights
above the plane of the Galaxy in Figure\,\ref{fig:mdf_z_kdgd_dcut}.

\section{Discussion} 
\label{sec:discussion} 

\subsection{The MDF for Different Spectral Types} 
\label{sec:st_variation} 

Past work by \citet{favata97} found a distribution of nearby K-dwarf
stars that was significantly more deficient in metal-poor stars than
their G-dwarf sample. They conjectured that this result could
originate from either a metallicity bias in their original catalog or
metal-enhanced cool star formation. However, later work by
\citet{rpmaciel98} found their K-dwarf sample in good agreement with
their G dwarfs. Similarly, \citet{kotoneva02} found a much broader
distribution of K dwarfs with respect to [Fe/H] than
\citet{favata97}. Likely, the narrow distribution of K dwarfs in
\citet{favata97} was due to small sample size and a tendency to target
brighter stars for spectroscopy.

We compare the MDF of the weighted SEGUE G and K dwarfs over the
overlapping distance range in Figure\,\ref{fig:overall_comp}.  Our
analysis confirms the work of \citet{rpmaciel98}, which found the
distribution of metallicity in G and K dwarfs in the solar
neighborhood to be in good agreement. For the overlap region for the G
and K dwarfs, i.e., between 1.59 and 1.84 kpc, both spectral types
peak at around [Fe/H] of $-$0.5, consistent with the peak metallicity
for the $\alpha$-enhanced sample identified by \citet{lee11b}. The
peak agreement of the two spectral types implies that the
chemical-enrichment and star-formation history for the G and K dwarfs
are quite similar.

The MDFs with respect to $|Z|$ of the two spectral types differ at the
extremes of the metallicity distribution
(Figure\,\ref{fig:mdf_z_kdgd_dcut}). For heights below 1 kpc, the K
dwarfs have a larger population of metal-rich stars. As discussed in
\S\,\ref{sec:evolutionary}, our color criteria for G dwarfs will
remove the most metal-rich stars with ages greater than 5 Gyr. Above 1
kpc, there are few stars at or above solar metallicity, so the
discrepancy between the two spectral types is no longer noticeable.

\begin{figure}
\centering \includegraphics[width=\textwidth]{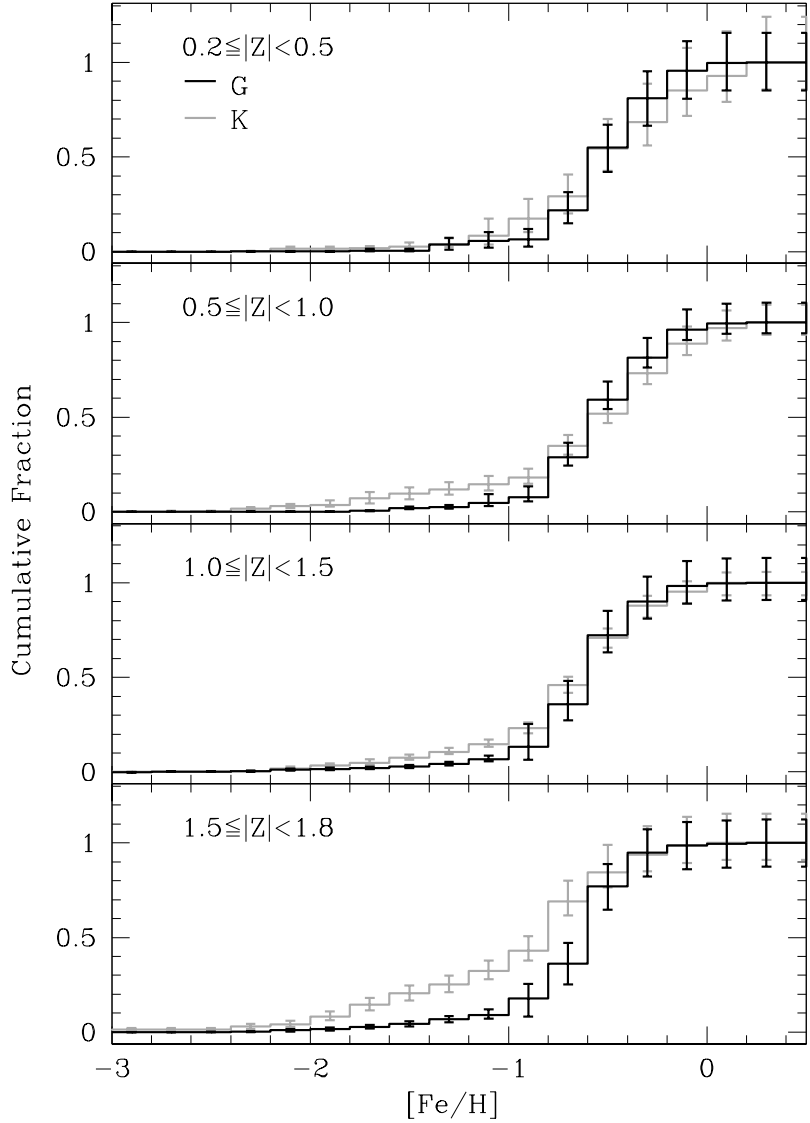}
\figcaption[]{The cumulative weighted MDF over the overlap distance
  range for the G (black) and K (gray) dwarfs at a range of $|Z|$,
  where $|Z|$ is the current distance from the plane in kpc. We
  combine the error bars for each individual metallicity bin (from our
  bootstrap and Monte Carlo analysis) together in quadrature to
  determine the uncertainty for each cumulative bin.  Above $|Z|$ of
  0.5 kpc, the K dwarfs show a more prominent metal-poor tail than the
  G dwarfs. Note that the fractions of the G- and K-dwarf sample below
  [Fe/H] of $-$1.0 are consistent until $|Z|\geq$1.5 kpc.
\label{fig:mdf_z_kdgd_cumu} } 
\end{figure}

When we examine the metal-poor end for the G- and K-dwarf samples at
different $|Z|$, we find a discrepancy
(Figure\,\ref{fig:mdf_z_kdgd_cumu}). For $|Z|$ distances below 1.5
kpc, the fractional contribution of stars with [Fe/H]$<-$1.0 agree
within 1$\sigma$ (Figure\,\ref{fig:mdf_z_kdgd_cumu}). The percentage
of stars below [Fe/H] of $-$1.0 is consistent for the two spectral
types up to $|Z|=$1.5 kpc. However, at $|Z|$ greater than 1.5 kpc, the
percentage of metal-poor ([Fe/H]$<-$1.0) K dwarfs is significantly
larger than the metal-poor G dwarfs. For $|Z|>$1.5 kpc, the fraction
of G dwarfs increases from $\sim$6\% at smaller $|Z|$ to
9$_{-5}^{+7}$\%. The K-dwarf percentage increases significantly more,
from around 15\% for 0.2$\leq |Z|<$1.5 kpc to 32$_{-12}^{+15}$\% above
$|Z|$ of 1.5 kpc (see Tables\,\ref{tab:gd_zbin_st},
\,\ref{tab:kd_zbin_st}, \,\ref{tab:gd_zbin_vc},
\,\ref{tab:kd_zbin_vc}). For smaller $|Z|$, the median metallicity for
both G and K dwarfs agree to within 0.04 dex. For the highest $|Z|$
bin, the median metallicity for the G dwarfs is $-$0.55, whereas the
K-dwarf median [Fe/H] is $-$0.74, much more metal-poor.

The origins of this discrepancy are currently unclear. The optimized
version of the SSPP ensures that the [Fe/H] estimates for the G and K
dwarfs are consistent with one another; there is no systematic offset
in SSPP metallicity between the two spectral types of this size
(Appendix\,\ref{sec:feh_det}). One possible cause of the MDF
differences at large $|Z|$ is our assumed binarity fraction. If there
is a larger probability of an undetected companion for K dwarfs than G
dwarfs, we will be underestimating the distance to more K stars. This
will cause more metal-poor stars to fall into the K-dwarf sample.

Examination of the G- and K-dwarf MDF at different heights above the
Galactic plane reveals that the K dwarfs are more metal-rich at low
heights and more metal-poor at large $|Z|$. This makes the total
K-dwarf MDF appear broader than the G-dwarf MDF over the overlapping
distance range. As the fraction below [Fe/H] of $-$1.0 from both
spectral types is consistent within 2$\sigma$, we leave a more
in-depth study of the metal-poor end of the G- and K-dwarf MDF for a
future paper.

\subsection{Extending the G-Dwarf Problem} 

\begin{figure}
\centering
\includegraphics[width=\textwidth]{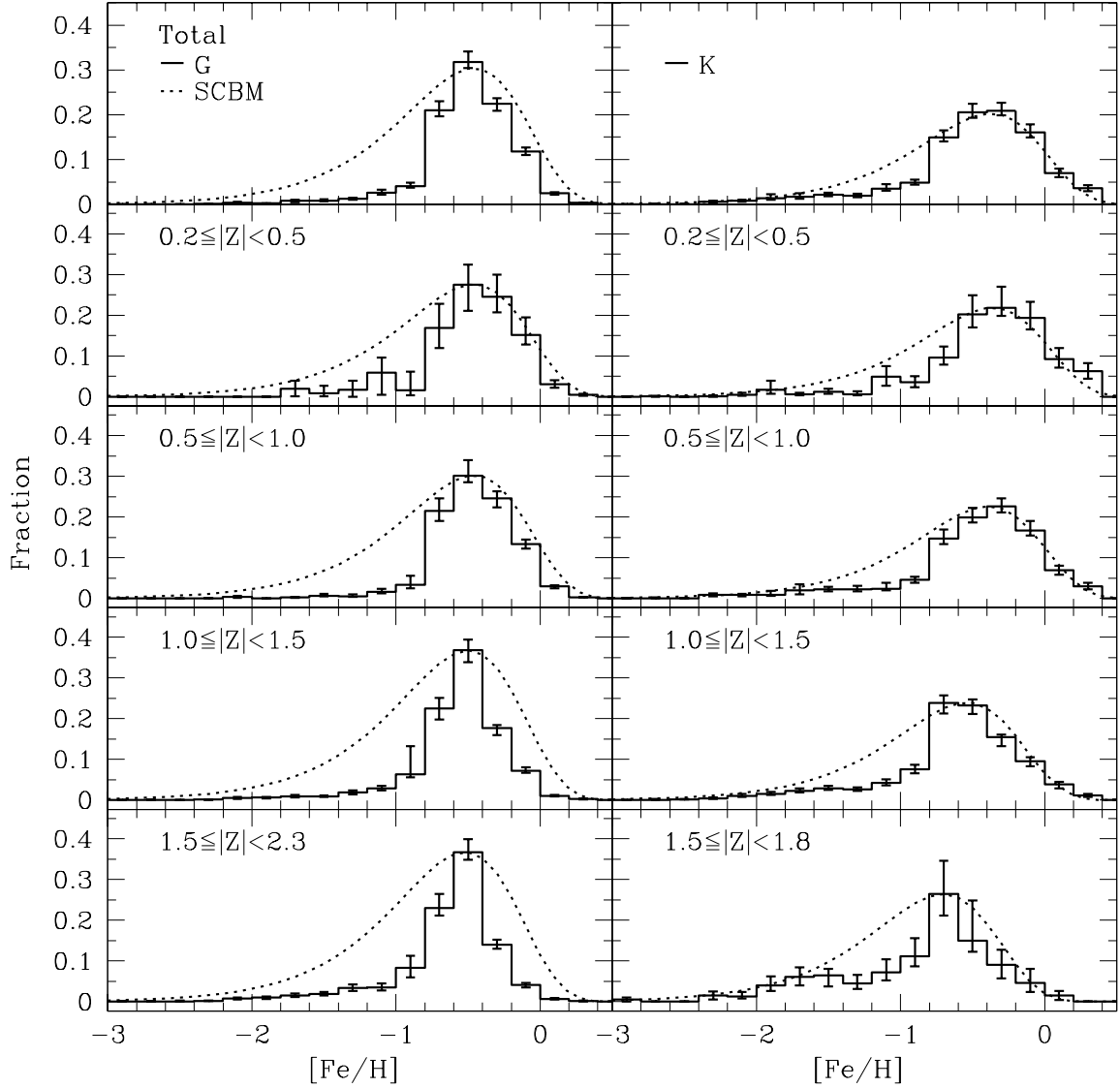}
\figcaption[mcp_simplez_0503.eps]{Comparison of the G- (left) and
  K-dwarf (right) sample in bins of $|Z|$ with a simple closed box
  model (dotted line) assuming instantaneous recycling
  \citep{schmidt}. The total metallicity distribution for each
  spectral type over its associated distance range is shown in the top
  row. The size of the discrepancy between the observed distribution
  and the model at the metal-poor end is consistent for G and K
  dwarfs. The G- and K-dwarf problem is evident throughout the Milky
  Way disk.
\label{fig:simple_model_zbin} 
}
\end{figure}

The original G-dwarf problem stemmed from the observational results
that a simple closed box model of chemical evolution over-predicted
the number of metal-poor stars. In the top row of
Figure\,\ref{fig:simple_model_zbin}, we compare our total MDF of G and
K dwarfs to a simple closed box model with instantaneous recycling,
based on \citet{pagel97}. As expected, we find that our empirical
distributions of both G and K dwarfs are deficient at the
low-metallicity end when compared to these models, confirming that the
G-dwarf problem persists in the large Galactic volume of our survey
and to lower temperatures. For the total sample, the fraction of K
stars with [Fe/H] less than $-$0.8 is $\sim$17$_{-3}^{+5}$\%, whereas
the model predicts 28\%.  Similarly, the model predicts 27\% of G
dwarfs should have [Fe/H]$< -$0.8, but we observe
$\sim$10$_{-2}^{+3}$\%.

The G-dwarf problem was originally reported for stars within the solar
neighborhood. Using the SEGUE sample, we find that the discrepancy
with the simple closed box model is evident throughout the disk
(Figure\,\ref{fig:simple_model_zbin}). The discrepancies are on the
order of 10--20\% between the observed distribution and the simple
closed box model, enhanced at larger distances from the plane, and
consistent between the spectral types.

\subsection{Comparison of the G and K Dwarf MDF with Previous Work}
\label{sec:comp_totmdf}

Our total MDF for G and K dwarfs covers a substantial portion of both
the thin and thick \emph{in-situ} Milky Way disk. As a consequence,
the peak metallicity for both spectral types is lower than previous
analyses, such as RAVE \citep{siebert11, coskunoglu11, bilir12} and
GCS \citep{holmberg07, casagrande11}, which are dominated by
thin-disk, more metal-rich stars. Similarly, the local MDFs for
smaller samples, such as \citet{rpmaciel98} and \citet{kotoneva02},
are more metal-rich than ours. These analyses have no stars in common
with the SEGUE sample and probe a lower $|Z|$ range. Although the
lowest $|Z|$ bin of G and K dwarfs ranges from 0.2 to 0.5 kpc, the
median distance from the Galactic plane for these SEGUE stars is
around 0.4 dex. Thus, our low $|Z|$ MDF is not directly comparable
with local samples, and we defer this analysis to a later paper.

\citet{allendeprieto06} measure the MDF of G-dwarf stars, identified
by effective temperature, between $|Z|$ of 1 and 3 kpc. This volume of
space is dominated by the thick disk. \citet{allendeprieto06} used the
SDSS calibration star sample to examine the MDF between 1 and 3 kpc, a
volume of space dominated by the thick disk. The size and homogeneity
of the data enabled measuremnts of the change in the observed sample
MDF with respect to distance from the plane. However, this sample
pre-dated the SEGUE survey, and the calibration star selection
introduced significant metallicity and evolutionary bias. The target
selection of their sample causes the MDF of \citet{allendeprieto06} to
drop off steeply above [Fe/H] of $-$0.4; it may also cause the MDF to
peak at around [Fe/H] of $-$0.7, more metal-poor than our MDF above
$|Z|$ of 1 kpc.

Whereas the GCS, RAVE, and \citet{allendeprieto06} sample probe
individual components of the disk, the photometric analyses of
\citet{ivezic08} and \citet{bond10} examine a larger volume of space,
from $|Z|$ of 0.8 to 7 kpc. At heights below 2 kpc, their MDF is
slightly more metal-poor than ours, by around 0.1 to 0.2 dex, perhaps
due to inaccuracies in their photometric metallicity calibration. The
contrasting MDFs of these different samples emphasizes that the
chemical structure of the disk varies significantly with respect to
spatial location. Only with a uniform, extensive, spectroscopic sample
such as SEGUE can we obtain accurate constraints on the chemistry of
the disk as a whole.

\subsubsection{\citet{katz11}}

Past work by \citet{katz11} sought to constrain the vertical
metallicity gradient of the thick disk alone. Their sample consists of
approximately 400 stars along two lines of sight; they calculate
atmospheric parameters using the ETOILE spectral matching
program. Utilizing a series of Besan\c{c}on Galaxy models, they
distinguish the thick-disk stars from the thin-disk population. 

\begin{figure} 
\centering
\includegraphics[width=\textwidth]{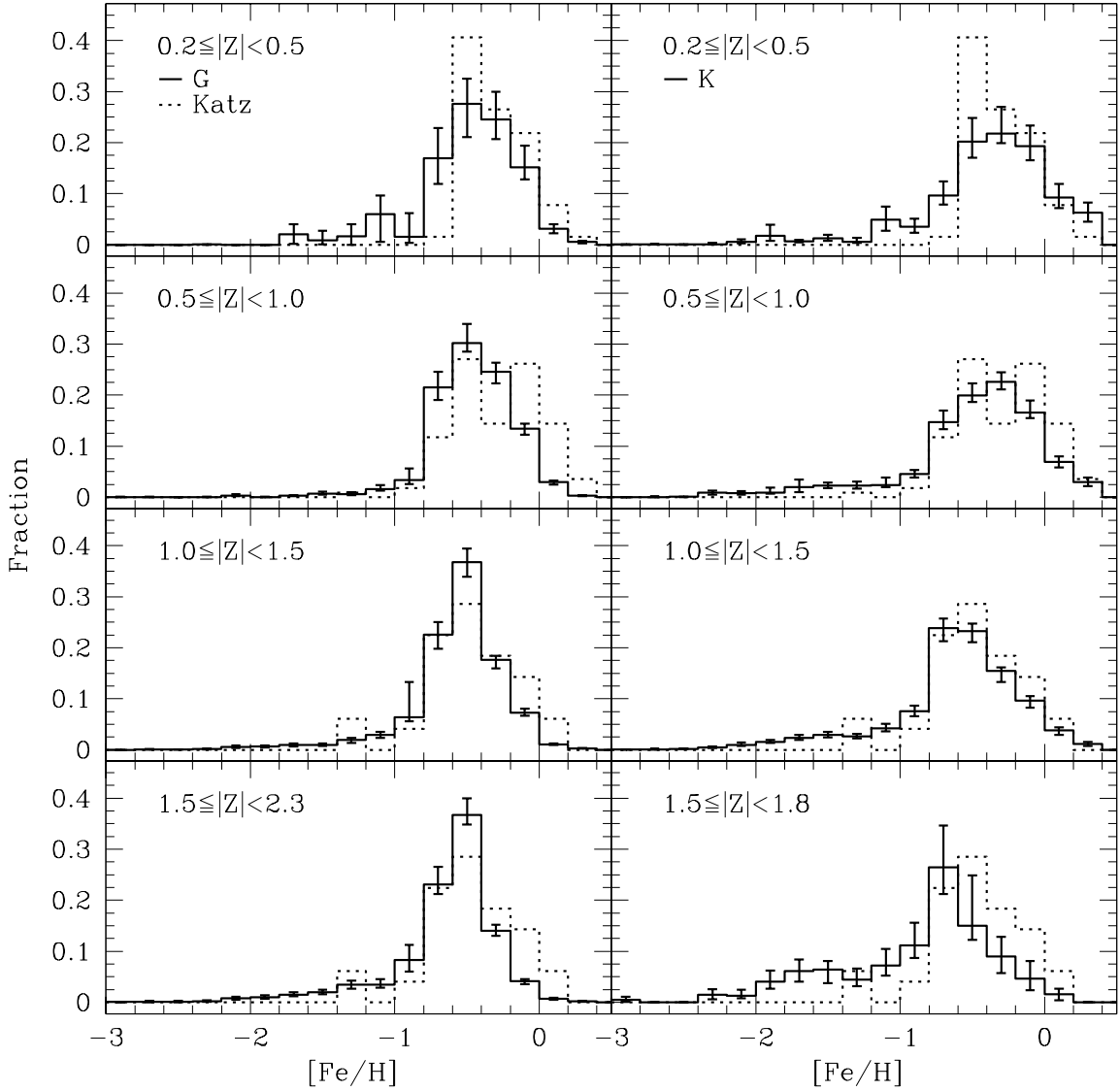}
\figcaption[mcp_katz_0503.eps]{Comparison of the SEGUE G- (left) and
  K-dwarf (right) MDFs with the distribution of \citet{katz11}. The
  SEGUE samples are represented by the solid line, they cover the
  spectral-type distance ranges. The \citet{katz11} distributions are
  plotted as the dashed lines.The MDF of the \citet{katz11} sample is
  typically more metal-rich than that of the G and K dwarfs.
\label{fig:katz_gdwarf} }
\end{figure}

The MDF of the \citet{katz11} sample is slightly more metal-rich than
our G- and K-dwarf sample at all $|Z|$
(Figure\,\ref{fig:katz_gdwarf}). However, the two samples show general
similarities to one another in structure and peak
metallicity. Although the \citet{katz11} sample probes beyond the
solar neighborhood, it is limited to only two lines of sight. Thus, it
is promising for their analysis that we see general agreement in the
metallicity distributions, but their limited sample size and volume
make it difficult to measure the variation in the MDF at different
locations in the disk.

\subsubsection{\citet{lee11b}}

\begin{figure} 
\centering
\includegraphics[width=\textwidth]{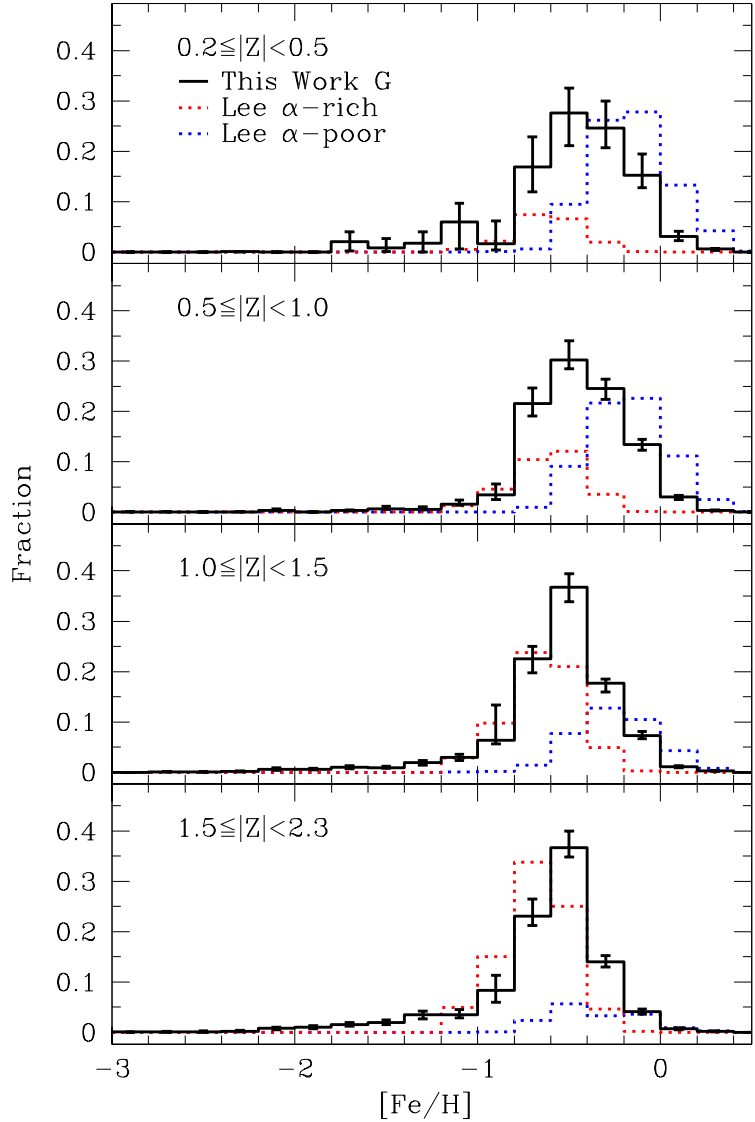}
\figcaption[zbp_glc_st_0.2_all_500_050312.eps]{Comparison of the SEGUE
  G-dwarf MDF over the spectral-type distance range with that of
  \citet{lee11b}. Their work is based on the SEGUE G-dwarf sample and
  divides their stars by [$\alpha$/Fe]. Our SEGUE G dwarfs are plotted
  in black. The blue dashed line is the \citet{lee11b} $\alpha$-poor
  sample, as a fraction of their entire sample of G dwarfs. The red
  dashed line is their $\alpha$-enhanced distribution and normalized
  in the same manner. The \citet{lee11b} sample has not been corrected
  for observational biases; thus, the two distributions are skewed to
  be more metal-poor. At small $|Z|$, our corrected distributions
  appears to sample both the $\alpha$ components. The MDFs of our
  corrected distributions shifts more metal-poor with increasing
  $|Z|$, better matching the $\alpha$-enhanced sample.
\label{fig:lee_gdwarf}}
\end{figure}

\citet{lee11b} use the [$\alpha$/Fe] ratios of the SEGUE G-dwarf stars
to chemically separate the thin- and thick-disk components. They use a
different sample than ours, extracting only stars specifically
targeted as G dwarfs with $S/N >$30, $\log g \geq$4.2, and distances
within 3 kpc. We compare our G-dwarf metallicity distribution over the
spectral distance range to the $\alpha$-poor and $\alpha$-enhanced
metallicity distributions in Figure\,\ref{fig:lee_gdwarf}. In this
figure, the two $\alpha$-separated components are normalized by the
total number of stars in the \citet{lee11b} sample. Farther away from
the plane of the Galaxy, the $\alpha$-rich component makes up a larger
part of the sample, indicating that this $\alpha$-rich population is
an important contribution to our MDF far from the plane.

At low heights above the plane, our observed MDFs contains a mix of
the $\alpha$-poor and $\alpha$-rich samples from \citet{lee11b}. At
$|Z|$ of 1.0 kpc, our MDF skews more metal-poor, matching the
\citet{lee11b} $\alpha$-rich sample better. This suggests that, at
these heights, the G- and K-dwarf sample is dominated by the
$\alpha$-enhanced disk stars. Based on this comparison with the
results of \citet{lee11b}, our observed vertical variation in the MDF
reflects the transition from a predominantly $\alpha$-poor population
at low $|Z|$ to one dominated by $\alpha$-rich stars at high $|Z|$, as
well as a change in mean metallicity with distance from the plane.

\subsection{Comparison with Galaxy Models} 
\label{sec:galmodel_mdfcomp} 

Models of chemical and dynamical evolution have evolved substantially
since the simple closed box model of \citet{schmidt}. Despite these
improvements, models of Galaxy development remain hamstrung by the
lack of extensive observational samples and are calibrated on samples
from the local volume. Until SEGUE, there were few samples available
to test and refine their predictions for the Milky Way properties far
from the plane of the Galaxy. With our G- and K-dwarf MDFs, we can
test existing Galaxy models and provide critical constraints on the
metallicity structure with respect to location, to improve the model
predictions and examine the star-formation history of the disk far
beyond the local volume. In particular, the range of $|Z|$ heights
makes the sample ideal for examining the transition between the thin
and thick disk.

We compare our MDFs with predictions from two Galaxy models. Those of
Sch{\"o}nrich \& Binney (SB, 2009a,b) combine both chemical and
dynamical evolution, whereas the TRILEGAL 1.4 model is based on
empirical relationships. The models differ importantly in their
modeling of the thick disk.  While the TRILEGAL model treats the thick
disk as a separate component of the Galaxy and allows different input
parameters for its spatial distribution and local density, the SB
models implement the radial migration of disk stars, forming a thick
disk in the solar neighborhood via outward migration of older, inner
disk stars. Thus, the SB model's metallicity distribution at high
$|Z|$ will be constrained by the chemistry of old, inner disk stars,
while TRILEGAL's thick disk metallicity distribution is fixed.  We are
also currently working on a comparison with the Besan\c{c}on Galaxy
model \citep{robin03} (Schlesinger et al., in preparation). We also
have provided numerical tables of our MDFs (Tables
\,\ref{tab:total_gkmdf}, \,\ref{tab:gd_zbin_st},
\,\ref{tab:kd_zbin_st}, \,\ref{tab:gd_zbin_vc},
\,\ref{tab:kd_zbin_vc}) for comparison with future Galaxy models
(e.g., \citealt{calura12}).

\subsubsection{The Sch{\"o}nrich \& Binney Galaxy Models} 
\label{sec:schoenrich_models} 

The models of \citet{schonrich09a, schonrich09b} rely on assumed
star-formation laws, nucleosynthetic yields, and a Salpeter initial
mass function \citep{salpeter55}.  These models base their
star-formation rate on the Kennicutt law \citep{kennicutt98};
\citet{schonrich09a, schonrich09b} model the expected nucleosynthetic
yields in conjunction with inflows, resulting in a predicted
age-metallicity relationship, rather than building off an empirical
version.  In addition, these models include dynamical behavior,
specifically, radial mixing, which alters the modeled chemical
structure. \citet{schonrich09b} compare the modeled vertical density
structure to that of \citet{juric08} and \citet{ivezic08}, finding
that the model accurately predicts the density structure of various
Galaxy components. Finally, these models are built using the BaSTI
isochrones \citep{pietr04}, which they artificially widen to simulate
the photometric errors of SDSS. Thus, the expected uncertainties
derived from applying our techniques on these models should accurately
reflect the photometric errors of the SEGUE sample. We do not factor
in the other uncertainties in the SEGUE sample, such as [Fe/H] and
distance, into the SB models.

The SB model has been calibrated in [Fe/H] with the Geneva-Copenhagen
survey \citep{nordstrom04, haywood08} and with updated parameters from
\citet{casagrande11}. \citet{schonrich09a} have shown that their model
accurately reflects the properties of stars in the solar
neighborhood. They compare their model predictions to other observed
samples, such as \citet{fuhrmann98}, \citet{bensby03},
\citet{reddy06}, and \citet{haywood08}, to constrain chemical
parameters. Although many of these samples contain thick-disk stars,
they are all limited to the solar neighborhood, which is thin-disk
dominated. The GCS sample is typically within 200 pc
\citep{casagrande11}. Similarly, the farthest distance probed by the
\citet{reddy06} sample, which was selected based on Hipparcos data, is
150 pc. Thus, although the SB Galaxy model predicts the structure of
the disk as a whole, it is only calibrated on the local volume.

\subsubsection{The TRILEGAL Galaxy Models} 
\label{sec:trilegal_models} 

Whereas the SB models simulate the disk's chemical and dynamical
evolution, the TRILEGAL 1.4 models are more empirical in scope,
utilizing a derived luminosity function, built from an assumed
star-formation rate, age-metallicity relation, and initial mass
function. These parameters vary slightly for different components of
the Galaxy.

The TRILEGAL 1.4 program allows us to simulate different lines of
sight and assume a range of Galaxy parameters. We manufacture a
TRILEGAL model for each SEGUE line of sight, using a limiting
magnitude of 23.0 in $r$, no dust extinction, and no binary
contamination over a circular 7 square degree area of the sky.  We
specify an exponential thin disk with a two-step star-formation
rate. The thin disk is assigned a scale length of 2800 pc and a scale
height which increases with stellar age, as described in
\citet{girardi05}. The assumed age-metallicity relationship for this
component is based on the empirical determination from \citet{rp00},
and includes $\alpha$-enhancement at lower [M/H]. The thick disk is
also modeled as an exponential, with a scale height of 800 pc and a
scale length of 2800 pc. In contrast to the thin disk, a constant
star-formation rate and a Z of 0.008 ([Fe/H]$=-$0.7 for an
$\alpha$-rich population) and $\sigma$[M/H] of 0.1 dex (which is
typically smaller than metallicity measurement errors) is assumed for
the thick disk. In addition to the thin and thick disk, we specify an
oblate spheroid halo and no central bulge. These properties are
combined with the Padova evolutionary tracks to produce a derived
luminosity function. Note that the TRILEGAL models predicts the total
metallicity, [M/H], rather than [Fe/H].

This model shows good agreement between the observed and predicted
star counts when compared to various stellar catalogs, such as 2MASS
and Hipparcos \citep{girardi05}. There has been little chemical
calibration of the TRILEGAL model.

\subsubsection{Accuracy of Galaxy Models} 
\label{sec:model_accuracy} 

\begin{figure} 
\centering
\includegraphics[width=\textwidth]{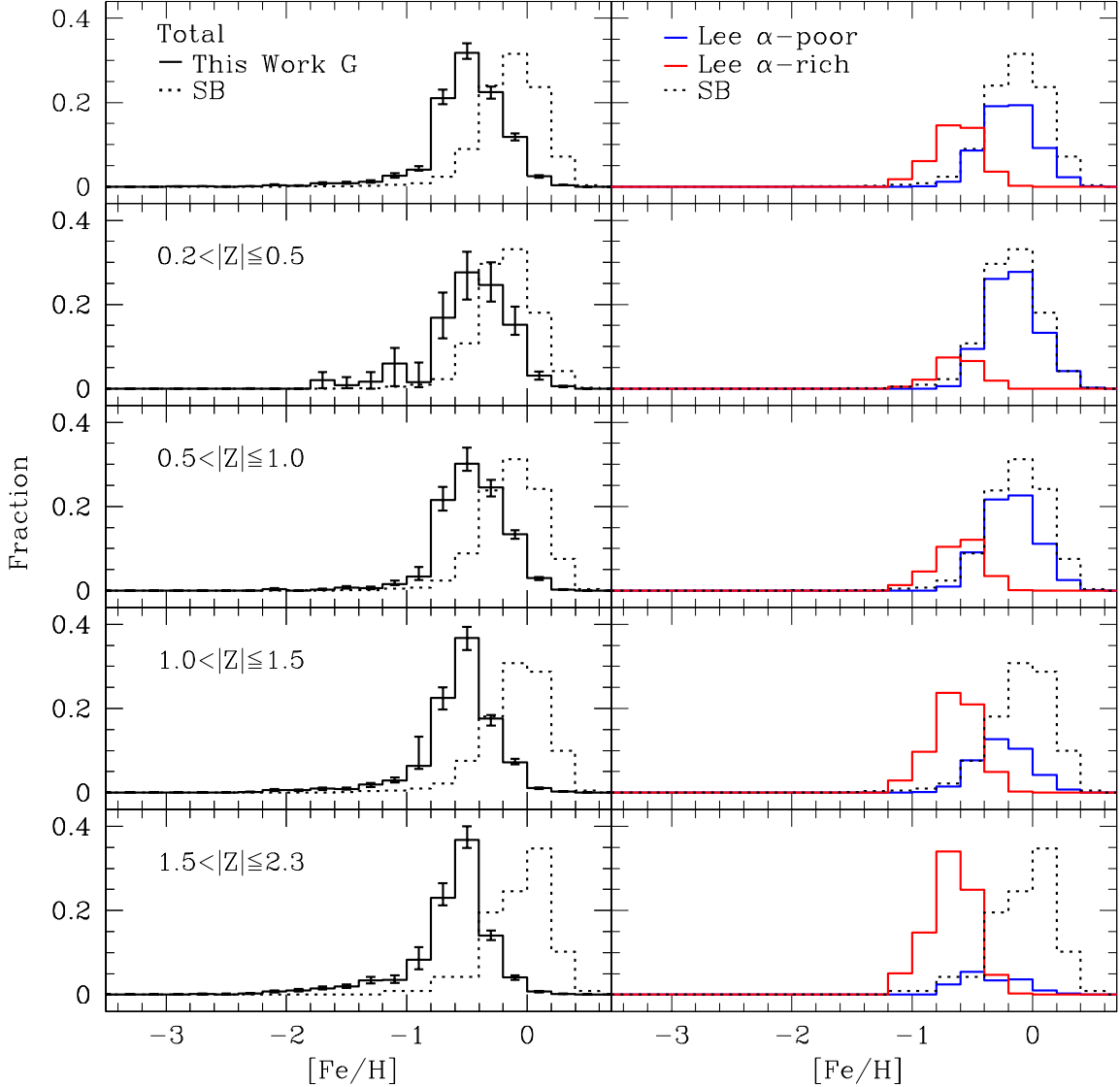}
\figcaption[ralph_mdf_comp_0503.eps]{Comparison of the distribution of
  G-dwarf stars in the SEGUE sample (solid black) to that in the SB
  Galaxy models (dashed black) in the left column. Both sample and
  model are limited to distances between 1.59 and 2.29 kpc. The SB
  Galaxy model is significantly more metal-rich than the SEGUE G-dwarf
  sample. This discrepancy is also seen for the K-dwarf sample. On the
  right side, we compare the G-dwarf distribution in the model (dashed
  black) to the $\alpha$-separated G-dwarf sample from
  \citet{lee11b}. The sample with enhanced [$\alpha$/Fe] is shown in
  red; that with low [$\alpha$/Fe] is plotted in blue. For $|Z|$ less
  than 1 kpc, the SB model is consistent with the $\alpha$-poor
  distribution. Above 1 kpc, the model is more metal-rich than this
  component. The general agreement with solely the $\alpha$-poor
  portion suggests that the SB models currently only recreate the
  chemistry of the this portion of the
  Galaxy. \label{fig:scho_model_comp} }
\end{figure}

\begin{figure} 
\centering
\includegraphics[width=\textwidth]{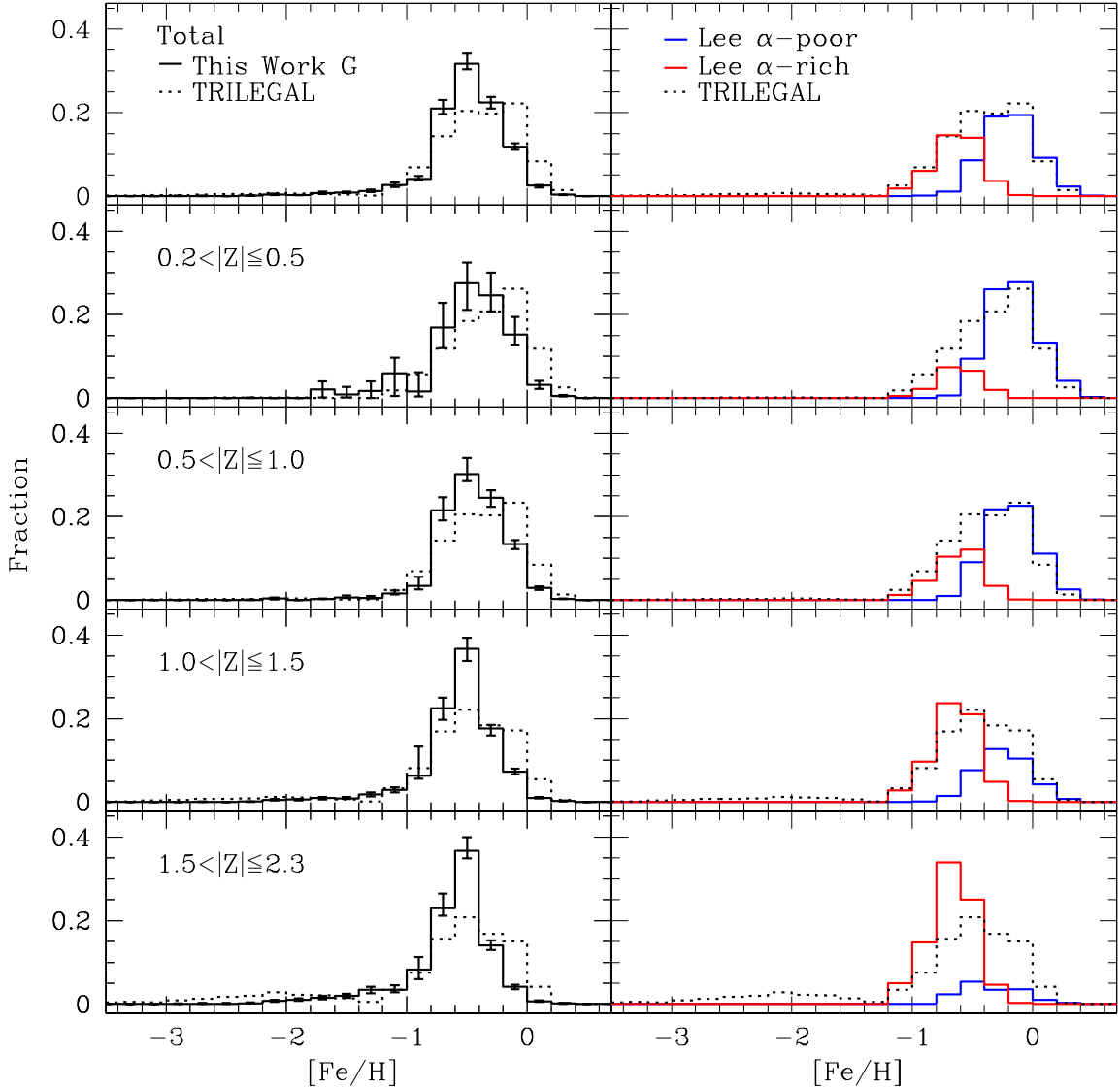}
\figcaption[trilegal_mdf_comp_0503.eps]{The same as
  Figure\,\ref{fig:scho_model_comp}, except now for the TRILEGAL
  models, rather than those of \citet{schonrich09a,
    schonrich09b}. Note that TRILEGAL simulates total metallicity,
  [M/H], rather than [Fe/H]. We assume a simple linear relationship
  between [Fe/H] and [$\alpha$/Fe] to adjust the model from [M/H] to
  [Fe/H]. This model predicts more metal-rich MDFs at all heights than
  our observed sample. Comparison with the $\alpha$-separated samples
  of \citet{lee11b} suggests that at high $|Z|$, TRILEGAL does not
  adequately model the $\alpha$-rich component. However, we can
  improve the agreement by adjusting the assumed [$\alpha$/Fe]-[Fe/H]
  relationship. The comparison of SEGUE K dwarfs with the TRILEGAL
  model shows similar behavior. \label{fig:tri_model_comp} }
\end{figure}

We compare the G-dwarf metallicity distribution to the SB and TRILEGAL
Galaxy models in Figures\,\ref{fig:scho_model_comp}
and\,\ref{fig:tri_model_comp}, respectively. For both models, we
combine all of the SEGUE-modeled lines of sight and extract all
simulated stars that fulfill the color, magnitude, and distance
criteria for our G-dwarf spectral type.

Both of these models are calibrated to past analyses, which are
typically limited to the local volume. Thus, we expect agreement
between the model and SEGUE sample close to the plane of the
Galaxy. However, Figures\,\ref{fig:scho_model_comp}
and\,\ref{fig:tri_model_comp} indicate that, although there is
substantial overlap in the range of [Fe/H], both models are more
metal-rich and sharply peaked than our observed sample. The observed
sample is broadened significantly by correlated errors in [Fe/H] and
distance, in addition to general uncertainites in [Fe/H] (see
\S\,\ref{sec:monte_carlo} and
Figure\,\ref{fig:correlated_mdf_err}). The models do not simulate
these uncertainties and thus exhibit a narrower MDF.

At higher distances from the plane, the metallicity discrepancy
between the SB model and the SEGUE sample increases. Whereas the SEGUE
distributions become more metal-poor with increasing $|Z|$, the models
remain at an [Fe/H] of $-$0.1 for SB.  In fact, the SB distributions
become slightly more metal rich at greater distances from the
plane. We find similar discrepancies for the K-dwarf sample. We
compare the SB modeled lines of sight with the $\alpha$-separated
stellar populations from \citet{lee11b} in
Figure\,\ref{fig:scho_model_comp}. The model agrees well with the
$\alpha$-poor distribution alone but are unable to accurately
represent the $\alpha$-rich component.

Rather than [Fe/H], the TRILEGAL model predicts [M/H], which includes
contributions from [$\alpha$/Fe]. The modeled [M/H] will thus be
systematically higher than our [Fe/H] at large $|Z|$, where
[$\alpha$/Fe] is expected to be large. We model the relationship
between [$\alpha$/Fe] and [Fe/H] using a simple linear relationship
based on the work of \citet{lee11b}, and remove the expected
[$\alpha$/Fe] contribution from the [M/H] value of each simulated
TRILEGAL star. We assume that stars of solar metallicity have
[$\alpha$/Fe] of 0. As [Fe/H] decreases, [$\alpha$/Fe] increases to
0.20 at [Fe/H]$=-$0.5, and 0.40 at [Fe/H]$=-$1.0. Without this
correction, the distribution is similar to that of SB, where the model
is systematically more metal rich than the observations at all
heights. When we account for [$\alpha$/Fe], the match between our
observed G-dwarf MDF and the model improves, however, TRILEGAL is
still offset in [Fe/H], by around 0.1--0.2 dex, at all heights
(Figure\,\ref{fig:tri_model_comp}). We find that the model does not
adequately simulate the structure of the $\alpha$-rich population at
large heights. In particular, from 1.5 to 2.3 kpc in $|Z|$, the
TRILEGAL distribution is very similar to that of the \citet{lee11b}
$\alpha$-poor component. Note that we can improve the match between
the simulation and observations by adjusting the contribution from
[$\alpha$/Fe].

The two models predict stars at large $|Z|$ to be more metal rich than
we observe. The current inability of these Galaxy models, in
particular the SB model, to predict the chemical structure of the disk
points to the danger of assuming that the local volume reflects the
properties over the span of the disk. Past analyses of the metallicity
distribution, used to calibrate these Galaxy models, were limited to
the solar neighborhood; they were dominated by thin-disk stars. While
these samples contain thick-disk stars passing through the local
volume, they may not provide a complete picture of the component's
structure. There appears to be a significant disagreement between the
picture of the disk structure from local kinematic samples and our
\emph{in-situ} observations. Thus, our unbiased metallicity
distributions of the SEGUE G and K dwarfs, and how they change with
respect to R and $|Z|$, provide critical constraints for refining and
improving Galaxy models, such that we gain a better understanding of
the different chemical and dynamical processes occuring in the disk
and no longer rely on local samples alone.

\section{Summary}

In this work we examine the variation in [Fe/H] of the SEGUE G- and
K-dwarf sample with respect to position in the Galaxy. These
distributions allow us to better constrain the chemical and dynamical
evolution processes in the disk. To ensure the accuracy of this
sample, we optimize the SSPP to better handle low-temperature dwarfs
and develop and test a procedure to account for target-selection
biases in the SEGUE sample, using SDSS photometry for each individual
line of sight. This technique corrects for the metal-poor bias in the
spectroscopic sample and is critical for accurately constraining the
metallicity structure of the Milky Way disk. For the first time, we
can utilize an unbiased sample of both G and K dwarfs with accurate
atmospheric parameters to examine the chemical structure over a large
volume of the disk.

Using distances estimated with an isochrone-matching technique,
accurate to $\sim$8\% for metal-poor and $\sim$18\% for metal-rich
stars, we determine the MDFs of the G- and K-dwarf samples
(Fig.\,\ref{fig:overall_comp}) and examine how the metallicity
structure of the disk varies with respect to position and spectral
type. Both the total MDFs and MDFs with respect to height above the
plane of the Galaxy for the G- and K-dwarf samples are in good
agreement with one another, implying that the two spectral types
undergo similar chemodynamical evolution over the span of the disk. We
compare both distributions to the simple closed box model, finding
that the G-dwarf problem persists to both lower temperatures and is
present throughout the volume of the Milky Way disk
(Fig.\,\ref{fig:simple_model_zbin}). This expands on previous work,
such as \citet{mould82}, \citet{favata97}, \citet{flynnmorell97},
\citet{rpmaciel98}, and \citet{kotoneva02}, which examined the solar
neighborhood. This is the first work where we can confirm that this
discrepancy extends past the solar neighborhood with a large and
unbiased stellar sample of two cool spectral types. In particular, all
previous analyses of K dwarfs were much more limited in sample size
and volume.

Analyses of the local volume find that the thin and thick disk are
distinct with respect to [$\alpha$/Fe]. As it is less likely than
kinematics to change significantly over time, stellar chemistry is a
useful diagnostic to disentangle the two components. Our analysis of
the G- and K-dwarf MDF finds that both spectral types have more
metal-poor stars as $|Z|$ increases (Fig.\,\ref{fig:mdf_z_kdgd_dcut},
Tables\,\ref{tab:gd_zbin_st}, \ref{tab:kd_zbin_st},
\ref{tab:gd_zbin_vc}, and\,\ref{tab:kd_zbin_vc}), implying similar
star-formation histories. Comparison with the [$\alpha$/Fe] analysis
of \citet{lee11b} suggests that our vertical metallicity structure
represents the transition between a population composed largely of
$\alpha$-poor stars at low $|Z|$ to predominantly $\alpha$-rich stars
at high $|Z|$. This change with respect to $|Z|$ is not predicted by
the SB or TRILEGAL Galaxy models. Both are more metal-rich than our
observed distribution. In particular, the SB model appears to simulate
only the $\alpha$-poor component of the Galaxy, suggesting that there
are discrepancies with the picture of the extended disk presented by
local samples and our own \emph{in-situ} observations. Thus, our
observed MDFs provide a valuable constraint on future Galaxy formation
and evolution models and strongly warn of the dangers of using
existing models to predict behavior far from the local volume.

\acknowledgements 

We thank J. Cheng for her valuable input on this work. K.S. and J.A.J
acknowledge support from NSF grant AST-0807997 and
AST-0607482. C.R. acknowledges support from the David and Lucille
Packard Foundation. H.L.M. acknowledges support from
AST-1009886. Y.S.L and T.C.B. acknowledge partial support from grant
PHY 08-22648: Physics Frontiers Center/Joint Institute for Nuclear
Astrophysics (JINA), awarded by the U.S. National Science Foundation.

Funding for SDSS-III has been provided by the Alfred P. Sloan
Foundation, the Participating Institutions, the National Science
Foundation, and the U.S. Department of Energy Office of Science. The
SDSS-III web site is http://www.sdss3.org/.

SDSS-III is managed by the Astrophysical Research Consortium for the
Participating Institutions of the SDSS-III Collaboration including the
University of Arizona, the Brazilian Participation Group, Brookhaven
National Laboratory, University of Cambridge, University of Florida,
the French Participation Group, the German Participation Group, the
Instituto de Astrofisica de Canarias, the Michigan State/Notre
Dame/JINA Participation Group, Johns Hopkins University, Lawrence
Berkeley National Laboratory, Max Planck Institute for Astrophysics,
New Mexico State University, New York University, Ohio State
University, Pennsylvania State University, University of Portsmouth,
Princeton University, the Spanish Participation Group, University of
Tokyo, University of Utah, Vanderbilt University, University of
Virginia, University of Washington, and Yale University.

\appendix

\section{Optimizing the SSPP for Cool Dwarf Stars}
\label{app:sspp_optimize} 

The SSPP analyzes a wide range of spectral types, and not all of the
techniques employed are appropriate for cool dwarfs. To ensure that
the SSPP provides accurate atmospheric parameters for G and K dwarfs,
we have developed an optimized version of the SSPP, carefully
selecting, and occasionally adjusting, the $\log g$ and [Fe/H]
techniques used in the analysis.

\subsection{Surface Gravity Determinations}
\label{sec:surface_gravity} 

\begin{figure}
\centering \includegraphics[width=\textwidth]{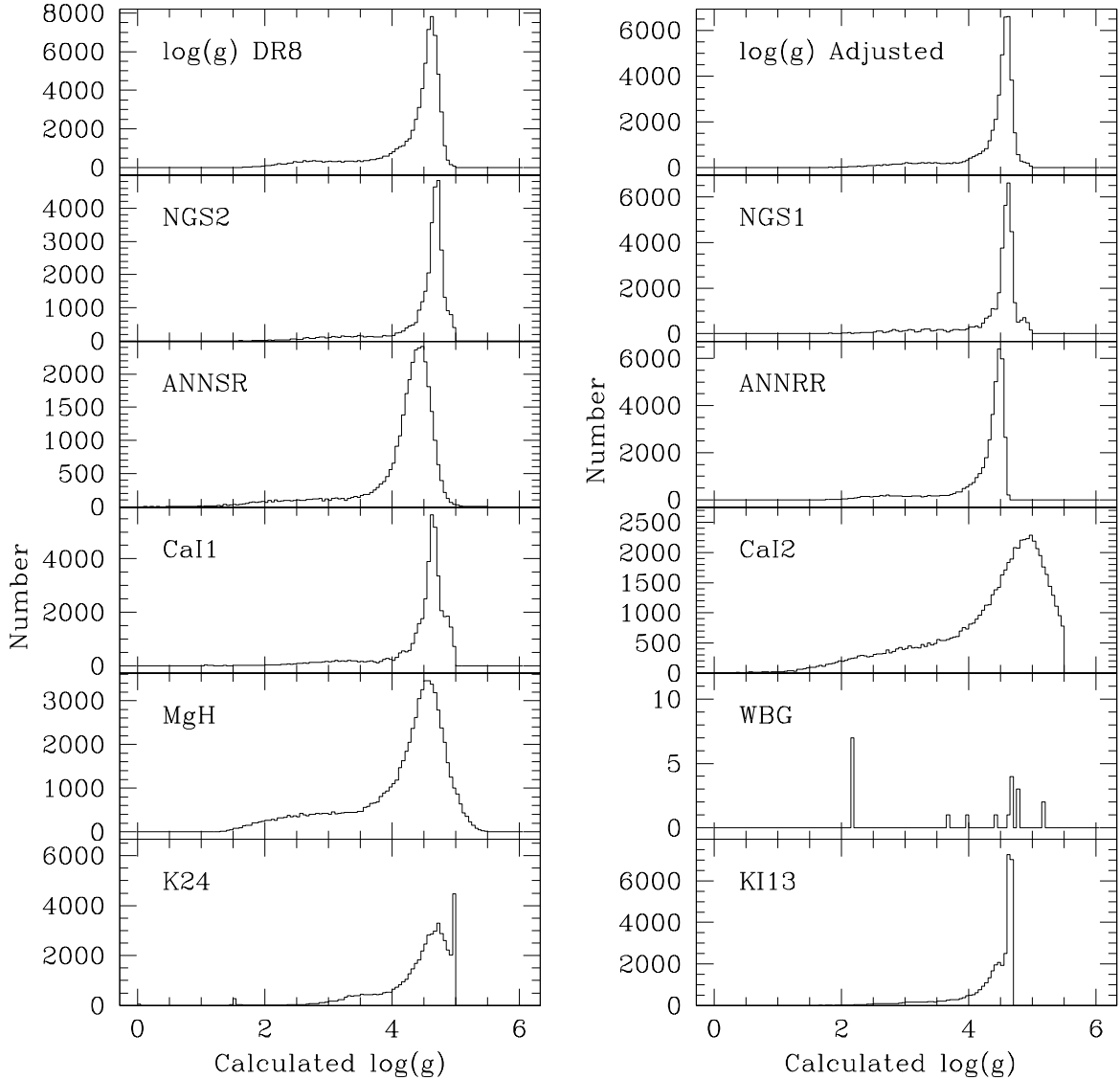}
\figcaption[logg_meth_1115.eps]{The surface gravities calculated using
  different methods in the SSPP for our sample of G and K dwarfs (see
  \citealt{lee08_I, lee08_II} for a description of methods). The
  original adopted $\log g$ distribution is in the top left. All of
  the distributions have a tail extending to $\log g$ values
  indicative of giants. However, this tail is significantly more
  pronounced for the CaI2 and MgH methods. Additionally, the width of
  the peaks for these two methods are larger than that of the others,
  making their estimate of $\log g$ generally more uncertain. The WBG
  method is, in general, unable to calculate a surface gravity for our
  targets. When the CaI2, MgH, and WBG techniques are removed, the low
  $\log g$ tail is diminished, and the peak of the distribution is
  narrower. This corrected $\log g$ is shown at the top right.
\label{fig:logg_meth_act} } 
\end{figure}

The SSPP uses ten methods to estimate surface gravities for stars,
each appropriate for a different range in $(g-r)_0$
\citep{lee08_I}. Although all of the techniques cover the range of
$(g-r)_0$ for G and K dwarfs, they are not all similarly accurate. As
shown in Fig.\,\ref{fig:logg_meth_act}, the CaI2 and MgH indicators
have large tails that extend to low values of $\log g$ not seen for
other indicators. Comparison between $\log g$ estimates from Kepler
and those from the SSPP indicate that the CaI2 and MgH techniques
return systematically low surface gravities. To confirm this, we
tested the different surface gravity indicators on synthetic spectra
(see \S\,\ref{sec:subgiants} for more information); while the majority
of the methods accurately reported the input $\log g$, both CaI2 and
MgH exhibited false low surface gravity tails. Examination of the SSPP
revealed that the implementation of these two techniques was flawed;
both CaI2 and MgH utilized an inaccurate conversion between $(g-r)$
and (B-V). Furthermore, we found that the relationship implemented in
the SSPP for these techniques was an outdated version, different than
that reported in \citet{morrison03}. In addition to removing CaI2 and
MgH, we also eliminate WBG technique from our $\log g$ estimate. As it
was designed for evolved stars, this indicator rarely works well for
the G- and K-dwarf sample.

\begin{figure} 
\centering \includegraphics[width=\textwidth]{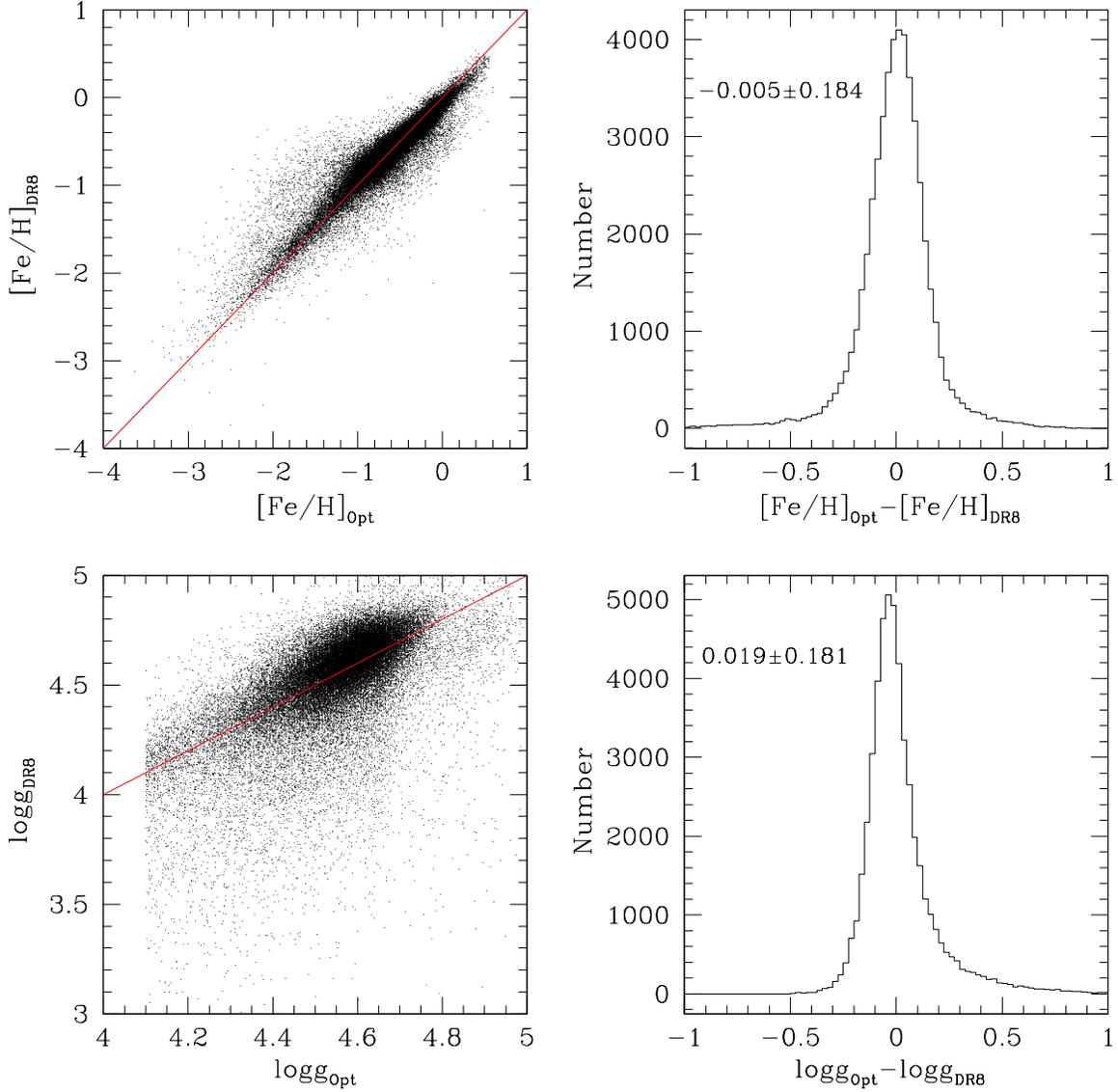}
\figcaption{A comparison of the atmospheric parameters determined by
  the optimized version of the SSPP with those from DR8. The left
  column directly compares the values, with the optimized version on
  the x axis and the DR8 on the y. The red line shows a one-to-one
  correlation. The right column shows the distribution of the
  difference between the two values for each parameter, with the mean
  and standard deviation of the difference noted in the top left
  corner. The top row is for [Fe/H] and bottom for $\log g$.  \label{fig:dr8_optimized_comp} }
\end{figure}

Using the CaI2, MgH, and WBG to determine $\log g$ will shift the
calculated values lower, potentially eliminating some dwarfs from our
sample by making their surface gravities indicative of evolved
stars. For our targets, we use $\log g$ values which do not take these
three methods into account. This tightens the distribution of surface
gravities, diminishing the population that artificially lies in the
low-valued tail (see Fig.\,\ref{fig:logg_meth_act}). On average,
removing CaI2, MgH, and WBG increases the surface gravities from DR8
by less than 0.1 dex (see Figure\,\ref{fig:dr8_optimized_comp}).

\subsection{[Fe/H] Determinations}
\label{sec:feh_det}

\begin{figure}
\centering \includegraphics[width=\textwidth]{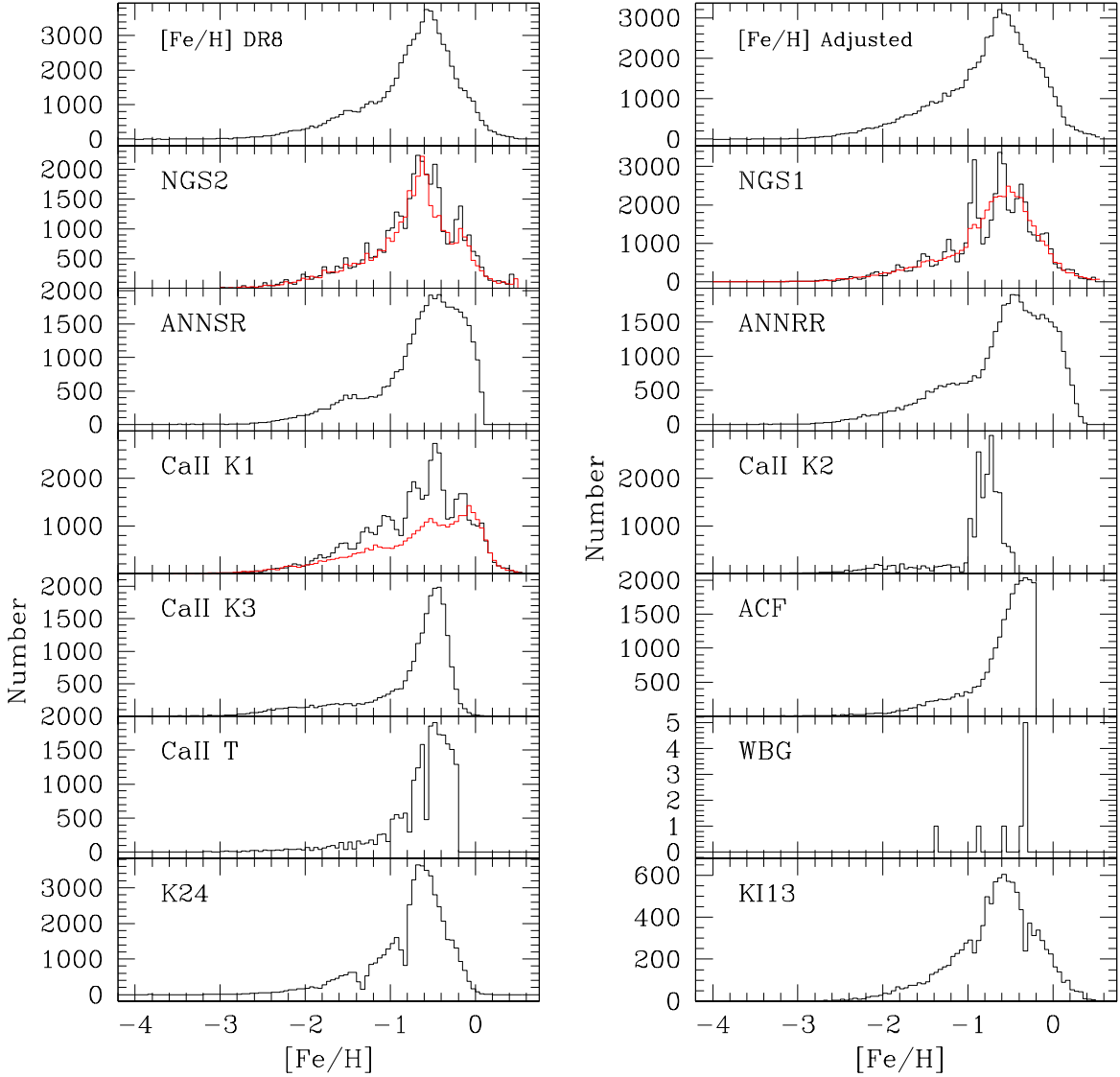}
\figcaption[feh_method_tester_0629.eps]{The metallicities calculated
  using different methods in the SSPP for our G- and K-dwarf
  sample. The original adopted [Fe/H] value is on the top left, with
  the adjusted distribution on the top right. For the different
  methods used in our optimized version of the SSPP, we plot the
  original distribution for the sample in black and the modified
  version in red. Note that for many methods, such as NGS1 and CaIIK1,
  the modified distribution is much smoother.  We have also eliminated
  a number of methods as they are either not applicable for our sample
  or do not cover the appropriate metallicity
  range. \label{fig:kd_fehmethod} }
\end{figure}

As with the surface-gravity estimates, not all of the
metallicity-estimation techniques available in the SSPP are accurate
for G and K dwarfs. For example, the CaIIK2 and WBG distributions are
erratic and spiky (Figure\,\ref{fig:kd_fehmethod}). Other techniques,
such as CaIIK3, ACF, ANNSR, and CaIIT, are limited in the metallicity
range they cover, only reaching as high as solar metallicity. This
limit will bias their measurements towards lower metallicities. We
exclude these six techniques from our [Fe/H] determinations.

Finally, there are other methods in the SSPP that appear to have
well-behaved distributions in [Fe/H] but are inappropriate for our
sample. Two of the Kurucz-model-based methods (KI13, K24) are designed
for stars with temperatures greater than 5000 K; there are numerous K
dwarfs for which this method is not accurate. The ANNRR method is also
problematic; this neural-network technique is based on outdated
parameters determined by the SSPP for DR7 \citep{abazajian09}. Thus,
our [Fe/H] estimates are based solely on NGS2, NGS1, and CaIIK1.

Although these three remaining methods cover the appropriate range of
parameter space, they still require adjustment. In particular, all
three show artificial peaks. These methods utilize a $\chi ^2$
minimization technique in T$_{\rm{eff}}$, $\log g$, and [Fe/H] to
match target spectra with various synthetic grids. These grids cover a
fairly narrow range of parameter space, resulting in the sharply
peaked structure. By modifying the $\chi ^2$ interpolation method in
parameter space from the [Fe/H] values reported in DR8
\citep{aihara11}, these peaks are smoothed, providing a more realistic
distribution in [Fe/H]. Figure\,\ref{fig:dr8_optimized_comp} indicates
that, on average, these modifications have very little effect on the
estimated [Fe/H]. These changes will be detailed in the upcoming SEGUE
DR9 paper, and are outside the scope of this work.

\section{Uncertainties in Calculated Distances}
\label{app:uncertainties} 

\subsection{Photometry} 

The photometric uncertainty of SDSS targets is typically 2-3\% for
each filter. Using the TRILEGAL simulations along SEGUE lines of
sight, we simulate the photometric uncertainties in SDSS based on a
Gaussian distribution. These changes factor in to distance
uncertainties twice; first, when we match each target to an isochrone
in $(g-r)_0$ and second, when we calculate the distance in each
individual filter. The change in distance with respect to photometric
uncertainty is typically less than 6\% over the entire metallicity
range.

\subsection{[Fe/H]}

There are two main components of uncertainty in the SSPP [Fe/H]
estimates. First, there is the error which reflects the variation from
the different techniques used by the SSPP to estimate [Fe/H]. This
uncertainty from the three methods we utilize has a mean value of
0.07$\pm$0.06 dex. Second is the uncertainty determined by comparing
with stars of known metallicity, from clusters and high-resolution
observations. This uncertainty varies with respect to $S/N$, ranging
from 0.24 dex for $S/N$ of 50 to 0.45 dex for $S/N$ of 10
\citep{lee08_I}.

For each SEGUE line of sight, we examine how the $S/N$ varies with
respect to magnitude, simulating the relationship for the associated
TRILEGAL model. Based on the assigned $S/N$, we estimate the expected
[Fe/H] uncertainty, combining it in quadrature with a simulation of
the pipeline [Fe/H] variation, convolved with a Gaussian. We then
compare the original distances calculated for the TRILEGAL stars to
those using [Fe/H] values adjusted for the simulated
uncertainty. Changes in [Fe/H] have a larger affect on distance
estimates for metal-rich stars ([Fe/H]$>-$1.0); these dominate the
uncertainty with a distance error of 15\%, whereas more metal-poor
stars have an error of around 4\%.

\subsection{[$\alpha$/Fe]} 

\citet{an09} assign each YREC isochrone a specific [$\alpha$/Fe] value
based upon its [Fe/H]. These values may not align with the
[$\alpha$/Fe] of the individual target, which will change the distance
estimates, especially for the metal-rich regime where small changes in
chemistry significantly change the isochrone. \citet{lee11a} found
that for a $S/N>$20, the SSPP can determine [$\alpha$/Fe] with an
uncertainty of approximately 0.1 dex. At lower $S/N$, the uncertainty
increases to 0.2 dex. To estimate the change in estimated distance
associated with a 0.2 dex uncertainty in [$\alpha$/Fe], we compare the
magnitudes of the Dartmouth isochrones at each value of [Fe/H] with
[$\alpha$/Fe] of 0 to those with [$\alpha$/Fe] of +0.2 and $-$0.2. We
must use the Dartmouth isochrones to estimate this effect, as the YREC
isochrones do not model a range of [$\alpha$/Fe] for each [Fe/H]. We
simulate the change in Dartmouth $ugriz$ at each [Fe/H] from a change
in [$\alpha$/Fe]; similar to the distance uncertainties from [Fe/H],
the change in distance is larger, approximately 10\%, at the
metal-rich end. Below [Fe/H] of $-$1.0, the $\sigma_{d([\alpha/Fe])}$
is approximately zero.

\subsection{Isochrone Choice} 
\label{sec:iso_choice} 

Each individual isochrone set will yield slightly different $ugriz$
values at the same $(g-r)$, resulting in systematic differences in
distance with respect to one another. The various assumptions
integrated into each of these different isochrone sets means that for
each $(g-r)$ color, there is a range of appropriate $r$ magnitude
about the YREC value \citep{gallart}. The difference between isochrone
sets is more extreme at the metal-rich end, making the uncertainty due
to isochrone choice larger than for metal-poor stars.

Distances calculated using YREC isochrones tend to be smaller than
those calculated via Dartmouth isochrones by around 6\% for stars with
[Fe/H]$\geq-$0.5; below this, the offset is around 3\%. In contrast,
both the BaSTI \citep{basti} and Padova \citep{girardi04} isochrone
sets will measure distances smaller than those from the YREC
isochrones, approximately 8\% for [Fe/H]$\geq$0. For more metal-poor
stars, the differences between BaSTI and YREC are around 1\%.

Changing which isochrone set we use will induce a systematic change in
our estimated distances. However, when we consider the many different
isochrones, choosing a particular set can make our distances smaller
$or$ larger. Thus, we treat isochrone choice as a random uncertainty
of approximately 7\% for high-metallicity stars ([Fe/H]$\geq-$0.5) and
3\% for low-metallicity stars.

\subsection{Age} 
\label{sec:age_effects} 

To calculate distances, we assume all stars have an age of 10 Gyr and
compare them to isochrones of the same age. However, the actual ages
of these stars are unknown. In particular, the most metal-rich stars
are likely to be significantly younger than this value, causing us to
systematically overestimate the distances to these stars.

To estimate the uncertainty from age, we utilize the modeled lines of
sight from the SB Galaxy models. These models simulate a wide range of
ages, from 2.5 to 11.5 Gyr using an age-metallicity relationship built
off of expected nucleosynthetic yields
(\S\,\ref{sec:schoenrich_models}). In contrast, the TRILEGAL Galaxy
model assumes an age-metallicity relationship from \citet{rp00}, and
contains stars from only 7 to 9 Gyr in age. For each modeled line of
sight, we calculate the distances for simulated G and K dwarfs using
10 Gyr YREC isochrones. We then calculate the distances for the same
sample using YREC isochrones that match the target in age, examining
the differences between the two values. For the youngest stars, from
2.5 to 5 Gyr, using 10 Gyr isochrones creates a systematic 3\%
overestimate of distances. This value decreases as the ages of the
stars increase. With respect to metallicity, this means that the
metal-rich stars, with [Fe/H]$>-$0.4, will be systematically offset in
distance from 1 to 3\%. More metal-poor stars are unaffected. As all
of our targets are on the main sequence, it is not surprising that we
see little effect with respect to age assumptions; at this point in
stellar evolution, isochrones of a given metallicity overlap quite
closely.

\subsection{Binarity} 
\label{sec:distance_binaries} 

An undetected companion will make an individual target appear brighter
and redder, causing us to underestimate the distance to the primary
star. We simulate a population of binaries, using techniques described
in \citet{schlesinger10}, to examine how much undetected binarity will
affect our distance calculations. We model a population of binaries,
using a Chabrier IMF \citep{chabrier03} for both the primary and
secondary stars, and determine the change in each $ugriz$ filter with
an undetected companion over a range of metallicity. We then match up
each star in the TRILEGAL modeled lines of sight with these $ugriz$
changes, comparing their original YREC distances to those estimated
with the companion present. Binarity will typically lead to a
systematic distance offset of 5\%. Metal-poor stars ([Fe/H]$<-$1.0)
have a slightly larger change in distance, around 7-8\%, whereas more
metal-rich stars are around 4\%. Although small changes in magnitude
have a larger effect on distance estimates for metal-rich stars, the
changes in magnitude from binarity are larger for metal-poor stars,
such that the change in distance is slightly larger. As approximately
65\% of all G and K stars are in binaries \citep{dm91}, this results
in a systematic offset to smaller distances of around 3\%.

In addition to affecting photometry, undetected pairs can also alter
our distance measurements via changes in the metallicity
determination. The expected [Fe/H] uncertainty at a $S/N$ of 25 is
$\pm$0.30 dex. \citet{schlesinger10} find that around 95\% of all
pairs will have a shift in [Fe/H] less than 0.15 dex. Thus, distance
uncertainties stemming from a secondary's effect on [Fe/H]
determinations are well within the error expected from the SSPP [Fe/H]
determination, and we do not take them into account separately.

\subsection{Summary of Random and Systematic Uncertainties in Distance}

We have explored the different uncertainties that factor into our
distance calculations using simulated lines of sight from TRILEGAL and
SB Galaxy models. The random uncertainty in distance stems from
photometry, [Fe/H] determinations, and [$\alpha$/Fe]
estimates. Photometry contributes a 6\% uncertainty over the entire
metallicity range; the chemistry of a star however, causes a different
amount of uncertainty for metal-rich and metal-poor stars. Metal-rich
stars will have a change in distance of 16\% from [Fe/H] and 10\% from
[$\alpha$/Fe] uncertainty. At the metal-poor end, these values are
smaller, around 4\% from [Fe/H] and 0\% from [$\alpha$/Fe]
uncertainty. In addition, we could use a number of different
isochrones to estimate our distances, adding an additional random
uncertainty in distance of 10\% for metal-rich and 3\% for metal-poor
stars. Combining these in quadrature, we estimate our random distance
uncertainty ranging from 15$-$18\% for [Fe/H]$\geq-$1.0 to 8\% for
stars with [Fe/H] below $-$2.0 (Fig.\,\ref{fig:prelim_dist}).

Our distance estimates also suffer from systematic uncertainties from
age assumptions and binary contamination. While the distance
uncertainties from age assumptions vary with metallicity, from 0 to
3\% from the metal-poor to metal-rich end, binarity results in a more
consistent offset of around 5\% over the full [Fe/H] range.

\subsection{Comparison with \citet{ivezic08} Photometric Parallax} 
\label{app:pp_comp} 

\citet{ivezic08} develop a photometric parallax relationship which
takes into account metallicity, building upon the work by
\citet{juric08}. The shape of the photometric relationship is based on
SDSS observations of five globular clusters: M2, M3, M5, M13, and
M15. These clusters range from 7 to 12 kpc in distance and $-$1.2 to
$-$2.1 in [Fe/H] \citep{harris96}. Using clusters from
\citet{vandenberg03}, which range from $-$2.5 to +0.12 in [Fe/H], they
then constrain the extent of magnitude shifts for different
metallicities. Combining the equations determined for different ranges
in $(g-i)$, their final relationship is as follows:
\begin{equation} 
M_r(g-i,[Fe/H]) = M_{r}^{0}(g-i)+\Delta M_{r}([Fe/H]) 
\label{eq:pprd}
\end{equation} 
\begin{equation}
M_{r}^{0}(g-i) = -5.06 + 14.32(g-i) - 12.97(g-i)^2 + 6.127(g-i)^3 - 1.267(g-r)^4 + 0.0967(g-i)^5 
\end{equation}
\begin{equation}
M_{r}([Fe/H]) = 4.50 - 1.11[Fe/H] - 0.18[Fe/H]^2
\end{equation}
\citet{ivezic08} specify that these equations are valid for dwarfs
with 0.2$<$($g-i$)$<$4.0. Less than 1\% of G and K dwarfs fall outside
of this color range. They expect uncertainties of around 5\% and
10-15\% from systematic and random errors, respectively
\citep{bond10}.

We compare our YREC isochrone distances to those calculated using
photometric parallax, finding an offset of approximately 8\%, with
YREC distances typically smaller. This result is due to an $\sim$0.2
mag offset between the YREC isochrones and photometric metallicity
relations. This offset varies with respect to metallicity: for
[Fe/H]$>$0, the change in distance is around 6\%, whereas for [Fe/H]
between $-$1.5 and $-$2.0, the change is minimal. Finally, for the
most metal-poor stars ([Fe/H]$<-$2.0), the photometric parallax
distances are actually smaller than those determined using YREC
isochrones, by around 7\%. To ensure our distance method does not
significantly change our results, we have also calculated the
metallicity distribution functions using photometric parallax
distances instead of those from YREC. There are only slight changes,
all of which are well within the expected bootstrap uncertainties for
the distributions and gradients.

\newpage

\clearpage

\newpage

\include{tables_mdfalone}

\end{document}

%% file: tables_mdfalone.tex
\begin{deluxetable}{cccc}
\tabletypesize{\small}
\tablewidth{0pt}
\tablecaption{Cluster Metallicities determined with the Optimized SSPP \label{tab:clusters_info}}
\tablehead{
\colhead{} & \multicolumn{3}{c}{[Fe/H]} \\
\colhead{Cluster} & \colhead{Literature} & \colhead{All} & \colhead{($g-r$)} }
\startdata
M92 & -2.25 & -2.23$\pm$0.22 &  -2.32$\pm$0.09  \\
M3  & -1.55 & -1.55$\pm$0.18 &  -1.58$\pm$0.16  \\
M71 & -0.79 & -0.78$\pm$0.12 &  -0.77$\pm$0.13  \\
NGC 2420 & -0.20\tablenotemark{a} & -0.28$\pm$0.13 &  -0.26$\pm$0.13  \\
NGC 2158 & -0.26 & -0.26$\pm$0.10 &  -0.30$\pm$0.09  \\    
M67 &  0.05\tablenotemark{b} &  0.06$\pm$0.07 &   0.03$\pm$0.10  \\
NGC 6791 &  0.31 &  0.40$\pm$0.12 &   0.36$\pm$0.09  \\
\enddata 
\tablenotetext{a}{\citet{jacobson11}}
\tablenotetext{b}{\citet{randich06}} 
\tablecomments{
The mean and $\sigma$ [Fe/H] values of cluster members based upon the
optimized version of the SSPP. All literature values are from
\citet{smolinski10} unless otherwise noted. The calculated
metallicities are similar to the literature values demonstrating that
this SSPP version accurately determines stellar parameters.  The
``($g-r$)'' column isolates members with colors in the appropriate
range for G and K dwarfs. These values are not significantly different
than those for the whole cluster sample. }
\end{deluxetable}

\begin{deluxetable}{cccccccc}
\tabletypesize{\small}
\tablewidth{0pt}
\tablecaption{Cluster Distances \label{tab:cluster_dist}}
\tablehead{
\colhead{Cluster} & 
\colhead{[Fe/H]} & 
\colhead{E(B-V)} & 
\colhead{Assumed Age} & 
\colhead{Literature} & 
\colhead{\citet{ivezic08} Photometric $\pi$} & 
\colhead{YREC}  \\
\colhead{} &
\colhead{} &
\colhead{} & 
\colhead{Gyr} & 
\colhead{kpc} & 
\colhead{kpc} & 
\colhead{kpc} }
\startdata
M13      & -1.54 & 0.017 & 14 & 7.7   & 7.71$\pm$0.78 & 8.01$\pm$1.01  \\
M67      & +0.02 & 0.032 & 4  & 0.91  & 0.93$\pm$0.10 & 0.87$\pm$0.12 \\
NGC 2420 & -0.44 & 0.041 & 2  & 3.09  & 2.73$\pm$0.26 & 2.39$\pm$0.19  \\
NGC 6791 & +0.30 & 0.117 & 4  & 4.10  & 3.83$\pm$0.69 & 3.47$\pm$0.52  \\ 
\enddata \tablecomments{ 
The parameters and distances determined for the test clusters. The
E(B-V) is based on \citet{sfd98}, as listed in \citet{lee08_II}. The
listed metallicity and literature distances are from \citet{harris96}
for the three globular clusters and WEBDA for the two open
clusters. Note that these metallicity values are slightly different
than those found using the SSPP listed in
Table\,\ref{tab:clusters_info}; the [Fe/H] listed in this table is estimated 
using the optimized version of the SSPP (\S\,\ref{app:sspp_optimize}). }
\end{deluxetable}

\begin{landscape}
\begin{deluxetable}{cc@{}lc@{}lc@{}lc@{}lc@{}lc@{}lc@{}lc@{}l}
\tabletypesize{\small}
\tablewidth{0pt}
\tablecaption{Total Metallicity Distribution Functions of G and K dwarfs over Different Distance Ranges \label{tab:total_gkmdf}}
\tablehead{
 & \multicolumn{8}{c}{G dwarfs} & \multicolumn{8}{c}{K dwarfs} \\
\colhead{[Fe/H]} & \multicolumn{4}{c}{1.59$\leq$d$\leq$2.29 kpc} & \multicolumn{4}{c}{1.59$\leq$d$\leq$1.84 kpc} & \multicolumn{4}{c}{1.18$\leq$d$\leq$1.84 kpc} & \multicolumn{4}{c}{1.59$\leq$d$\leq$1.84 kpc} \\
 & \multicolumn{2}{c}{Raw} & \multicolumn{2}{c}{Weighted} &  \multicolumn{2}{c}{Raw} & \multicolumn{2}{c}{Weighted} & \multicolumn{2}{c}{Raw} & \multicolumn{2}{c}{Weighted} &  \multicolumn{2}{c}{Raw} & \multicolumn{2}{c}{Weighted} }
\startdata
\multirow{2}{*}{-3.3} & \multirow{2}{*}{0} & $+$0 & \multirow{2}{*}{0} & $+$0 & \multirow{2}{*}{0} & $+$0 & \multirow{2}{*}{0} & $+$0 &  					        \multirow{2}{*}{1} & $+$1 & \multirow{2}{*}{7} & $+$6 & \multirow{2}{*}{0} & $+$0 & \multirow{2}{*}{0} & $+$0 \\ 							     
  &  & $-$0  &  & $-$0  &  & $-$0  &  & $-$0 &  																         & $-$1  &  & $-$7  &  & $-$0  &  & $-$0 \\ 																     
\multirow{2}{*}{-3.1} & \multirow{2}{*}{2} & $+$1 & \multirow{2}{*}{1} & $+$1 & \multirow{2}{*}{0} & $+$0 & \multirow{2}{*}{0} & $+$0 &  					        \multirow{2}{*}{6} & $+$2 & \multirow{2}{*}{25} & $+$49 & \multirow{2}{*}{4} & $+$2 & \multirow{2}{*}{18} & $+$50 \\ 							     
  &  & $-$1  &  & $-$1  &  & $-$0  &  & $-$0 &  																         & $-$3  &  & $-$13  &  & $-$2  &  & $-$11 \\ 																     
\multirow{2}{*}{-2.9} & \multirow{2}{*}{3} & $+$2 & \multirow{2}{*}{15} & $+$15 & \multirow{2}{*}{0} & $+$0 & \multirow{2}{*}{0} & $+$0 &  					        \multirow{2}{*}{5} & $+$3 & \multirow{2}{*}{22} & $+$17 & \multirow{2}{*}{2} & $+$2 & \multirow{2}{*}{6} & $+$8 \\ 							     
  &  & $-$2  &  & $-$10  &  & $-$0  &  & $-$0 &  																         & $-$2  &  & $-$10  &  & $-$1  &  & $-$4 \\ 																     
\multirow{2}{*}{-2.7} & \multirow{2}{*}{8} & $+$3 & \multirow{2}{*}{26} & $+$10 & \multirow{2}{*}{4} & $+$2 & \multirow{2}{*}{10} & $+$6 &  					        \multirow{2}{*}{16} & $+$5 & \multirow{2}{*}{58} & $+$83 & \multirow{2}{*}{3} & $+$4 & \multirow{2}{*}{18} & $+$75 \\ 							     
  &  & $-$3  &  & $-$13  &  & $-$2  &  & $-$7 &  																         & $-$4  &  & $-$21  &  & $-$1  &  & $-$10 \\ 																     
\multirow{2}{*}{-2.5} & \multirow{2}{*}{9} & $+$4 & \multirow{2}{*}{9} & $+$7 & \multirow{2}{*}{4} & $+$2 & \multirow{2}{*}{4} & $+$3 &  					        \multirow{2}{*}{24} & $+$5 & \multirow{2}{*}{57} & $+$22 & \multirow{2}{*}{6} & $+$3 & \multirow{2}{*}{18} & $+$17 \\ 							     
  &  & $-$3  &  & $-$3  &  & $-$2  &  & $-$3 &  																         & $-$5  &  & $-$17  &  & $-$2  &  & $-$9 \\ 																     
\multirow{2}{*}{-2.3} & \multirow{2}{*}{26} & $+$7 & \multirow{2}{*}{38} & $+$24 & \multirow{2}{*}{12} & $+$3 & \multirow{2}{*}{15} & $+$7 &  					        \multirow{2}{*}{50} & $+$11 & \multirow{2}{*}{507} & $+$206 & \multirow{2}{*}{20} & $+$6 & \multirow{2}{*}{303} & $+$167 \\ 						     
  &  & $-$5  &  & $-$11  &  & $-$4  &  & $-$6 &  																         & $-$6  &  & $-$149  &  & $-$4  &  & $-$136 \\ 															     
\multirow{2}{*}{-2.1} & \multirow{2}{*}{59} & $+$14 & \multirow{2}{*}{190} & $+$80 & \multirow{2}{*}{20} & $+$6 & \multirow{2}{*}{33} & $+$33 &  				        \multirow{2}{*}{94} & $+$15 & \multirow{2}{*}{673} & $+$224 & \multirow{2}{*}{47} & $+$8 & \multirow{2}{*}{487} & $+$153 \\ 						     
  &  & $-$6  &  & $-$53  &  & $-$4  &  & $-$13 &  																         & $-$9  &  & $-$150  &  & $-$7  &  & $-$167 \\ 															     
\multirow{2}{*}{-1.9} & \multirow{2}{*}{94} & $+$16 & \multirow{2}{*}{110} & $+$28 & \multirow{2}{*}{25} & $+$8 & \multirow{2}{*}{24} & $+$17 &  				        \multirow{2}{*}{148} & $+$23 & \multirow{2}{*}{1151} & $+$780 & \multirow{2}{*}{65} & $+$11 & \multirow{2}{*}{298} & $+$669 \\ 					     
  &  & $-$10  &  & $-$18  &  & $-$4  &  & $-$6 &  																         & $-$12  &  & $-$238  &  & $-$9  &  & $-$52 \\ 															     
\multirow{2}{*}{-1.7} & \multirow{2}{*}{138} & $+$21 & \multirow{2}{*}{388} & $+$134 & \multirow{2}{*}{50} & $+$12 & \multirow{2}{*}{95} & $+$47 &  				        \multirow{2}{*}{177} & $+$27 & \multirow{2}{*}{1492} & $+$670 & \multirow{2}{*}{62} & $+$14 & \multirow{2}{*}{873} & $+$413 \\ 					     
  &  & $-$12  &  & $-$126  &  & $-$8  &  & $-$26 &  																         & $-$13  &  & $-$440  &  & $-$6  &  & $-$455 \\ 															     
\multirow{2}{*}{-1.5} & \multirow{2}{*}{187} & $+$32 & \multirow{2}{*}{425} & $+$151 & \multirow{2}{*}{65} & $+$15 & \multirow{2}{*}{165} & $+$70 &  				        \multirow{2}{*}{230} & $+$28 & \multirow{2}{*}{1880} & $+$346 & \multirow{2}{*}{100} & $+$13 & \multirow{2}{*}{770} & $+$268 \\ 					     
  &  & $-$14  &  & $-$68  &  & $-$8  &  & $-$47 &  																         & $-$18  &  & $-$306  &  & $-$13  &  & $-$181 \\ 															     
\multirow{2}{*}{-1.3} & \multirow{2}{*}{276} & $+$37 & \multirow{2}{*}{611} & $+$169 & \multirow{2}{*}{93} & $+$16 & \multirow{2}{*}{238} & $+$120 &  				        \multirow{2}{*}{237} & $+$32 & \multirow{2}{*}{1637} & $+$472 & \multirow{2}{*}{101} & $+$13 & \multirow{2}{*}{836} & $+$324 \\ 					     
  &  & $-$18  &  & $-$113  &  & $-$11  &  & $-$94 &  																         & $-$14  &  & $-$248  &  & $-$10  &  & $-$194 \\ 															     
\multirow{2}{*}{-1.1} & \multirow{2}{*}{535} & $+$42 & \multirow{2}{*}{1324} & $+$326 & \multirow{2}{*}{183} & $+$20 & \multirow{2}{*}{390} & $+$135 &  			        \multirow{2}{*}{426} & $+$45 & \multirow{2}{*}{3074} & $+$884 & \multirow{2}{*}{165} & $+$20 & \multirow{2}{*}{1168} & $+$593 \\ 					     
  &  & $-$33  &  & $-$241  &  & $-$17  &  & $-$90 &  																         & $-$19  &  & $-$420  &  & $-$11  &  & $-$230 \\ 															     
\multirow{2}{*}{-0.9} & \multirow{2}{*}{978} & $+$128 & \multirow{2}{*}{2083} & $+$404 & \multirow{2}{*}{359} & $+$110 & \multirow{2}{*}{622} & $+$218 &  			        \multirow{2}{*}{681} & $+$65 & \multirow{2}{*}{4253} & $+$545 & \multirow{2}{*}{267} & $+$25 & \multirow{2}{*}{2069} & $+$335 \\ 					     
  &  & $-$36  &  & $-$163  &  & $-$22  &  & $-$82 &  																         & $-$25  &  & $-$407  &  & $-$19  &  & $-$379 \\ 															     
\multirow{2}{*}{-0.7} & \multirow{2}{*}{1700} & $+$102 & \multirow{2}{*}{1.072$\times$10$^{4}$} & $+$1027 & \multirow{2}{*}{645} & $+$66 & \multirow{2}{*}{3618} & $+$540 &  	        \multirow{2}{*}{908} & $+$50 & \multirow{2}{*}{1.309$\times$10$^{4}$} & $+$1351 & \multirow{2}{*}{333} & $+$23 & \multirow{2}{*}{5941} & $+$708 \\ 			     
  &  & $-$56  &  & $-$707  &  & $-$30  &  & $-$411 &  																         & $-$32  &  & $-$848  &  & $-$20  &  & $-$888 \\ 															     
\multirow{2}{*}{-0.5} & \multirow{2}{*}{1550} & $+$77 & \multirow{2}{*}{1.62$\times$10$^{4}$} & $+$1195 & \multirow{2}{*}{605} & $+$42 & \multirow{2}{*}{5860} & $+$594 &  	        \multirow{2}{*}{884} & $+$44 & \multirow{2}{*}{1.793$\times$10$^{4}$} & $+$1666 & \multirow{2}{*}{311} & $+$24 & \multirow{2}{*}{7023} & $+$1228 \\ 			     
  &  & $-$42  &  & $-$691  &  & $-$23  &  & $-$437 &  																         & $-$27  &  & $-$1001  &  & $-$16  &  & $-$567 \\ 															     
\multirow{2}{*}{-0.3} & \multirow{2}{*}{1045} & $+$24 & \multirow{2}{*}{1.146$\times$10$^{4}$} & $+$655 & \multirow{2}{*}{437} & $+$20 & \multirow{2}{*}{3972} & $+$473 &  	        \multirow{2}{*}{698} & $+$25 & \multirow{2}{*}{1.83$\times$10$^{4}$} & $+$1536 & \multirow{2}{*}{230} & $+$14 & \multirow{2}{*}{6368} & $+$1208 \\ 			     
  &  & $-$43  &  & $-$776  &  & $-$21  &  & $-$331 &  																         & $-$28  &  & $-$884  &  & $-$15  &  & $-$503 \\ 															     
\multirow{2}{*}{-0.1} & \multirow{2}{*}{720} & $+$27 & \multirow{2}{*}{6007} & $+$447 & \multirow{2}{*}{352} & $+$16 & \multirow{2}{*}{2326} & $+$309 &  			        \multirow{2}{*}{470} & $+$22 & \multirow{2}{*}{1.404$\times$10$^{4}$} & $+$1540 & \multirow{2}{*}{130} & $+$16 & \multirow{2}{*}{4785} & $+$1167 \\ 			     
  &  & $-$28  &  & $-$378  &  & $-$24  &  & $-$170 &  																         & $-$23  &  & $-$987  &  & $-$9  &  & $-$527 \\ 															     
\multirow{2}{*}{0.1} & \multirow{2}{*}{330} & $+$19 & \multirow{2}{*}{1258} & $+$115 & \multirow{2}{*}{152} & $+$13 & \multirow{2}{*}{548} & $+$59 &  				        \multirow{2}{*}{176} & $+$15 & \multirow{2}{*}{6102} & $+$910 & \multirow{2}{*}{58} & $+$5 & \multirow{2}{*}{2475} & $+$515 \\ 					     
  &  & $-$32  &  & $-$186  &  & $-$16  &  & $-$114 &  																         & $-$14  &  & $-$746  &  & $-$11  &  & $-$587 \\ 															     
\multirow{2}{*}{0.3} & \multirow{2}{*}{102} & $+$9 & \multirow{2}{*}{168} & $+$30 & \multirow{2}{*}{45} & $+$8 & \multirow{2}{*}{60} & $+$27 &  				        \multirow{2}{*}{87} & $+$7 & \multirow{2}{*}{3119} & $+$597 & \multirow{2}{*}{16} & $+$7 & \multirow{2}{*}{1115} & $+$458 \\ 						     
  &  & $-$15  &  & $-$28  &  & $-$7  &  & $-$11 &  																         & $-$13  &  & $-$592  &  & $-$3  &  & $-$326 \\ 															     
\multirow{2}{*}{0.5} & \multirow{2}{*}{1} & $+$1 & \multirow{2}{*}{0} & $+$0 & \multirow{2}{*}{1} & $+$1 & \multirow{2}{*}{0} & $+$0 &  					        \multirow{2}{*}{0} & $+$0 & \multirow{2}{*}{0} & $+$0 & \multirow{2}{*}{0} & $+$0 & \multirow{2}{*}{0} & $+$0 \\ 							     
  &  & $-$1  &  & $-$0  &  & $-$1  &  & $-$0 &  																         & $-$0  &  & $-$0  &  & $-$0  &  & $-$0 \\ 																     
\enddata 
\tablecomments{
The number of G and K dwarfs in each metallicity bin using different
distance limits. The Raw numbers are the original spectroscopic
sample; the Weighted are once we have applied our corrections that
accounting for SEGUE target selection effects. The listed errors reflect
the total uncertainties in each bin determined by combining a bootstrap analysis
over 500 iterations with Monte Carlos over various properties (\S\,\ref{sec:monte_carlo}). }
\end{deluxetable}
\end{landscape}

\begin{deluxetable}{cc@{}lc@{}lc@{}lc@{}lc@{}lc@{}lc@{}lc@{}l}
\tabletypesize{\scriptsize}
\tablewidth{0pt}
\tablecaption{The Metallicity Distribution Functions of G dwarfs over Different Ranges of $|Z|$ for 1.59$\leq$d$\leq$2.29 kpc. \label{tab:gd_zbin_st}}
\tablehead{
\colhead{[Fe/H]} & \multicolumn{4}{c}{0.0$\leq |Z| <$0.5 kpc} & \multicolumn{4}{c}{0.5$\leq |Z| <$1.0 kpc} & \multicolumn{4}{c}{1.0$\leq |Z| <$1.5 kpc} & \multicolumn{4}{c}{1.5$\leq |Z| <$2.5 kpc}  \\
 & \multicolumn{2}{c}{Raw} & \multicolumn{2}{c}{Weighted} &  \multicolumn{2}{c}{Raw} & \multicolumn{2}{c}{Weighted} & \multicolumn{2}{c}{Raw} & \multicolumn{2}{c}{Weighted} 
 &  \multicolumn{2}{c}{Raw} & \multicolumn{2}{c}{Weighted} }
\startdata
\multirow{2}{*}{-3.3} & \multirow{2}{*}{0} & $+$0 & \multirow{2}{*}{0} & $+$0 & \multirow{2}{*}{0} & $+$0 & \multirow{2}{*}{0} & $+$0 & \multirow{2}{*}{0} & $+$0 & \multirow{2}{*}{0} & $+$0 & \multirow{2}{*}{0} & $+$0 & \multirow{2}{*}{0} & $+$0 \\ 
&  & $-$0 &  & $-$0 &  & $-$0 &  & $-$0&  & $-$0 &  & $-$0 &  & $-$0 &  & $-$0 \\ 
\multirow{2}{*}{-3.1} & \multirow{2}{*}{0} & $+$0 & \multirow{2}{*}{0} & $+$0 & \multirow{2}{*}{1} & $+$1 & \multirow{2}{*}{1} & $+$1 & \multirow{2}{*}{0} & $+$0 & \multirow{2}{*}{0} & $+$0 & \multirow{2}{*}{1} & $+$1 & \multirow{2}{*}{0} & $+$0 \\ 
&  & $-$0 &  & $-$0 &  & $-$1 &  & $-$1&  & $-$0 &  & $-$0 &  & $-$1 &  & $-$0 \\ 
\multirow{2}{*}{-2.9} & \multirow{2}{*}{0} & $+$0 & \multirow{2}{*}{0} & $+$0 & \multirow{2}{*}{1} & $+$1 & \multirow{2}{*}{12} & $+$13 & \multirow{2}{*}{0} & $+$0 & \multirow{2}{*}{0} & $+$0 & \multirow{2}{*}{2} & $+$1 & \multirow{2}{*}{3} & $+$2 \\ 
&  & $-$0 &  & $-$0 &  & $-$1 &  & $-$13&  & $-$0 &  & $-$0 &  & $-$2 &  & $-$2 \\ 
\multirow{2}{*}{-2.7} & \multirow{2}{*}{0} & $+$0 & \multirow{2}{*}{0} & $+$0 & \multirow{2}{*}{2} & $+$1 & \multirow{2}{*}{13} & $+$12 & \multirow{2}{*}{4} & $+$3 & \multirow{2}{*}{9} & $+$7 & \multirow{2}{*}{2} & $+$2 & \multirow{2}{*}{4} & $+$5 \\ 
&  & $-$0 &  & $-$0 &  & $-$1 &  & $-$9&  & $-$2 &  & $-$6 &  & $-$1 &  & $-$3 \\ 
\multirow{2}{*}{-2.5} & \multirow{2}{*}{1} & $+$1 & \multirow{2}{*}{0} & $+$0 & \multirow{2}{*}{2} & $+$2 & \multirow{2}{*}{2} & $+$2 & \multirow{2}{*}{3} & $+$2 & \multirow{2}{*}{3} & $+$3 & \multirow{2}{*}{3} & $+$3 & \multirow{2}{*}{3} & $+$7 \\ 
&  & $-$1 &  & $-$0 &  & $-$2 &  & $-$2&  & $-$2 &  & $-$2 &  & $-$2 &  & $-$2 \\ 
\multirow{2}{*}{-2.3} & \multirow{2}{*}{1} & $+$1 & \multirow{2}{*}{3} & $+$4 & \multirow{2}{*}{3} & $+$3 & \multirow{2}{*}{10} & $+$24 & \multirow{2}{*}{10} & $+$3 & \multirow{2}{*}{17} & $+$7 & \multirow{2}{*}{12} & $+$5 & \multirow{2}{*}{8} & $+$5 \\ 
&  & $-$1 &  & $-$2 &  & $-$2 &  & $-$6&  & $-$4 &  & $-$7 &  & $-$4 &  & $-$4 \\ 
\multirow{2}{*}{-2.1} & \multirow{2}{*}{1} & $+$1 & \multirow{2}{*}{0} & $+$0 & \multirow{2}{*}{7} & $+$4 & \multirow{2}{*}{103} & $+$75 & \multirow{2}{*}{20} & $+$7 & \multirow{2}{*}{58} & $+$27 & \multirow{2}{*}{31} & $+$8 & \multirow{2}{*}{29} & $+$9 \\ 
&  & $-$1 &  & $-$0 &  & $-$3 &  & $-$52&  & $-$4 &  & $-$22 &  & $-$9 &  & $-$10 \\ 
\multirow{2}{*}{-1.9} & \multirow{2}{*}{0} & $+$0 & \multirow{2}{*}{0} & $+$0 & \multirow{2}{*}{8} & $+$4 & \multirow{2}{*}{10} & $+$6 & \multirow{2}{*}{37} & $+$9 & \multirow{2}{*}{63} & $+$20 & \multirow{2}{*}{49} & $+$10 & \multirow{2}{*}{37} & $+$11 \\ 
&  & $-$0 &  & $-$0 &  & $-$3 &  & $-$4&  & $-$9 &  & $-$19 &  & $-$13 &  & $-$11 \\ 
\multirow{2}{*}{-1.7} & \multirow{2}{*}{3} & $+$2 & \multirow{2}{*}{146} & $+$140 & \multirow{2}{*}{31} & $+$8 & \multirow{2}{*}{88} & $+$39 & \multirow{2}{*}{52} & $+$13 & \multirow{2}{*}{98} & $+$34 & \multirow{2}{*}{52} & $+$14 & \multirow{2}{*}{55} & $+$17 \\ 
&  & $-$2 &  & $-$135 &  & $-$8 &  & $-$25&  & $-$10 &  & $-$31 &  & $-$7 &  & $-$12 \\ 
\multirow{2}{*}{-1.5} & \multirow{2}{*}{1} & $+$3 & \multirow{2}{*}{58} & $+$135 & \multirow{2}{*}{33} & $+$12 & \multirow{2}{*}{201} & $+$139 & \multirow{2}{*}{62} & $+$17 & \multirow{2}{*}{94} & $+$30 & \multirow{2}{*}{91} & $+$18 & \multirow{2}{*}{72} & $+$18 \\ 
&  & $-$1 &  & $-$50 &  & $-$5 &  & $-$48&  & $-$8 &  & $-$18 &  & $-$17 &  & $-$16 \\ 
\multirow{2}{*}{-1.3} & \multirow{2}{*}{2} & $+$2 & \multirow{2}{*}{121} & $+$166 & \multirow{2}{*}{32} & $+$11 & \multirow{2}{*}{170} & $+$130 & \multirow{2}{*}{99} & $+$19 & \multirow{2}{*}{196} & $+$45 & \multirow{2}{*}{143} & $+$22 & \multirow{2}{*}{126} & $+$28 \\ 
&  & $-$2 &  & $-$121 &  & $-$6 &  & $-$50&  & $-$24 &  & $-$57 &  & $-$28 &  & $-$30 \\ 
\multirow{2}{*}{-1.1} & \multirow{2}{*}{6} & $+$3 & \multirow{2}{*}{428} & $+$267 & \multirow{2}{*}{87} & $+$28 & \multirow{2}{*}{478} & $+$226 & \multirow{2}{*}{203} & $+$30 & \multirow{2}{*}{303} & $+$63 & \multirow{2}{*}{239} & $+$69 & \multirow{2}{*}{127} & $+$38 \\ 
&  & $-$5 &  & $-$389 &  & $-$13 &  & $-$89&  & $-$37 &  & $-$59 &  & $-$41 &  & $-$24 \\ 
\multirow{2}{*}{-0.9} & \multirow{2}{*}{7} & $+$19 & \multirow{2}{*}{115} & $+$332 & \multirow{2}{*}{121} & $+$74 & \multirow{2}{*}{1012} & $+$660 & \multirow{2}{*}{367} & $+$403 & \multirow{2}{*}{651} & $+$718 & \multirow{2}{*}{483} & $+$171 & \multirow{2}{*}{303} & $+$108 \\ 
&  & $-$5 &  & $-$91 &  & $-$26 &  & $-$257&  & $-$31 &  & $-$72 &  & $-$136 &  & $-$85 \\ 
\multirow{2}{*}{-0.7} & \multirow{2}{*}{25} & $+$5 & \multirow{2}{*}{1217} & $+$428 & \multirow{2}{*}{286} & $+$30 & \multirow{2}{*}{6443} & $+$922 & \multirow{2}{*}{638} & $+$62 & \multirow{2}{*}{2319} & $+$258 & \multirow{2}{*}{751} & $+$110 & \multirow{2}{*}{839} & $+$126 \\ 
&  & $-$7 &  & $-$359 &  & $-$25 &  & $-$741&  & $-$67 &  & $-$284 &  & $-$57 &  & $-$68 \\ 
\multirow{2}{*}{-0.5} & \multirow{2}{*}{35} & $+$5 & \multirow{2}{*}{1984} & $+$355 & \multirow{2}{*}{358} & $+$32 & \multirow{2}{*}{9024} & $+$1142 & \multirow{2}{*}{660} & $+$44 & \multirow{2}{*}{3784} & $+$271 & \multirow{2}{*}{497} & $+$35 & \multirow{2}{*}{1338} & $+$118 \\ 
&  & $-$7 &  & $-$464 &  & $-$16 &  & $-$496&  & $-$42 &  & $-$297 &  & $-$24 &  & $-$68 \\ 
\multirow{2}{*}{-0.3} & \multirow{2}{*}{40} & $+$7 & \multirow{2}{*}{1769} & $+$389 & \multirow{2}{*}{350} & $+$15 & \multirow{2}{*}{7346} & $+$538 & \multirow{2}{*}{421} & $+$15 & \multirow{2}{*}{1815} & $+$85 & \multirow{2}{*}{234} & $+$17 & \multirow{2}{*}{512} & $+$43 \\ 
&  & $-$6 &  & $-$277 &  & $-$24 &  & $-$668&  & $-$32 &  & $-$175 &  & $-$16 &  & $-$38 \\ 
\multirow{2}{*}{-0.1} & \multirow{2}{*}{40} & $+$7 & \multirow{2}{*}{1093} & $+$307 & \multirow{2}{*}{318} & $+$17 & \multirow{2}{*}{4012} & $+$298 & \multirow{2}{*}{258} & $+$22 & \multirow{2}{*}{746} & $+$85 & \multirow{2}{*}{104} & $+$9 & \multirow{2}{*}{152} & $+$15 \\ 
&  & $-$6 &  & $-$173 &  & $-$21 &  & $-$341&  & $-$15 &  & $-$59 &  & $-$13 &  & $-$22 \\ 
\multirow{2}{*}{0.1} & \multirow{2}{*}{22} & $+$5 & \multirow{2}{*}{224} & $+$68 & \multirow{2}{*}{169} & $+$12 & \multirow{2}{*}{888} & $+$90 & \multirow{2}{*}{104} & $+$12 & \multirow{2}{*}{109} & $+$19 & \multirow{2}{*}{35} & $+$8 & \multirow{2}{*}{26} & $+$8 \\ 
&  & $-$5 &  & $-$66 &  & $-$18 &  & $-$151&  & $-$14 &  & $-$17 &  & $-$7 &  & $-$6 \\ 
\multirow{2}{*}{0.3} & \multirow{2}{*}{9} & $+$2 & \multirow{2}{*}{40} & $+$12 & \multirow{2}{*}{52} & $+$7 & \multirow{2}{*}{88} & $+$32 & \multirow{2}{*}{32} & $+$7 & \multirow{2}{*}{30} & $+$7 & \multirow{2}{*}{9} & $+$4 & \multirow{2}{*}{8} & $+$4 \\ 
&  & $-$4 &  & $-$17 &  & $-$10 &  & $-$18&  & $-$9 &  & $-$10 &  & $-$4 &  & $-$4 \\ 
\multirow{2}{*}{0.5} & \multirow{2}{*}{0} & $+$0 & \multirow{2}{*}{0} & $+$0 & \multirow{2}{*}{1} & $+$1 & \multirow{2}{*}{1} & $+$1 & \multirow{2}{*}{0} & $+$0 & \multirow{2}{*}{0} & $+$0 & \multirow{2}{*}{0} & $+$0 & \multirow{2}{*}{0} & $+$0 \\ 
&  & $-$0 &  & $-$0 &  & $-$1 &  & $-$1&  & $-$0 &  & $-$0 &  & $-$0 &  & $-$0 \\ 
\enddata 
\tablecomments{
The G-dwarf raw and weighted metallicity distribution function for different ranges of $|Z|$. This is for the G dwarf sample limited to distances between 1.59 and 2.29 kpc. }
\end{deluxetable}

\begin{deluxetable}{cc@{}lc@{}lc@{}lc@{}lc@{}lc@{}lc@{}lc@{}l}
\tabletypesize{\scriptsize}
\tablewidth{0pt}
\tablecaption{The Metallicity Distribution Functions of K dwarfs over Different ranges of $|Z|$ for 1.18$\leq$d$\leq$1.84 kpc. \label{tab:kd_zbin_st}}
\tablehead{
\colhead{[Fe/H]} & \multicolumn{4}{c}{0.0$\leq |Z| <$0.5 kpc} & \multicolumn{4}{c}{0.5$\leq |Z| <$1.0 kpc} & \multicolumn{4}{c}{1.0$\leq |Z| <$1.5 kpc} & \multicolumn{4}{c}{1.5$\leq |Z| <$2.5 kpc}  \\
 & \multicolumn{2}{c}{Raw} & \multicolumn{2}{c}{Weighted} &  \multicolumn{2}{c}{Raw} & \multicolumn{2}{c}{Weighted} & \multicolumn{2}{c}{Raw} & \multicolumn{2}{c}{Weighted} 
 &  \multicolumn{2}{c}{Raw} & \multicolumn{2}{c}{Weighted} }
\startdata
\multirow{2}{*}{-3.3} & \multirow{2}{*}{0} & $+$0 & \multirow{2}{*}{0} & $+$0 & \multirow{2}{*}{1} & $+$1 & \multirow{2}{*}{7} & $+$6 & \multirow{2}{*}{0} & $+$0 & \multirow{2}{*}{0} & $+$0 & \multirow{2}{*}{0} & $+$0 & \multirow{2}{*}{0} & $+$0 \\ 
&  & $-$0 &  & $-$0 &  & $-$1 &  & $-$7&  & $-$0 &  & $-$0 &  & $-$0 &  & $-$0 \\ 
\multirow{2}{*}{-3.1} & \multirow{2}{*}{0} & $+$0 & \multirow{2}{*}{0} & $+$0 & \multirow{2}{*}{1} & $+$1 & \multirow{2}{*}{6} & $+$46 & \multirow{2}{*}{3} & $+$2 & \multirow{2}{*}{9} & $+$10 & \multirow{2}{*}{2} & $+$1 & \multirow{2}{*}{10} & $+$8 \\ 
&  & $-$0 &  & $-$0 &  & $-$1 &  & $-$6&  & $-$2 &  & $-$6 &  & $-$2 &  & $-$8 \\ 
\multirow{2}{*}{-2.9} & \multirow{2}{*}{1} & $+$1 & \multirow{2}{*}{10} & $+$10 & \multirow{2}{*}{0} & $+$0 & \multirow{2}{*}{0} & $+$0 & \multirow{2}{*}{2} & $+$2 & \multirow{2}{*}{6} & $+$7 & \multirow{2}{*}{2} & $+$2 & \multirow{2}{*}{6} & $+$7 \\ 
&  & $-$1 &  & $-$10 &  & $-$0 &  & $-$0&  & $-$1 &  & $-$4 &  & $-$1 &  & $-$5 \\ 
\multirow{2}{*}{-2.7} & \multirow{2}{*}{1} & $+$1 & \multirow{2}{*}{21} & $+$32 & \multirow{2}{*}{3} & $+$2 & \multirow{2}{*}{21} & $+$79 & \multirow{2}{*}{12} & $+$3 & \multirow{2}{*}{16} & $+$20 & \multirow{2}{*}{0} & $+$0 & \multirow{2}{*}{0} & $+$0 \\ 
&  & $-$1 &  & $-$15 &  & $-$2 &  & $-$11&  & $-$4 &  & $-$6 &  & $-$0 &  & $-$0 \\ 
\multirow{2}{*}{-2.5} & \multirow{2}{*}{1} & $+$1 & \multirow{2}{*}{4} & $+$5 & \multirow{2}{*}{8} & $+$2 & \multirow{2}{*}{27} & $+$18 & \multirow{2}{*}{15} & $+$5 & \multirow{2}{*}{26} & $+$11 & \multirow{2}{*}{0} & $+$0 & \multirow{2}{*}{0} & $+$0 \\ 
&  & $-$1 &  & $-$3 &  & $-$4 &  & $-$14&  & $-$4 &  & $-$9 &  & $-$0 &  & $-$0 \\ 
\multirow{2}{*}{-2.3} & \multirow{2}{*}{2} & $+$2 & \multirow{2}{*}{21} & $+$67 & \multirow{2}{*}{21} & $+$6 & \multirow{2}{*}{406} & $+$186 & \multirow{2}{*}{23} & $+$7 & \multirow{2}{*}{61} & $+$35 & \multirow{2}{*}{4} & $+$2 & \multirow{2}{*}{19} & $+$13 \\ 
&  & $-$1 &  & $-$14 &  & $-$5 &  & $-$168&  & $-$4 &  & $-$14 &  & $-$2 &  & $-$12 \\ 
\multirow{2}{*}{-2.1} & \multirow{2}{*}{7} & $+$3 & \multirow{2}{*}{136} & $+$125 & \multirow{2}{*}{26} & $+$6 & \multirow{2}{*}{373} & $+$162 & \multirow{2}{*}{50} & $+$10 & \multirow{2}{*}{148} & $+$73 & \multirow{2}{*}{11} & $+$4 & \multirow{2}{*}{16} & $+$15 \\ 
&  & $-$3 &  & $-$75 &  & $-$5 &  & $-$128&  & $-$7 &  & $-$29 &  & $-$4 &  & $-$6 \\ 
\multirow{2}{*}{-1.9} & \multirow{2}{*}{7} & $+$4 & \multirow{2}{*}{442} & $+$590 & \multirow{2}{*}{51} & $+$14 & \multirow{2}{*}{423} & $+$445 & \multirow{2}{*}{74} & $+$13 & \multirow{2}{*}{234} & $+$63 & \multirow{2}{*}{16} & $+$5 & \multirow{2}{*}{51} & $+$27 \\ 
&  & $-$3 &  & $-$237 &  & $-$7 &  & $-$107&  & $-$9 &  & $-$33 &  & $-$5 &  & $-$17 \\ 
\multirow{2}{*}{-1.7} & \multirow{2}{*}{9} & $+$4 & \multirow{2}{*}{171} & $+$77 & \multirow{2}{*}{52} & $+$11 & \multirow{2}{*}{898} & $+$643 & \multirow{2}{*}{99} & $+$15 & \multirow{2}{*}{346} & $+$90 & \multirow{2}{*}{17} & $+$6 & \multirow{2}{*}{77} & $+$30 \\ 
&  & $-$3 &  & $-$61 &  & $-$7 &  & $-$448&  & $-$13 &  & $-$57 &  & $-$4 &  & $-$26 \\ 
\multirow{2}{*}{-1.5} & \multirow{2}{*}{12} & $+$4 & \multirow{2}{*}{337} & $+$167 & \multirow{2}{*}{81} & $+$14 & \multirow{2}{*}{1019} & $+$292 & \multirow{2}{*}{115} & $+$17 & \multirow{2}{*}{442} & $+$83 & \multirow{2}{*}{22} & $+$4 & \multirow{2}{*}{82} & $+$21 \\ 
&  & $-$4 &  & $-$149 &  & $-$12 &  & $-$230&  & $-$13 &  & $-$71 &  & $-$7 &  & $-$34 \\ 
\multirow{2}{*}{-1.3} & \multirow{2}{*}{12} & $+$4 & \multirow{2}{*}{155} & $+$200 & \multirow{2}{*}{72} & $+$14 & \multirow{2}{*}{1026} & $+$375 & \multirow{2}{*}{130} & $+$18 & \multirow{2}{*}{400} & $+$75 & \multirow{2}{*}{23} & $+$8 & \multirow{2}{*}{56} & $+$28 \\ 
&  & $-$4 &  & $-$65 &  & $-$9 &  & $-$264&  & $-$16 &  & $-$67 &  & $-$5 &  & $-$16 \\ 
\multirow{2}{*}{-1.1} & \multirow{2}{*}{27} & $+$7 & \multirow{2}{*}{1274} & $+$674 & \multirow{2}{*}{162} & $+$33 & \multirow{2}{*}{1078} & $+$641 & \multirow{2}{*}{204} & $+$25 & \multirow{2}{*}{630} & $+$131 & \multirow{2}{*}{33} & $+$10 & \multirow{2}{*}{91} & $+$42 \\ 
&  & $-$6 &  & $-$558 &  & $-$24 &  & $-$193&  & $-$24 &  & $-$81 &  & $-$7 &  & $-$24 \\ 
\multirow{2}{*}{-0.9} & \multirow{2}{*}{34} & $+$12 & \multirow{2}{*}{921} & $+$400 & \multirow{2}{*}{225} & $+$32 & \multirow{2}{*}{2050} & $+$371 & \multirow{2}{*}{374} & $+$44 & \multirow{2}{*}{1141} & $+$164 & \multirow{2}{*}{48} & $+$19 & \multirow{2}{*}{141} & $+$57 \\ 
&  & $-$6 &  & $-$305 &  & $-$25 &  & $-$310&  & $-$44 &  & $-$152 &  & $-$6 &  & $-$31 \\ 
\multirow{2}{*}{-0.7} & \multirow{2}{*}{54} & $+$11 & \multirow{2}{*}{2520} & $+$703 & \multirow{2}{*}{321} & $+$29 & \multirow{2}{*}{6642} & $+$1004 & \multirow{2}{*}{485} & $+$27 & \multirow{2}{*}{3591} & $+$286 & \multirow{2}{*}{48} & $+$10 & \multirow{2}{*}{336} & $+$104 \\ 
&  & $-$6 &  & $-$474 &  & $-$15 &  & $-$632&  & $-$37 &  & $-$380 &  & $-$8 &  & $-$66 \\ 
\multirow{2}{*}{-0.5} & \multirow{2}{*}{88} & $+$15 & \multirow{2}{*}{5272} & $+$1208 & \multirow{2}{*}{410} & $+$24 & \multirow{2}{*}{8963} & $+$1062 & \multirow{2}{*}{359} & $+$17 & \multirow{2}{*}{3509} & $+$212 & \multirow{2}{*}{27} & $+$9 & \multirow{2}{*}{190} & $+$126 \\ 
&  & $-$8 &  & $-$823 &  & $-$18 &  & $-$570&  & $-$24 &  & $-$326 &  & $-$4 &  & $-$35 \\ 
\multirow{2}{*}{-0.3} & \multirow{2}{*}{100} & $+$15 & \multirow{2}{*}{5677} & $+$1357 & \multirow{2}{*}{378} & $+$20 & \multirow{2}{*}{1.018$\times$10$^{4}$} & $+$857 & \multirow{2}{*}{202} & $+$12 & \multirow{2}{*}{2325} & $+$105 & \multirow{2}{*}{18} & $+$3 & \multirow{2}{*}{115} & $+$48 \\ 
&  & $-$8 &  & $-$503 &  & $-$19 &  & $-$674&  & $-$17 &  & $-$321 &  & $-$6 &  & $-$42 \\ 
\multirow{2}{*}{-0.1} & \multirow{2}{*}{77} & $+$8 & \multirow{2}{*}{5042} & $+$1049 & \multirow{2}{*}{265} & $+$17 & \multirow{2}{*}{7486} & $+$1064 & \multirow{2}{*}{121} & $+$12 & \multirow{2}{*}{1449} & $+$139 & \multirow{2}{*}{7} & $+$2 & \multirow{2}{*}{58} & $+$44 \\ 
&  & $-$10 &  & $-$745 &  & $-$17 &  & $-$528&  & $-$11 &  & $-$203 &  & $-$3 &  & $-$28 \\ 
\multirow{2}{*}{0.1} & \multirow{2}{*}{33} & $+$5 & \multirow{2}{*}{2412} & $+$693 & \multirow{2}{*}{93} & $+$13 & \multirow{2}{*}{3096} & $+$522 & \multirow{2}{*}{47} & $+$7 & \multirow{2}{*}{574} & $+$95 & \multirow{2}{*}{3} & $+$1 & \multirow{2}{*}{20} & $+$14 \\ 
&  & $-$6 &  & $-$539 &  & $-$9 &  & $-$451&  & $-$8 &  & $-$136 &  & $-$2 &  & $-$14 \\ 
\multirow{2}{*}{0.3} & \multirow{2}{*}{16} & $+$4 & \multirow{2}{*}{1631} & $+$513 & \multirow{2}{*}{55} & $+$4 & \multirow{2}{*}{1317} & $+$415 & \multirow{2}{*}{16} & $+$6 & \multirow{2}{*}{171} & $+$64 & \multirow{2}{*}{0} & $+$0 & \multirow{2}{*}{0} & $+$0 \\ 
&  & $-$4 &  & $-$468 &  & $-$13 &  & $-$323&  & $-$4 &  & $-$60 &  & $-$0 &  & $-$0 \\ 
\multirow{2}{*}{0.5} & \multirow{2}{*}{0} & $+$0 & \multirow{2}{*}{0} & $+$0 & \multirow{2}{*}{0} & $+$0 & \multirow{2}{*}{0} & $+$0 & \multirow{2}{*}{0} & $+$0 & \multirow{2}{*}{0} & $+$0 & \multirow{2}{*}{0} & $+$0 & \multirow{2}{*}{0} & $+$0 \\ 
&  & $-$0 &  & $-$0 &  & $-$0 &  & $-$0&  & $-$0 &  & $-$0 &  & $-$0 &  & $-$0 \\ 
\enddata 
\tablecomments{
The K-dwarf raw and weighted metallicity distribution function for different ranges of $|Z|$. This is for the distances between 1.18 and 1.84 kpc. }
\end{deluxetable}

\begin{deluxetable}{cc@{}lc@{}lc@{}lc@{}lc@{}lc@{}lc@{}lc@{}l}
\tabletypesize{\scriptsize} 
\tablewidth{0pt}
\tablecaption{The Metallicity Distribution Functions of G dwarfs over Different Ranges of $|Z|$ for 1.59$\leq$d$\leq$1.84 kpc. \label{tab:gd_zbin_vc}}
\tablehead{
\colhead{[Fe/H]} & \multicolumn{4}{c}{0.0$\leq |Z| <$0.5 kpc} & \multicolumn{4}{c}{0.5$\leq |Z| <$1.0 kpc} & \multicolumn{4}{c}{1.0$\leq |Z| <$1.5 kpc} & \multicolumn{4}{c}{1.5$\leq |Z| <$2.5 kpc}  \\
 & \multicolumn{2}{c}{Raw} & \multicolumn{2}{c}{Weighted} &  \multicolumn{2}{c}{Raw} & \multicolumn{2}{c}{Weighted} & \multicolumn{2}{c}{Raw} & \multicolumn{2}{c}{Weighted} 
 &  \multicolumn{2}{c}{Raw} & \multicolumn{2}{c}{Weighted} }
\startdata
\multirow{2}{*}{-3.3} & \multirow{2}{*}{0} & $+$0 & \multirow{2}{*}{0} & $+$0 & \multirow{2}{*}{0} & $+$0 & \multirow{2}{*}{0} & $+$0 & \multirow{2}{*}{0} & $+$0 & \multirow{2}{*}{0} & $+$0 & \multirow{2}{*}{0} & $+$0 & \multirow{2}{*}{0} & $+$0  \\
		      &  & $-$0 &  & $-$0 &  & $-$0 &  & $-$0&  & $-$0 &  & $-$0 &  & $-$0 &  & $-$0  \\
\multirow{2}{*}{-3.1} & \multirow{2}{*}{0} & $+$0 & \multirow{2}{*}{0} & $+$0 & \multirow{2}{*}{0} & $+$0 & \multirow{2}{*}{0} & $+$0 & \multirow{2}{*}{0} & $+$0 & \multirow{2}{*}{0} & $+$0 & \multirow{2}{*}{0} & $+$0 & \multirow{2}{*}{0} & $+$0  \\
		      &  & $-$0 &  & $-$0 &  & $-$0 &  & $-$0&  & $-$0 &  & $-$0 &  & $-$0 &  & $-$0  \\
\multirow{2}{*}{-2.9} & \multirow{2}{*}{0} & $+$0 & \multirow{2}{*}{0} & $+$0 & \multirow{2}{*}{0} & $+$0 & \multirow{2}{*}{0} & $+$0 & \multirow{2}{*}{0} & $+$0 & \multirow{2}{*}{0} & $+$0 & \multirow{2}{*}{0} & $+$0 & \multirow{2}{*}{0} & $+$0  \\
		      &  & $-$0 &  & $-$0 &  & $-$0 &  & $-$0&  & $-$0 &  & $-$0 &  & $-$0 &  & $-$0  \\
\multirow{2}{*}{-2.7} & \multirow{2}{*}{0} & $+$0 & \multirow{2}{*}{0} & $+$0 & \multirow{2}{*}{1} & $+$1 & \multirow{2}{*}{4} & $+$4 & \multirow{2}{*}{2} & $+$2 & \multirow{2}{*}{6} & $+$6 & \multirow{2}{*}{1} & $+$1 & \multirow{2}{*}{0} & $+$0  \\
		      &  & $-$0 &  & $-$0 &  & $-$1 &  & $-$5&  & $-$1 &  & $-$5 &  & $-$1 &  & $-$0  \\
\multirow{2}{*}{-2.5} & \multirow{2}{*}{1} & $+$1 & \multirow{2}{*}{0} & $+$0 & \multirow{2}{*}{1} & $+$2 & \multirow{2}{*}{1} & $+$3 & \multirow{2}{*}{1} & $+$1 & \multirow{2}{*}{3} & $+$3 & \multirow{2}{*}{1} & $+$2 & \multirow{2}{*}{0} & $+$2  \\
		      &  & $-$1 &  & $-$0 &  & $-$1 &  & $-$1&  & $-$1 &  & $-$3 &  & $-$1 &  & $-$0  \\
\multirow{2}{*}{-2.3} & \multirow{2}{*}{1} & $+$1 & \multirow{2}{*}{3} & $+$4 & \multirow{2}{*}{2} & $+$1 & \multirow{2}{*}{3} & $+$5 & \multirow{2}{*}{6} & $+$3 & \multirow{2}{*}{9} & $+$6 & \multirow{2}{*}{3} & $+$3 & \multirow{2}{*}{0} & $+$2  \\
		      &  & $-$1 &  & $-$3 &  & $-$2 &  & $-$2&  & $-$3 &  & $-$5 &  & $-$2 &  & $-$0  \\
\multirow{2}{*}{-2.1} & \multirow{2}{*}{1} & $+$1 & \multirow{2}{*}{0} & $+$0 & \multirow{2}{*}{0} & $+$0 & \multirow{2}{*}{0} & $+$0 & \multirow{2}{*}{10} & $+$5 & \multirow{2}{*}{27} & $+$17 & \multirow{2}{*}{9} & $+$3 & \multirow{2}{*}{6} & $+$3 \\ 
		      &  & $-$1 &  & $-$0 &  & $-$0 &  & $-$0&  & $-$3 &  & $-$16 &  & $-$7 &  & $-$5  \\
\multirow{2}{*}{-1.9} & \multirow{2}{*}{0} & $+$0 & \multirow{2}{*}{0} & $+$0 & \multirow{2}{*}{6} & $+$3 & \multirow{2}{*}{6} & $+$5 & \multirow{2}{*}{12} & $+$6 & \multirow{2}{*}{14} & $+$11 & \multirow{2}{*}{7} & $+$4 & \multirow{2}{*}{5} & $+$3  \\
		      &  & $-$0 &  & $-$0 &  & $-$3 &  & $-$3&  & $-$4 &  & $-$6 &  & $-$4 &  & $-$3  \\
\multirow{2}{*}{-1.7} & \multirow{2}{*}{2} & $+$2 & \multirow{2}{*}{16} & $+$36 & \multirow{2}{*}{16} & $+$5 & \multirow{2}{*}{54} & $+$26 & \multirow{2}{*}{21} & $+$13 & \multirow{2}{*}{18} & $+$18 & \multirow{2}{*}{11} & $+$7 & \multirow{2}{*}{8} & $+$5  \\
		      &  & $-$2 &  & $-$15 &  & $-$7 &  & $-$27&  & $-$7 &  & $-$7 &  & $-$5 &  & $-$3  \\
\multirow{2}{*}{-1.5} & \multirow{2}{*}{0} & $+$0 & \multirow{2}{*}{0} & $+$0 & \multirow{2}{*}{19} & $+$11 & \multirow{2}{*}{130} & $+$81 & \multirow{2}{*}{24} & $+$14 & \multirow{2}{*}{24} & $+$17 & \multirow{2}{*}{22} & $+$9 & \multirow{2}{*}{11} & $+$5  \\
		      &  & $-$0 &  & $-$0 &  & $-$6 &  & $-$60&  & $-$4 &  & $-$6 &  & $-$10 &  & $-$5  \\
\multirow{2}{*}{-1.3} & \multirow{2}{*}{2} & $+$1 & \multirow{2}{*}{121} & $+$132 & \multirow{2}{*}{17} & $+$15 & \multirow{2}{*}{44} & $+$65 & \multirow{2}{*}{44} & $+$10 & \multirow{2}{*}{57} & $+$23 & \multirow{2}{*}{30} & $+$13 & \multirow{2}{*}{16} & $+$8 \\  
		      &  & $-$2 &  & $-$98 &  & $-$5 &  & $-$16&  & $-$12 &  & $-$20 &  & $-$9 &  & $-$5  \\
\multirow{2}{*}{-1.1} & \multirow{2}{*}{1} & $+$2 & \multirow{2}{*}{79} & $+$115 & \multirow{2}{*}{41} & $+$90 & \multirow{2}{*}{208} & $+$458 & \multirow{2}{*}{97} & $+$60 & \multirow{2}{*}{86} & $+$57 & \multirow{2}{*}{44} & $+$40 & \multirow{2}{*}{16} & $+$16 \\ 
		      &  & $-$1 &  & $-$100 &  & $-$27 &  & $-$145&  & $-$32 &  & $-$30 &  & $-$27 &  & $-$10  \\
\multirow{2}{*}{-0.9} & \multirow{2}{*}{4} & $+$13 & \multirow{2}{*}{26} & $+$127 & \multirow{2}{*}{48} & $+$51 & \multirow{2}{*}{295} & $+$317 & \multirow{2}{*}{194} & $+$345 & \multirow{2}{*}{242} & $+$431 & \multirow{2}{*}{113} & $+$97 & \multirow{2}{*}{59} & $+$51 \\  
		      &  & $-$4 &  & $-$33 &  & $-$19 &  & $-$127&  & $-$196 &  & $-$248 &  & $-$121 &  & $-$63 \\ 
\multirow{2}{*}{-0.7} & \multirow{2}{*}{14} & $+$4 & \multirow{2}{*}{599} & $+$294 & \multirow{2}{*}{131} & $+$24 & \multirow{2}{*}{2077} & $+$491 & \multirow{2}{*}{321} & $+$40 & \multirow{2}{*}{814} & $+$112 & \multirow{2}{*}{179} & $+$71 & \multirow{2}{*}{127} & $+$52 \\  
		      &  & $-$5 &  & $-$229 &  & $-$16 &  & $-$361&  & $-$63 &  & $-$172 &  & $-$54 &  & $-$39   \\
\multirow{2}{*}{-0.5} & \multirow{2}{*}{21} & $+$4 & \multirow{2}{*}{1271} & $+$284 & \multirow{2}{*}{158} & $+$22 & \multirow{2}{*}{2986} & $+$561 & \multirow{2}{*}{308} & $+$33 & \multirow{2}{*}{1332} & $+$155 & \multirow{2}{*}{118} & $+$14 & \multirow{2}{*}{281} & $+$34  \\
		      &  & $-$6 &  & $-$416 &  & $-$10 &  & $-$230&  & $-$19 &  & $-$113 &  & $-$15 &  & $-$38  \\
\multirow{2}{*}{-0.3} & \multirow{2}{*}{20} & $+$5 & \multirow{2}{*}{997} & $+$297 & \multirow{2}{*}{151} & $+$14 & \multirow{2}{*}{2190} & $+$382 & \multirow{2}{*}{203} & $+$12 & \multirow{2}{*}{652} & $+$61 & \multirow{2}{*}{63} & $+$9 & \multirow{2}{*}{124} & $+$19  \\
		      &  & $-$4 &  & $-$244 &  & $-$11 &  & $-$214&  & $-$18 &  & $-$65 &  & $-$8 &  & $-$20  \\
\multirow{2}{*}{-0.1} & \multirow{2}{*}{23} & $+$6 & \multirow{2}{*}{562} & $+$245 & \multirow{2}{*}{159} & $+$10 & \multirow{2}{*}{1454} & $+$164 & \multirow{2}{*}{144} & $+$11 & \multirow{2}{*}{297} & $+$28 & \multirow{2}{*}{26} & $+$6 & \multirow{2}{*}{27} & $+$10  \\
		      &  & $-$4 &  & $-$115 &  & $-$18 &  & $-$162&  & $-$16 &  & $-$40 &  & $-$6 &  & $-$7 \\ 
\multirow{2}{*}{0.1}  & \multirow{2}{*}{17} & $+$3 & \multirow{2}{*}{163} & $+$40 & \multirow{2}{*}{70} & $+$11 & \multirow{2}{*}{315} & $+$60 & \multirow{2}{*}{56} & $+$7 & \multirow{2}{*}{52} & $+$11 & \multirow{2}{*}{9} & $+$4 & \multirow{2}{*}{5} & $+$4  \\
		      &  & $-$5 &  & $-$65 &  & $-$10 &  & $-$71&  & $-$11 &  & $-$12 &  & $-$4 &  & $-$2 \\ 
\multirow{2}{*}{0.3}  & \multirow{2}{*}{3} & $+$2 & \multirow{2}{*}{6} & $+$12 & \multirow{2}{*}{23} & $+$5 & \multirow{2}{*}{36} & $+$22 & \multirow{2}{*}{16} & $+$6 & \multirow{2}{*}{12} & $+$5 & \multirow{2}{*}{3} & $+$1 & \multirow{2}{*}{4} & $+$2  \\
		      &  & $-$2 &  & $-$3 &  & $-$5 &  & $-$10&  & $-$5 &  & $-$4 &  & $-$3 &  & $-$3  \\
\multirow{2}{*}{0.5}  & \multirow{2}{*}{0} & $+$0 & \multirow{2}{*}{0} & $+$0 & \multirow{2}{*}{1} & $+$1 & \multirow{2}{*}{1} & $+$1 & \multirow{2}{*}{0} & $+$0 & \multirow{2}{*}{0} & $+$0 & \multirow{2}{*}{0} & $+$0 & \multirow{2}{*}{0} & $+$0  \\
		      &  & $-$0 &  & $-$0 &  & $-$1 &  & $-$1&  & $-$0 &  & $-$0 &  & $-$0 &  & $-$0  \\
\enddata
\tablecomments{
The G-dwarf raw and weighted metallicity distribution function for different ranges of $|Z|$. This is for the distances between 1.59 and 1.84 kpc. }
\end{deluxetable}

\begin{deluxetable}{cc@{}lc@{}lc@{}lc@{}lc@{}lc@{}lc@{}lc@{}l}
\tabletypesize{\scriptsize}
\tablewidth{0pt}
\tablecaption{The Metallicity Distribution Functions of K dwarfs over Different ranges of $|Z|$ for 1.59$\leq$d$\leq$1.84 kpc. \label{tab:kd_zbin_vc}}
\tablehead{
\colhead{[Fe/H]} & \multicolumn{4}{c}{0.0$\leq |Z| <$0.5 kpc} & \multicolumn{4}{c}{0.5$\leq |Z| <$1.0 kpc} & \multicolumn{4}{c}{1.0$\leq |Z| <$1.5 kpc} & \multicolumn{4}{c}{1.5$\leq |Z| <$2.5 kpc}  \\
 & \multicolumn{2}{c}{Raw} & \multicolumn{2}{c}{Weighted} &  \multicolumn{2}{c}{Raw} & \multicolumn{2}{c}{Weighted} & \multicolumn{2}{c}{Raw} & \multicolumn{2}{c}{Weighted} 
 &  \multicolumn{2}{c}{Raw} & \multicolumn{2}{c}{Weighted} }
\startdata
\multirow{2}{*}{-3.3}& \multirow{2}{*}{0} & $+$0 & \multirow{2}{*}{0} & $+$0 & \multirow{2}{*}{0} & $+$0 & \multirow{2}{*}{0} & $+$0 & \multirow{2}{*}{0} & $+$0 & \multirow{2}{*}{0} & $+$0 & \multirow{2}{*}{0} & $+$0 & \multirow{2}{*}{0} & $+$0 \\ 							
		     &  & $-$0 &  & $-$0 &  & $-$0 &  & $-$0&  & $-$0 &  & $-$0 &  & $-$0 &  & $-$0 \\ 																									
\multirow{2}{*}{-3.1}& \multirow{2}{*}{0} & $+$0 & \multirow{2}{*}{0} & $+$0 & \multirow{2}{*}{1} & $+$1 & \multirow{2}{*}{6} & $+$49 & \multirow{2}{*}{1} & $+$2 & \multirow{2}{*}{1} & $+$5 & \multirow{2}{*}{2} & $+$1 & \multirow{2}{*}{10} & $+$9 \\ 							
		     &  & $-$0 &  & $-$0 &  & $-$1 &  & $-$6&  & $-$1 &  & $-$1 &  & $-$2 &  & $-$8 \\ 																									
\multirow{2}{*}{-2.9}& \multirow{2}{*}{0} & $+$0 & \multirow{2}{*}{0} & $+$0 & \multirow{2}{*}{0} & $+$0 & \multirow{2}{*}{0} & $+$0 & \multirow{2}{*}{0} & $+$0 & \multirow{2}{*}{0} & $+$0 & \multirow{2}{*}{2} & $+$2 & \multirow{2}{*}{6} & $+$7 \\ 							
		     &  & $-$0 &  & $-$0 &  & $-$0 &  & $-$0&  & $-$0 &  & $-$0 &  & $-$1 &  & $-$5 \\ 																									
\multirow{2}{*}{-2.7}& \multirow{2}{*}{0} & $+$0 & \multirow{2}{*}{0} & $+$0 & \multirow{2}{*}{2} & $+$1 & \multirow{2}{*}{17} & $+$86 & \multirow{2}{*}{1} & $+$3 & \multirow{2}{*}{1} & $+$19 & \multirow{2}{*}{0} & $+$0 & \multirow{2}{*}{0} & $+$0 \\ 							
		     &  & $-$0 &  & $-$0 &  & $-$2 &  & $-$13&  & $-$1 &  & $-$1 &  & $-$0 &  & $-$0 \\ 																									
\multirow{2}{*}{-2.5}& \multirow{2}{*}{0} & $+$0 & \multirow{2}{*}{0} & $+$0 & \multirow{2}{*}{1} & $+$1 & \multirow{2}{*}{8} & $+$13 & \multirow{2}{*}{5} & $+$2 & \multirow{2}{*}{11} & $+$8 & \multirow{2}{*}{0} & $+$0 & \multirow{2}{*}{0} & $+$0 \\ 							
		     &  & $-$0 &  & $-$0 &  & $-$1 &  & $-$7&  & $-$3 &  & $-$6 &  & $-$0 &  & $-$0 \\ 																									
\multirow{2}{*}{-2.3}& \multirow{2}{*}{1} & $+$1 & \multirow{2}{*}{3} & $+$35 & \multirow{2}{*}{8} & $+$4 & \multirow{2}{*}{267} & $+$171 & \multirow{2}{*}{7} & $+$5 & \multirow{2}{*}{14} & $+$25 & \multirow{2}{*}{4} & $+$2 & \multirow{2}{*}{19} & $+$14 \\ 						
		     &  & $-$1 &  & $-$3 &  & $-$3 &  & $-$148&  & $-$3 &  & $-$6 &  & $-$2 &  & $-$12 \\ 																									
\multirow{2}{*}{-2.1}& \multirow{2}{*}{2} & $+$1 & \multirow{2}{*}{104} & $+$73 & \multirow{2}{*}{11} & $+$3 & \multirow{2}{*}{279} & $+$130 & \multirow{2}{*}{24} & $+$6 & \multirow{2}{*}{90} & $+$70 & \multirow{2}{*}{10} & $+$4 & \multirow{2}{*}{15} & $+$14 \\ 					
		     &  & $-$2 &  & $-$82 &  & $-$4 &  & $-$134&  & $-$6 &  & $-$26 &  & $-$4 &  & $-$6 \\ 																									
\multirow{2}{*}{-1.9}& \multirow{2}{*}{0} & $+$0 & \multirow{2}{*}{0} & $+$0 & \multirow{2}{*}{18} & $+$5 & \multirow{2}{*}{142} & $+$398 & \multirow{2}{*}{32} & $+$7 & \multirow{2}{*}{106} & $+$50 & \multirow{2}{*}{15} & $+$5 & \multirow{2}{*}{50} & $+$26 \\ 						
		     &  & $-$0 &  & $-$0 &  & $-$5 &  & $-$49&  & $-$6 &  & $-$25 &  & $-$5 &  & $-$18 \\ 																									
\multirow{2}{*}{-1.7}& \multirow{2}{*}{1} & $+$1 & \multirow{2}{*}{23} & $+$33 & \multirow{2}{*}{13} & $+$5 & \multirow{2}{*}{667} & $+$420 & \multirow{2}{*}{31} & $+$9 & \multirow{2}{*}{105} & $+$76 & \multirow{2}{*}{17} & $+$5 & \multirow{2}{*}{77} & $+$29 \\ 					
		     &  & $-$1 &  & $-$19 &  & $-$4 &  & $-$473&  & $-$6 &  & $-$25 &  & $-$4 &  & $-$27 \\ 																									
\multirow{2}{*}{-1.5}& \multirow{2}{*}{2} & $+$2 & \multirow{2}{*}{65} & $+$115 & \multirow{2}{*}{31} & $+$5 & \multirow{2}{*}{456} & $+$237 & \multirow{2}{*}{46} & $+$9 & \multirow{2}{*}{175} & $+$50 & \multirow{2}{*}{21} & $+$4 & \multirow{2}{*}{74} & $+$23 \\ 					
		     &  & $-$2 &  & $-$44 &  & $-$9 &  & $-$176&  & $-$11 &  & $-$48 &  & $-$7 &  & $-$29 \\ 																								
\multirow{2}{*}{-1.3}& \multirow{2}{*}{3} & $+$2 & \multirow{2}{*}{88} & $+$165 & \multirow{2}{*}{19} & $+$5 & \multirow{2}{*}{469} & $+$271 & \multirow{2}{*}{57} & $+$8 & \multirow{2}{*}{223} & $+$50 & \multirow{2}{*}{22} & $+$8 & \multirow{2}{*}{56} & $+$27 \\ 					
		     &  & $-$2 &  & $-$65 &  & $-$5 &  & $-$183&  & $-$11 &  & $-$56 &  & $-$5 &  & $-$16 \\ 																								
\multirow{2}{*}{-1.1}& \multirow{2}{*}{5} & $+$9 & \multirow{2}{*}{316} & $+$627 & \multirow{2}{*}{51} & $+$12 & \multirow{2}{*}{489} & $+$500 & \multirow{2}{*}{78} & $+$12 & \multirow{2}{*}{275} & $+$86 & \multirow{2}{*}{31} & $+$10 & \multirow{2}{*}{89} & $+$39 \\ 					
		     &  & $-$4 &  & $-$327 &  & $-$13 &  & $-$162&  & $-$13 &  & $-$52 &  & $-$6 &  & $-$23 \\ 																								
\multirow{2}{*}{-0.9}& \multirow{2}{*}{11} & $+$4 & \multirow{2}{*}{657} & $+$355 & \multirow{2}{*}{59} & $+$14 & \multirow{2}{*}{710} & $+$280 & \multirow{2}{*}{154} & $+$29 & \multirow{2}{*}{571} & $+$129 & \multirow{2}{*}{43} & $+$21 & \multirow{2}{*}{131} & $+$66 \\ 				
		     &  & $-$5 &  & $-$390 &  & $-$14 &  & $-$188&  & $-$38 &  & $-$153 &  & $-$5 &  & $-$29 \\ 																								
\multirow{2}{*}{-0.7}& \multirow{2}{*}{10} & $+$3 & \multirow{2}{*}{844} & $+$360 & \multirow{2}{*}{98} & $+$11 & \multirow{2}{*}{3229} & $+$625 & \multirow{2}{*}{180} & $+$17 & \multirow{2}{*}{1547} & $+$207 & \multirow{2}{*}{45} & $+$9 & \multirow{2}{*}{322} & $+$97 \\ 				
		     &  & $-$3 &  & $-$379 &  & $-$10 &  & $-$658&  & $-$18 &  & $-$218 &  & $-$7 &  & $-$67 \\ 																								
\multirow{2}{*}{-0.5}& \multirow{2}{*}{20} & $+$6 & \multirow{2}{*}{1826} & $+$737 & \multirow{2}{*}{115} & $+$15 & \multirow{2}{*}{3322} & $+$901 & \multirow{2}{*}{150} & $+$10 & \multirow{2}{*}{1688} & $+$161 & \multirow{2}{*}{26} & $+$9 & \multirow{2}{*}{188} & $+$119 \\ 				
		     &  & $-$4 &  & $-$578 &  & $-$9 &  & $-$363&  & $-$15 &  & $-$220 &  & $-$3 &  & $-$35 \\ 																								
\multirow{2}{*}{-0.3}& \multirow{2}{*}{20} & $+$7 & \multirow{2}{*}{1003} & $+$920 & \multirow{2}{*}{118} & $+$10 & \multirow{2}{*}{4101} & $+$703 & \multirow{2}{*}{74} & $+$8 & \multirow{2}{*}{1150} & $+$127 & \multirow{2}{*}{18} & $+$3 & \multirow{2}{*}{115} & $+$47 \\ 				
		     &  & $-$3 &  & $-$212 &  & $-$10 &  & $-$435&  & $-$9 &  & $-$210 &  & $-$6 &  & $-$42 \\ 																								
\multirow{2}{*}{-0.1}& \multirow{2}{*}{12} & $+$5 & \multirow{2}{*}{1192} & $+$739 & \multirow{2}{*}{75} & $+$10 & \multirow{2}{*}{3033} & $+$784 & \multirow{2}{*}{36} & $+$9 & \multirow{2}{*}{502} & $+$112 & \multirow{2}{*}{7} & $+$2 & \multirow{2}{*}{58} & $+$38 \\ 					
		     &  & $-$3 &  & $-$355 &  & $-$8 &  & $-$373&  & $-$5 &  & $-$93 &  & $-$3 &  & $-$30 \\ 																								
\multirow{2}{*}{0.1} & \multirow{2}{*}{7} & $+$3 & \multirow{2}{*}{554} & $+$483 & \multirow{2}{*}{30} & $+$4 & \multirow{2}{*}{1588} & $+$316 & \multirow{2}{*}{18} & $+$3 & \multirow{2}{*}{314} & $+$93 & \multirow{2}{*}{3} & $+$1 & \multirow{2}{*}{20} & $+$14 \\ 					
		     &  & $-$3 &  & $-$238 &  & $-$7 &  & $-$499&  & $-$5 &  & $-$108 &  & $-$2 &  & $-$15 \\ 																								
\multirow{2}{*}{0.3} & \multirow{2}{*}{4} & $+$2 & \multirow{2}{*}{516} & $+$367 & \multirow{2}{*}{11} & $+$3 & \multirow{2}{*}{590} & $+$286 & \multirow{2}{*}{1} & $+$4 & \multirow{2}{*}{9} & $+$73 & \multirow{2}{*}{0} & $+$0 & \multirow{2}{*}{0} & $+$0 \\ 						
		     &  & $-$2 &  & $-$259 &  & $-$3 &  & $-$241&  & $-$1 &  & $-$6 &  & $-$0 &  & $-$0 \\ 																									
\multirow{2}{*}{0.5} & \multirow{2}{*}{0} & $+$0 & \multirow{2}{*}{0} & $+$0 & \multirow{2}{*}{0} & $+$0 & \multirow{2}{*}{0} & $+$0 & \multirow{2}{*}{0} & $+$0 & \multirow{2}{*}{0} & $+$0 & \multirow{2}{*}{0} & $+$0 & \multirow{2}{*}{0} & $+$0 \\ 							
		     &  & $-$0 &  & $-$0 &  & $-$0 &  & $-$0&  & $-$0 &  & $-$0 &  & $-$0 &  & $-$0 \\ 																									
\enddata 
\tablecomments{
The K-dwarf raw and weighted metallicity distribution function for different ranges of $|Z|$. This is for the distances between 1.59 and 1.84 kpc.}
\end{deluxetable}